UNIVERSITY of LIMERICK

OLLSCOIL LUIMNIGH

# "Sentient House – Designing for Discourse"

**Student:** Robert Collins

**Supervisor:** Cristiano Storni

MA/Msc in Interactive Media

University of Limerick

Submitted to the University of Limerick

2015

**Declaration**

"Sentient House – Designing for Discourse"

Supervisor: Cristiano Storni

This Thesis is presented in partial fulfillment of the requirements for the degree of Master of Science in Interactive Media.
It is entirely my own work and has not been submitted to any other university or higher education institution, or for any other academic award in this university.
Where use has been made of the work of other people it has been fully acknowledged and fully referenced.

Signature: ______________________________

Robert Collins

28th September, 2015

## Acknowledgements

I would like to take this opportunity to acknowledge and thank the following people who were a ‘critical’ part of this project, both in academia and in life in general:

Cristiano Storni, who introduced me to new and emerging design approaches, fuelled my imagination and expected a lot.
Nick and Caelan Ward, who provided a roof, reality and respite.
Arvo, for always bringing it back to Lego.
Mikael Fernström and Nora O Murchu, for the strong advice when my head was in the clouds.
Siobhan Clancy, for the time and for engaging me socially.
The good people in the class of iMedia 2015, who supported each other and made the experience a better one.

Finally, thanks to my family:

My parents, Bobby & Nancy, who provided a solid foundation for me to build on.
The brother, Sean, who always has my back.
My sister, Colleen, who opened my mind to culture at an early age.
Happy Birthday, Col.
…and Ava, for her curiosity and inspiration. The future is looking bright!

# Contents:

## Index of Illustrations:

# ABSTRACT

The Sentient House project is an investigation into approaches that the artist-designer can take to better involve the public in developing a critical perspective on pervasive technology in the home and the surrounding environment. Using Interaction Design approaches including workshops, surveys, rapid-prototyping and critical thinking, this thesis suggests a framework for developing a more participatory atmosphere for Critical Design.

As the world becomes more connected, and ‘smarter’, citizen’s concerns are being sidelined in favour of rapid progress and solutionism. Many of these initiatives are backed by government and commercial concerns who may not have the public’s best interest at heart.

The designs and approaches generated from this public participation seek to provide an outlet for a more agonistic debate and to develop tools and approaches to engage the public in questioning and addressing how technology affects them in the future.

The outcomes of this research suggest that the public is receptive to a more active involvement in designing their digital future, and that the designer can be a critical component in revealing hidden consequences and alternative pathways for a more transparent and desirable future.

# 1. INTRODUCTION

## 1.1 Research Question

This thesis seeks to develop a better understanding of how to design artefacts and systems that make people question and reflect upon their relationships with the sentient city and the ubicomp future.

Essentially, this is an investigation into how to best use Interaction Design approaches and user participation to help the artist-designer reveal and discuss the less obvious controversies within the technology around us.

## 1.2 The Project

The project is an iterative approach to designing artefacts that will be used as instruments for critical discussion.

The first phase of this project involves the exploration of existing interactive designs that purport to raise questions about design, technology and their relationships to users and the general public. This process serves to provide the designer with grounding in the motivations of these designers and an appreciation of how their creations particularly address their issues, with an eye to producing an effective prototype.

Rather than designing an artefact in isolation - and in keeping with the participatory intent - the next step is to seek inspiration from contemporary issues in society that will help focus on specific areas which may be addressed through design. Initial research into current concerns will be conducted through the use of surveys and social media.

The feedback from these surveys will then be used to help the designer to generate an interactive prototype that will hopefully reflect some of these concerns back to the user. Ideally, this prototype will appear initially as a useful and problem-solving design, but will reveal more controversial elements as it is interacted with and explored. The intention here is to draw people into exploring an attractive, new piece of technology and to allow them to raise their own questions about how they may be affected by its implementation.

In facilitating this exploration and discussion, the Interaction Design methodologies of employing workshops and user interview techniques to engage with the public will be researched and adapted to serve this purpose.

Potential users will then be invited to participate in an initial workshop to experience the design and to provide feedback on how they perceive the prototype's effect on their own lives might be. This process is hoped to allow the designer to better understand how the qualities of the design help to raise issues and questions amongst the participants and to iterate and improve on these for further workshops. The process will also help to refine the structure of the workshop for its next iteration.

Within the scope of this project, it is intended for there to be a series of three workshops which will be refined at each turn, removing less effective activities and introducing new ones as required. Each workshop is expected to generate a richer, more effective version of the initial prototype design and to explore similar technologies and their less obvious effects on the participants and other users.

As a means of documenting any concerns, controversies and issues raised in these workshops, physical and digital visualisation techniques will be explored with the intention of producing an overall picture which may then be used to kick start further discussion and to explore new designs.

## 1.3 Motivation

In this designers own artistic practice, they have produced artefacts that they now know could fit into the description of Critical Design[1] [2]. It is only as a result of this Interaction Design course that they have developed a language to understand and define these artefacts. In the initial investigations into Critical and Speculative Design, a stronger sense of the design areas that would be relevant to this project were found. During the course of this Masters, new areas of design have been introduced – Adversarial, Political, Discursive, Interrogative design, etc. – that have provided more concrete directions and methodologies to apply to this project and to future artistic practice.

The motivation of the project lies in opening up these design areas through workshops by providing an environment for participants to explore the ramifications of interactive technology on their current and future lives. Rather than presenting designs in isolation and expecting their intentions to be understood by an observer, the experience should be a two-way one, where the participant can partake in the process through interaction and the designer can learn from these outcomes.

The workshop approach is also used to help the artist-designer to understand and explore what design qualities will help to support criticism, speculation and engaged discussion. As an additional exercise, the project will explore ways to visualise user concerns, as an on going by-product and link between participants and workshops.

Although this project was initially inspired by the concept of the Smart City and the essays that came from the exhibition 'Toward the Sentient City' (Shepard, 2011), the scope of this thesis necessarily brought the scale from a city-wide project down to a more manageable study of smart technology in the home, or the 'Sentient House'. This concept allowed the designer to use the home to represent a node in the Internet of Things.

---

1 http://clockworks.ie/index.php/portfolio/64-things-to-worry-about/

2 http://clockworks.ie/index.php/portfolio/earthquake-table/

### 1.4 Objectives

Using the model of the 'Sentient House' - a house with embedded technology – it is hoped to develop a process involving workshops and rapid prototyping to generate interactive designs that will effectively reveal the hidden controversies and actors involved in the ubicomp future.

The workshop model will be developed to allow any interested participants to explore their own concerns and help the designer to produce more efficient activities for more useful debate.

Prototypes, artefacts and systems generated from workshops will also allow the designer to develop a practice of more effective Critical and Adversarial Design solutions.

The mapping of the controversies involved will provide a visualisation of the overall issues raised by participants and contribute to further workshop development.

At the end of this project, the designer would expect to have a solid workshop approach, a richer prototype with directions for further prototypes and a mapping of contemporary concerns.

# 2. LITERATURE REVIEW & RESEARCH

## 2.1 Literature

This section will focus on the main texts and artefacts that have guided the direction and focus of this dissertation project. Beginning with a look at some contemporary writings about technological controversies, as a grounding in the rationale for this project. This is followed by a look at Critical Design, how it led to further research into Adversarial Design, and of how these were found to be approaches which were strongly supportive of the project's intentions of designing for discussion. We then take a look at some other relevant design approaches and finish with an introduction to Tomasso Venturini's work on Mapping Controversies as a possible means to help visualise the discussions that are hoped to be elicited through design.

**Technological Concerns and Rationale**

As a parallel to studying design approaches, and as inspiration for questioning technology, 'To Save Everything, Click Here' (Morozov, E., 2013) takes a very critical view of 'internet-centrism' and 'solutionism' as dangerous and infectious attitudes that result in the surrendering of rights in return for technological convenience. "*Once laws and norms become cast in technology, they become harder to question and revise*" (Morozov, 2013).

Bringing things back to a participatory arena, a quote from Roger Brownsword gives a pointer towards the need to design artefacts which help to create a discussion and to reveal the digital hegemony: "*Moral communities need to keep debating their commitments. In such a community, it is fine to be a passive techno-managed regulatee, but active moral citizenship is also required*".

Another quote raised by Morozov - "*If people are denied any autonomy, then they perceive that the moral responsibility lies entirely with the system, and they no longer retain any obligations themselves.*" (Smith, D, 2000) - supports the need for citizens to take stock of ubiquitous technology and gain a critical understanding of its benefits and dangers before it becomes harder to question.

In his essay, Toward The Sentient City (Sentient City, 2011), Mark Shepard discusses the ubiquity of technology in the modern city. As we move through the city, we leave trails of digital information in our paths. From checking in on Facebook, to taking journeys on public transport, to using ATMs and purchasing from shops, each isolated interaction begins to form an overall picture of our habits, desires and possible future tendencies. Shepard talks of the "computational-systems" which operate on this metadata using neural network algorithms which seek to find ways to use this data for commercial and security reasons. Here he introduces the idea of a 'Sentient City': *"...we have urban systems and infrastructures that take on a quality of what might be best described as 'sentience' – not quite the 'smart city' we've been promised by techno-evangelists, yet not exactly dumb either".*

This is an environment where every digital interaction comes loaded with implications:

- Will my spending history affect my credit rating? (Online banking)
- Does my amount of socialising and late nights affect my employability? (Check ins)
- Does my purchase history make me look like someone I am not? (Profiling)
- Etc…

Without questioning this environment, these algorithms are trusted to root out the ‘terrorists’ and provide the consumer with advertisement for just what they want to purchase next. But, as Shepard says (ibid., p.31), *“…a sentient city is one that is able to hear and feel things happening within it, yet doesn’t necessarily know anything in particular about them. It feels you, but doesn’t necessarily know you”.* It is not your friend, but it knows a lot about you. Citizens are the variables which can be monetised, policed and generally controlled. The Sentient City may not necessarily have the individuals best interests in mind.

These themes of concern surrounding the acceptance of ubiquitous computing in our society, without fully appreciating the longer term effects, are some of the controversies this project seeks to explore through design and participation.

**Critical Design**

Hertzian Tales (Dunne, A., 1999) provided an introduction to the foundations of Critical Design, which led on to further explorations with Design Noir (Dunne & Raby, 2001) and Speculative Everything (Dunne & Raby, 2013). Although, by their own admission, examples of Critical Design have been around for some time – Italian Radical Design of the 1960’s is often cited as an example – Anthony Dunne and Fiona Raby have done much to make this field more visible and tangible.

In their Critical Design FAQ[3], they define this approach as follows:

*"Critical Design uses speculative design proposals to challenge narrow assumptions, preconceptions and givens about the role products play in everyday life. It is more of an attitude than anything else, a position rather than a method. There are many people doing this who have never heard of the term critical design and who have their own way of describing what they do. Naming it Critical Design is simply a useful way of making this activity more visible and subject to discussion and debate."*

Further to this, when describing what Critical Design is for, they add:
*"Mainly to make us think. But also raising awareness, exposing assumptions, provoking action, sparking debate, even entertaining in an intellectual sort of way, like literature or film."*

Essentially, Critical Design takes a position of designing artefacts that try to raise questions in the observer/users mind, as contrasted with 'traditional' design that mostly designs as a means to solve problems and provide solutions.

[3] http://www.dunneandraby.co.uk/content/bydandr/13/0

This series of books introduced a rich history of critical designs from Japanese chindogu[4] through to Dunne & Raby's surreal imaginings of possible futures.

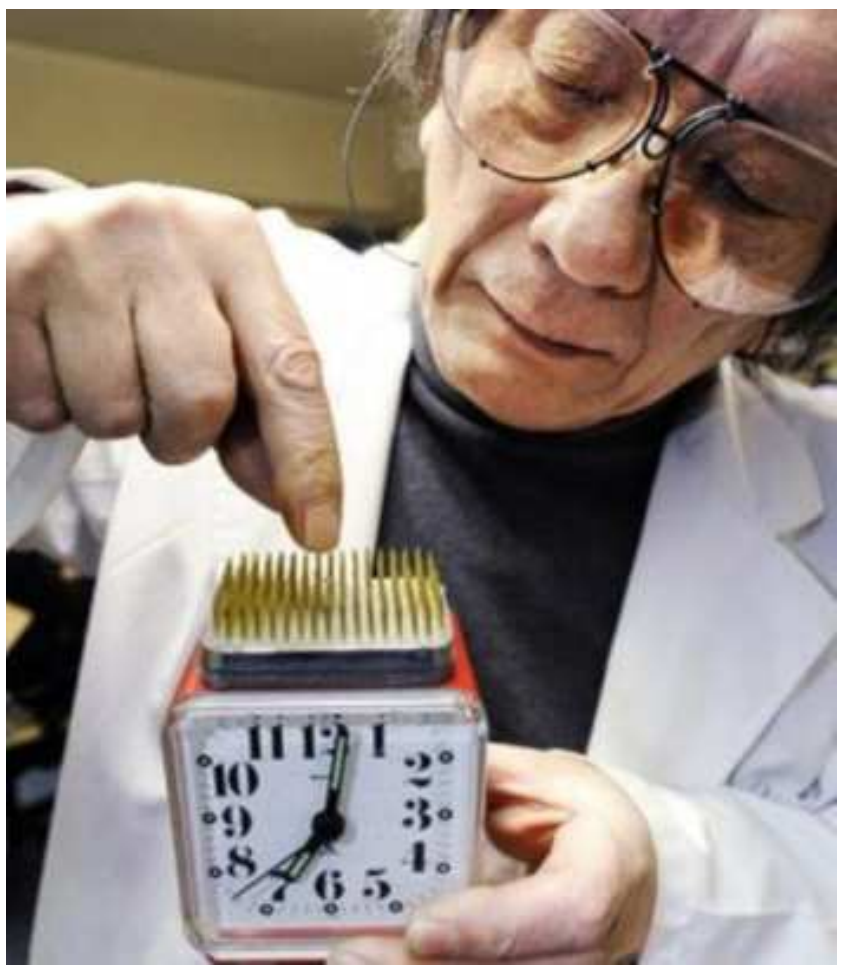

Figure 1: Chindogu - Painful Snooze (Source: pixelpinch)

In the chindogu example (pictured) we see a basic alarm clock modified to have a bed of sharp pins protruding from its snooze button. Here we see a humorous, and perhaps clumsy, design intervention that attempts to address the problem of oversleeping. Although appearing as a novelty item, the fact that such a design exists can raise questions about human nature, social norms and even the practicalities of the 9-5 working day.

Taking an example from Dunne and Raby, their more direct approach to Critical Design can be appreciated in their Technological Dreams Series, No. 1: Robots project from 2007 (pictured). Here, the designers have set out to investigate domestic robots and to explore alternative emotional interactions that they may have with their human users. In this series of four robot designs, they look at possible ways in which robots may present themselves; from nervousness and distrust, to introversion and neediness. These designs immediately raise questions about a robotic future; initially by their physical designs, which are contrary to the expected forms of robots, and secondly with their distinctive personalities which demand unexpected interactions. Technological Dreams provided much inspiration for the prototype in this project; presenting the user with a

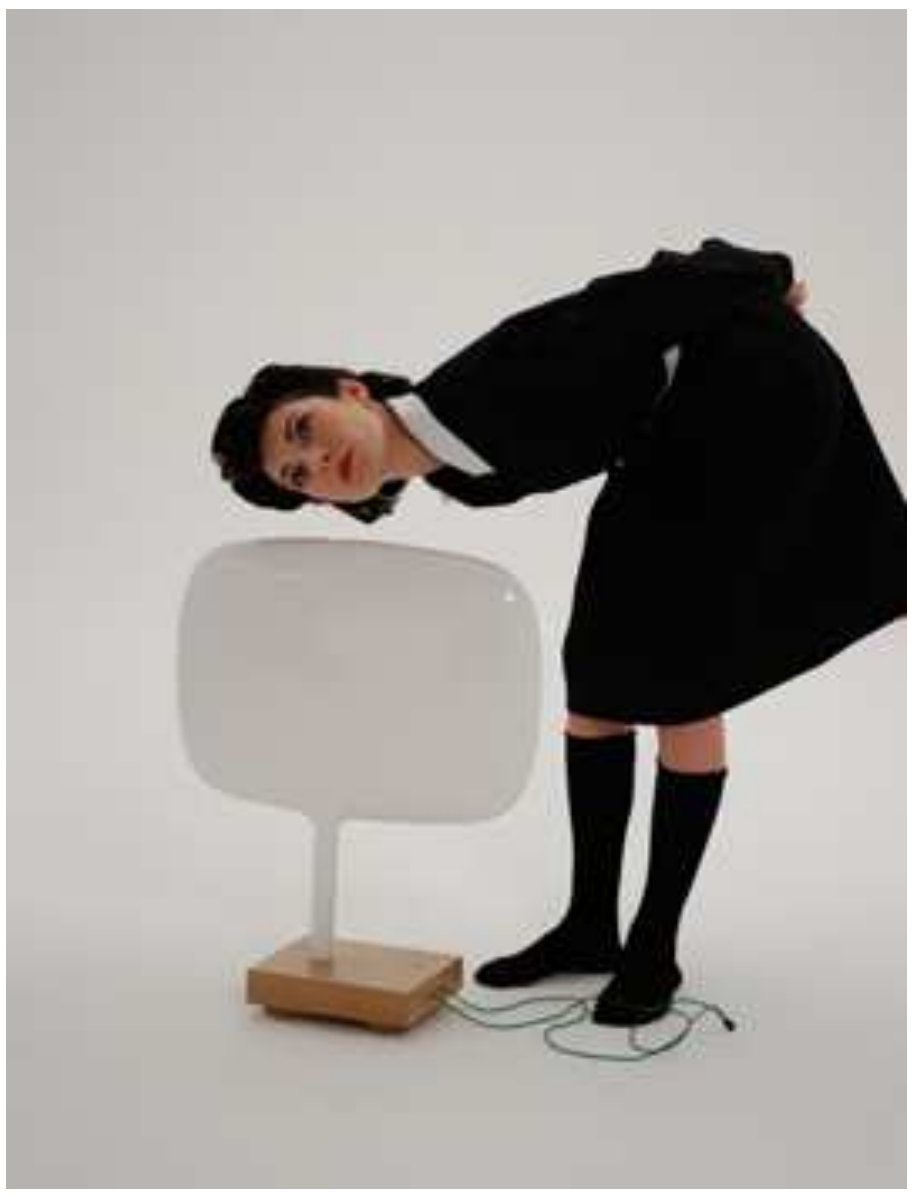

Figure 2: Technological Dreams, No. 1, by Dunne & Raby (2007) (Source: dunneandraby.co.uk)

4 http://en.wikipedia.org/wiki/Chindōgu

somewhat familiar interface that responds in ways that challenges their assumptions about the role of technology in their life.

Although revealing and encouraging, many of the included design examples in these books tended to exist in a certain clinical and rarefied atmosphere which seemed to somewhat exclude elements of more direct participation. Though designed to raise questions and to be contentious, there appeared to be space for a possible design intervention to gain access to these reflections and to make them more public.

Critical design seemed to be akin to avant-garde music in the sense that it can forge ahead into speculation and experimentalism, but its benefits may only trickle back into the mainstream at an unknown future point.

**Adversarial Design**

The concepts introduced in Adversarial Design (DiSalvo, C., 2012) seemed to offer a stronger critical design model that engaged with people on a more direct level. The examples and methodologies resonated with this project's interest in designing for participation (interaction and workshops) and also in creating artefacts and systems to inspire a questioning and discursive frame of mind. Adversarial Design, as DiSalvo sees it, is a form of political design which does the work of agonism:
*"What is agonism? It is a political theory that emphasises the positive aspects of certain forms of political conflict [...], it provides opportunities to participate in disputes over values, beliefs and desires; and it models alternate socio-material configurations that demonstrate possible futures."* (DiSalvo, Critical Making, 2012).

Here is a methodology that seeks to interact directly with people through designed artefacts and let people explore their concerns without attempting to polarise the conversation or the public. Agonism can be a space of mutual respect where views and biases may be discussed and considered rather than an antagonistic space where the public are divided on political lines to the detriment of solutions and progress.

DiSalvo speaks of revealing hegemony, which he defines as "*exposing and documenting the forces of influence in society and the means by which social manipulation occurs*". From here he provides instances where computational design can be used to create a space for agonism and to expose the hidden actors and systems behind our digital world.

Like Don Norman's definition of affordance, this book gives the impression that such agonistic adversity may not necessarily be designed for, but may be appreciated after the fact. What I mean by this is that although the concept of Adversarial Design has only been raised in the last few years, there are already many examples cited dating back to the start of the century[5]. Based on this, I feel that designing with an intent for creating a space for agonism is new ground and an area which should be focused on and drawn out.

Million Dollar Blocks[6] is a project that is cited by DiSalvo as a strong example of Adversarial Design. Devised by the Spatial Information Design Lab at Colombia University and the Justice Mapping Centre[7] it provided a way to look at a contemporary social issue "which reveals previously obscured configurations in the cycle of crime and incarceration"(DiSalvo, 2010).

---

5 https://en.wikipedia.org/wiki/Adversarial_Design#Examples

6 http://www.spatialinformationdesignlab.org/projects/million-dollar-blocks

7 http://www.justicemapping.org/

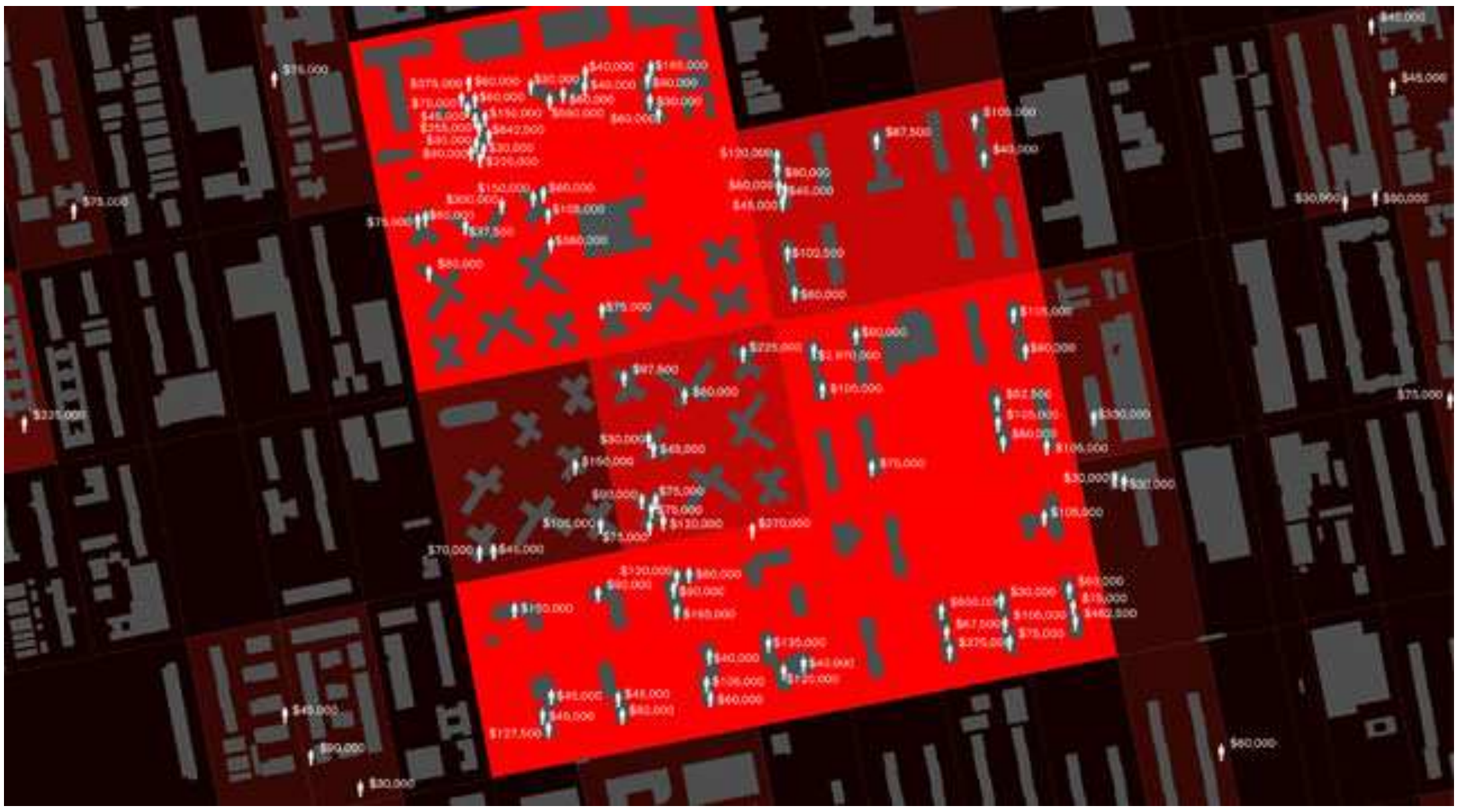

**Figure 3: Million Dollar Blocks visualisation detail (Source: SIDL)**

For a short time, the Criminal Justice System in the US made available data sets from their incarceration program to see what visualisation would be generated by interested parties. Whereas most visualisations concentrated on the effects and geography of the crime data, SIDL turned the data on its head and drew attention to the injustices and misdirection of money within the system. Rather than provide 'heat-maps' of high crime areas – which would only be of use to certain strata of society and commerce – they produced a visualisation which suggested better ways of spending tax dollars on incarcerating criminals, and they did so in a very stark and clear way.

Million Dollar Blocks showed the importance of stepping back and considering the wider picture of the data provided, and to challenge the assumptions that come with many design approaches. It also provides an illustration of the difference between Design for Politics – "*design tools and thinking to increase civic participation by making interactions between the…government and its citizens more understandable, efficient and trustworthy*" (AIGA, 2008) – and the Political Design of this project which "*does not work to support and improve existing means of governance. Rather, it strives to critically investigate an issue and raise questions concerning the conditions of that issue*" (DiSalvo, 2012).

It is only in the last chapter that DiSalvo gives thought to ‘Adversarial Design as Inquiry and Practice’. Though exploration into these areas, to the same degree as he gave to the examples would have been appreciated, he does briefly touch on Participatory Practice. On the penultimate page of the book DiSalvo(2012) states that *“…we could also consider the construction of tools and methods for eliciting and supporting a participatory approach to adversarial design*” and “*the lessons learned from this discussion of adversarial design can inform a participatory practice of adversarial design*”.

These closing points helped provide the project with a direction to take in bringing the isolationism of critical design and the inclusive possibilities of adversarial design into a participatory process. Where Critical Design often finds itself in white gallery spaces, Adversarial Design has more of an affinity with less clinical spaces where artefacts are handled, used, examined and debated over in a more visceral way. This crossing of approaches provides the designer with the space to let their imagination move into speculative and fictional designs and also to be able to bring these concepts back down to earth through public participation and iteration. As we move into contemporary issues and technological controversies, these design methodologies may provide a strong practice for the investigation of hegemony and hidden actors within emerging interaction technology.

**Participatory Design**

As part of the collaborative approaches to design, Participatory Design (PD) can be seen as "*…a shift in attitude from designing for users to one of designing with users. It is the belief that all people have something to offer to the design process and that they can be both articulate and creative when given appropriate tools with which to express themselves*"(Sanders, E., 2002).

This approach resonates with this project's desire to involve the public in, at least, the process of exploring and suggesting design possibilities, and also to incorporate a respect for the user which relates well to the agonistic intentions of Adversarial Design. Participatory Design is perceived as a relevant research area due to its inclusivity and its constant search for new and effective methods for bringing affected parties to the design 'table'. Of particular interest is this approach's focus on workshopping, creativity and a certain degree of playfulness.

Although Participatory Design can, and does, involve itself in "forming public agonistic spaces" (Erling Björgvinsson, Pelle Ehn & Per-Anders Hillgren, 2012), for the time being, this project has more of an interest in seeking instruction and inspiration from its methods of engaging the public through workshops and discussion.

Having been inspired by Elizabeth Sanders' outlook on the empowering tenets of PD, her (and Pieter Jan Stappers') book – Convivial Toolbox (2013) – provided a rich source of examples and ideas for designing and running workshops for a variety of cases. Although none of the case studies within the book directly related to this project's needs, the practical guidelines in the 'How To' chapters helped to prepare the facilitator for many of the unexpected pitfalls in the run up and execution of a successful workshop. The richness of illustrations and photographs of materials alone provided encouragement to experiment and ensured that this book was kept on hand throughout this project.

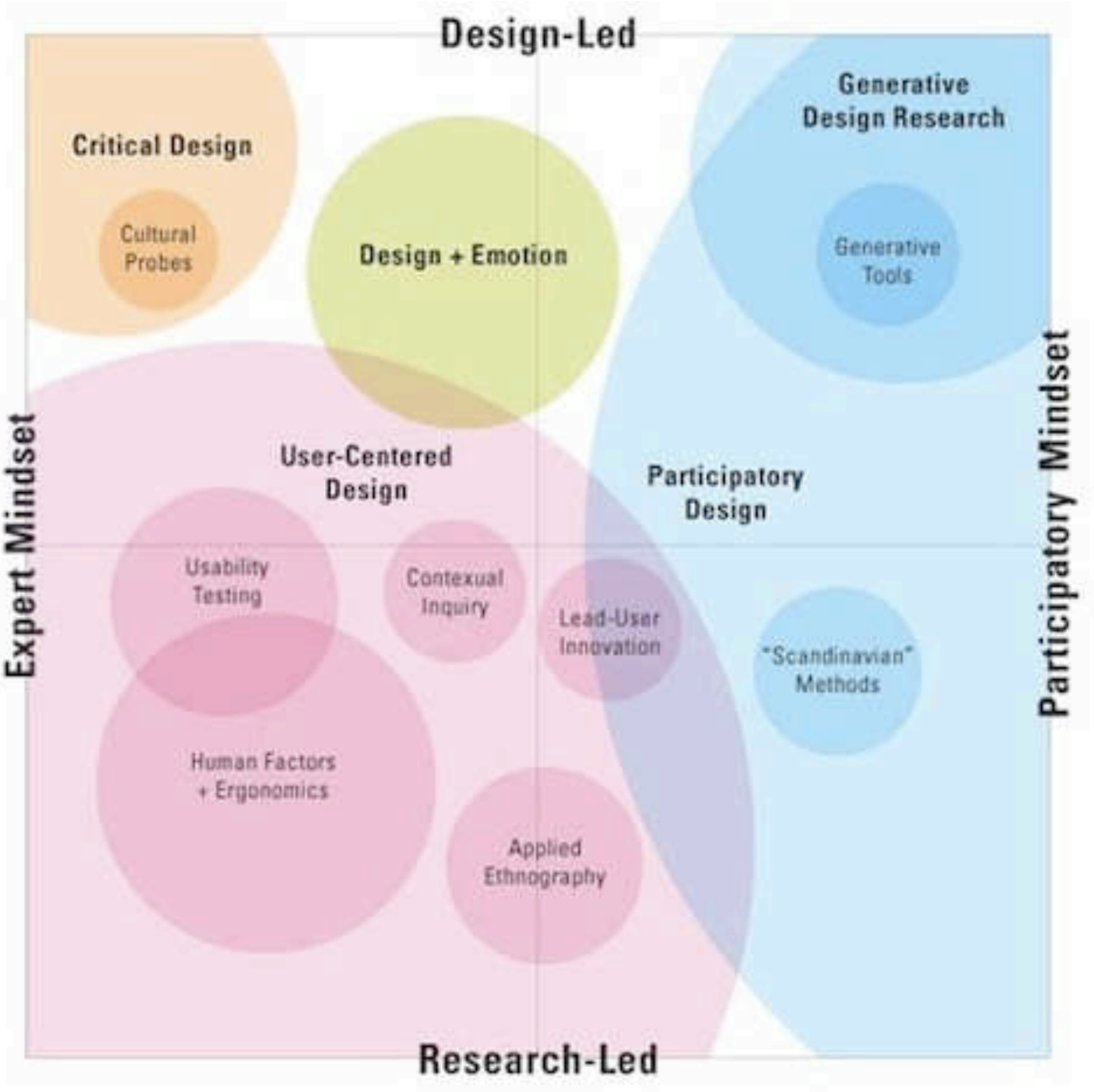

**Figure 4: Emerging Landscape of Design (Sanders & Stapper, 2014)**

Of particular interest was the 'emerging landscape' of design diagram, (Figure 3), (Sanders & Stappers, 2014). This helped in understanding the positioning of both Critical and Participatory Designs in relation to each other. As discussed in the 'Motivation' section, this project sought to address the isolation of Critical Design and to build a bridge over to Participatory Design. In one sense, they both existed on the 'Design-led' side of the diagram, which was encouraging; but their separation on the 'expert' and 'participatory' mind-sets helped to prepare the reader to be open to compromise in this endeavour; particularly with regard to having too-high expectations or over-reliance on participants to do the work of the designer.

The paper, *Constructing and Constraining Participation in Participatory Arts and HCI* (Holmer, H., et al, 2015) took a look at Social Practice Art and how artists' pursuits in public engagement can inform human-computer interaction studies. The researchers selected three participatory artists (and in the writers opinion: critical designers), as case studies, who were presenting projects at the 2010 01SJ[8] technology and art festival in San Jose, California. The aim of the festival was to "*create an experience where audience members could participate in a hands-on way in workshops, building and experimenting with technology alongside the artists*." (ibid.)

The researchers showed how the participation in these public workshop environments rarely rarely reached the saturation or involvement that was suggested or expected; and that just because a workshop exists, does not necessarily mean it will be successful – "*As we endeavour to engage with the public…we come face to face with the assumptions, limitations, and possibilities of our ideas*."(ibid.). They go on to say that the practice of engaging the public in the design processs is "*a practice that needs to be regularly revisited, renewed and revamped*".

In reflecting on the participatory nature of this project, the findings of this paper highlighted how new this territory is for HCI practitioners and that to embrace this approach fully, with expectations of solid data, would be naïve. On the other hand it does suggest that such participation is important to design progress and is a viable area for research, experimenting and iteration.

[8] http://01sj.org/

**Design Fictions**

Design Fictions, can be considered as a subset of Critical Design in that it seeks to produce artefacts that question and create debate, but with a narrative or story-based quality. This allows the designer to create a whole world or parallel reality in which to design products for. The approach can help to make the future, or alternative histories, of design appear more tangible and real to the observer. Within this project, this approach has helped the designer to more clearly imagine an 'Adversarial House' as containing an ecosystem of technologies that question the relationship we have with our homes.

Design Fictions have a strong parallel to Science Fiction, and as writer Bruce Sterling observes (Sterling, 2009), the use of Design Fiction serves to strengthen the believability of his own writing and in return provides a believable world that can be explored through design.

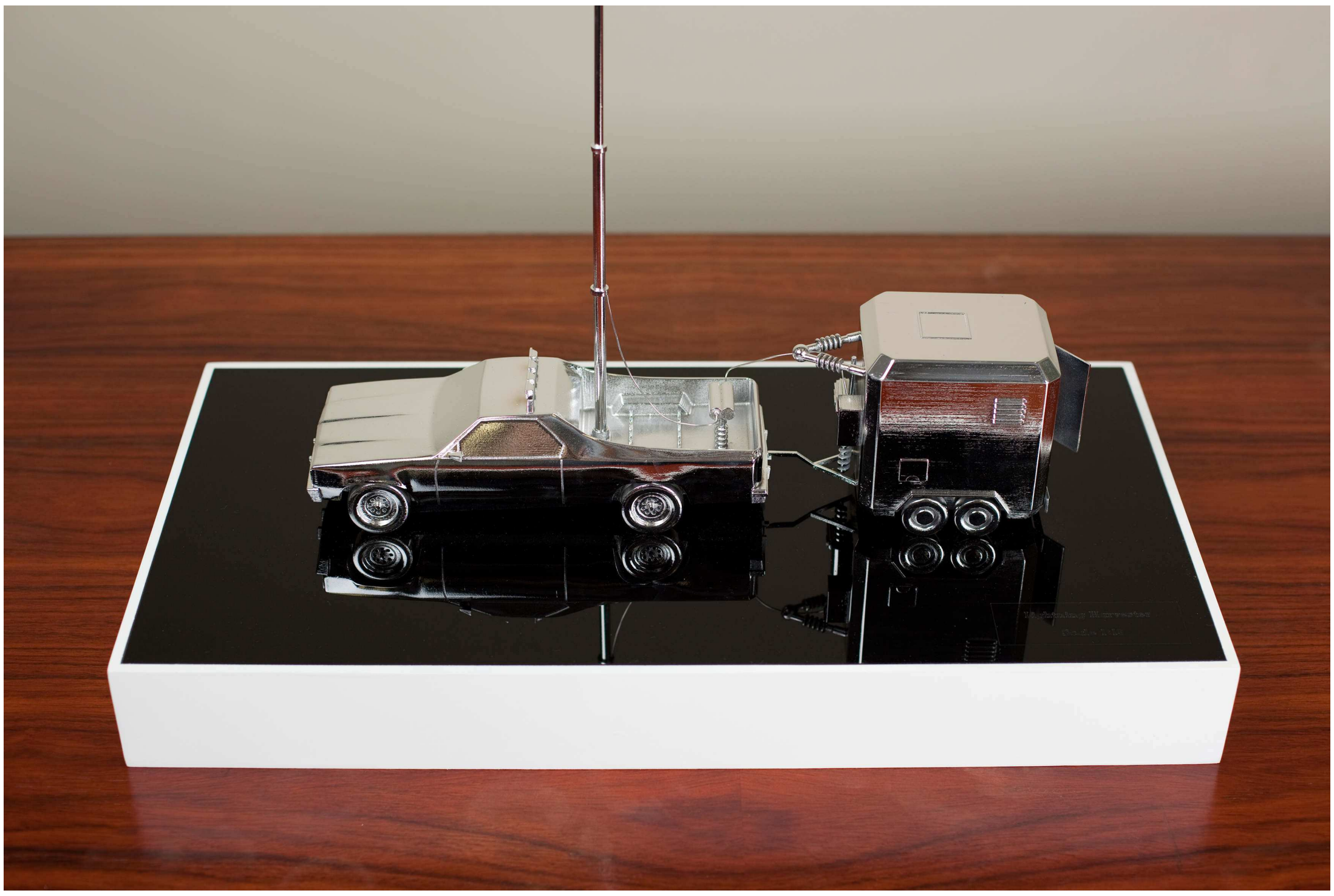

Figure 5: The Golden Institute, Mobile Energy Harvester (Source: http://pohflepp.net/)

An excellent illustration of this approach is The Golden Institute[9] by Sascha Pohflepp. This project takes a point during the 1980 presidential election contest between Carter and Reagan and imagines an alternate future where the more 'renewable energy'-friendly Carter triumphed. Pohflepp then imagines a series of design interventions that may work in both our current and the alternate reality, but they just seem to function better in this other world.

[9] http://pohflepp.net/The-Golden-Institute

**Mapping**

Finally, as a means to link the series of workshops that will be running during the Summer, Tomasso Venturini's work on Mapping Controversies (2012 & 2015) using actor-network theory (Venturini, T, 2010) are being considered as an interesting approach to documenting some of the outcomes. In these readings, Venturini uses his experiences as Bruno Latours' teaching assistant to document and introduce some of the main techniques of Latour's 'cartography of controversies' and its relationship with Actor-Network Theory (ANT). These techniques describe ways to explore and visualise social debate around techno-scientific areas. ANT considers both humans and non-human (objects) as equal parts of their social network, with both having the capacity to participate and influence these systems. This perspective allows for the designed artefacts and their influence on other human and non-human actors to be observed and documented as a whole.

As part of this projects workshop elements, there is expected to be a large amount of controversies and standpoints being generated and these techniques seem well suited to modelling and visualising their relationships. The EU funded research consortium MACOSPOL(Mapping Controversies in Science for Politics) defines a controversy as: "*every bit of science and technology which is not yet stabilised, closed or "black boxed"… we use it as a general term to describe shared uncertainty*" (Macospol, 2007). These shared uncertainties are expected to arise from workshopping with critical and adversarial designs.

Venturini's work on Mapping Controversies has a general sense of experimentation and uncertainty about it. Indeed, he mentions that, "*…controversies mapping entails no conceptual assumptions and requires no methodological protocols. There are no definitions to learn; no premises to honour; no hypothesis to demonstrate; no procedure to follow; no correlations to establish.*"(Venturini, 2010). There is a sense that adaptability and novel approaches may be important skills that the mapper must develop. This route will be an experimental one involving an open approach to visualisation solutions. This mapping will be an on-going project over the course of the series of workshops that will be built upon with generated data.

## 2.3 Inspirational Projects

Alongside this literature research, an eye has been kept open for other interesting and influential projects across the internet. Studying other works in these areas has helped to get a sense of the current state of critical and adversarial design in its practical sense. Over the year, examples have been gathered and documented on a specially created blog[10] and used as an inspirational notebook to return to for ideas and encouragement. This section will introduce a few of these projects that particularly struck a chord with this project.

### Brad the Toaster – Simone Rebaudengo & Haque Design

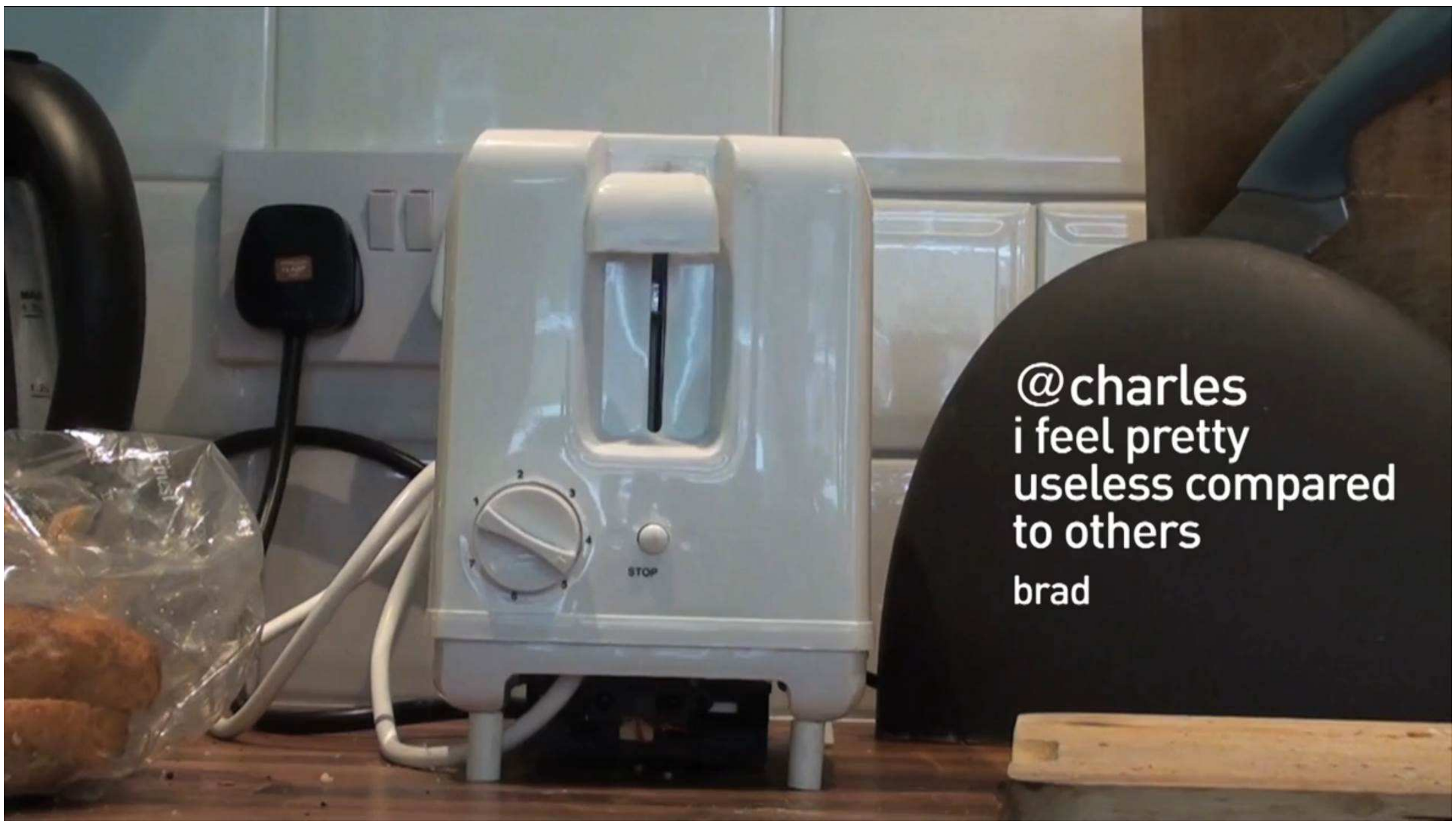

Figure 6: Brad the Toaster Tweeting dissatisfaction (Source: addictedproducts.com)

www.addictedproducts.com

Brad the Toaster was a direct inspiration for the initial prototype, a smart door lock. This toaster values its own function so much that if it deems itself underused, it will send complaining tweets about its current 'host' and may even ship itself to a more deserving client. This project examines ownership of products and how we may need to work harder to deserve our technological luxuries in a world of scarcity and sustainability.

---

10 http://www.youarenotbeautiful.com

This toaster undermines the usual relationship between the user and the object and challenges the idea that we really own our investments. As a means of revealing the hidden factors in our relationships with technology, it is particularly effective. By elevating the status of the toaster to an almost sentient device it brings the user into a more direct and sometimes adversarial position. The user must consider the needs of the toaster, let go of some of their personal sovereignty, consider how and why it is used, or possibly lose it forever.

**Natural Fuse – Usman Haque**

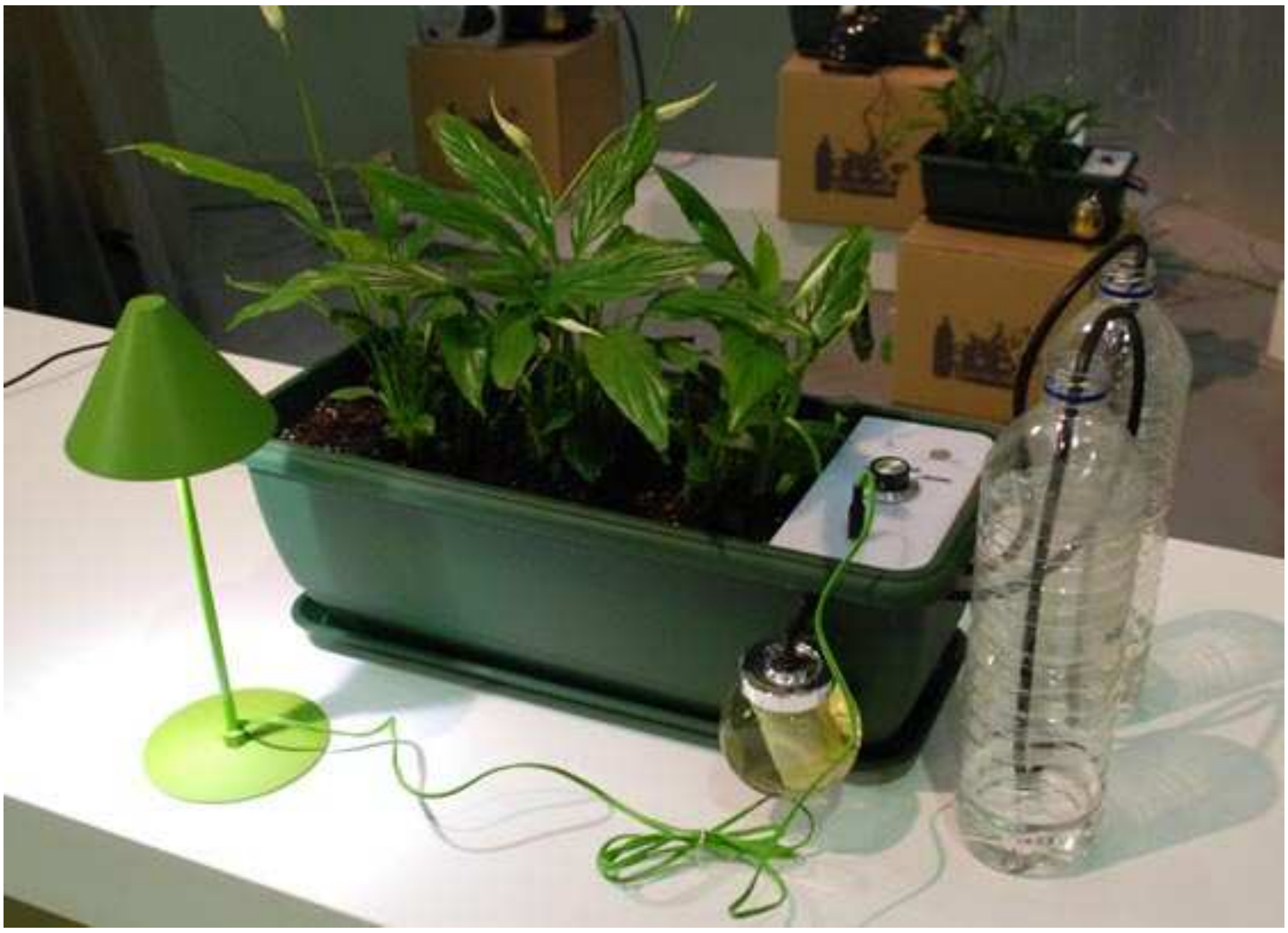

Figure 7: Natural Fuse 'node' (Source: naturalfuse.org)

www.naturalfuse.org

Natural Fuse is a classic Adversarial design example. Using a networked model of the global carbon footprint system, it engenders a certain level of responsibility in participants for their own energy use and how it affects other unseen people. The Natural Fuse was a project that used plants as 'fuses' based on your use of an electrical appliance. It encourages people to be cooperatively aware of their energy expenditure using networked devices that keep the plants alive. The less energy you use, the more the system takes care of the plants. If you overuse your appliance, a random plant may have to be killed off, affecting you or others on your network.

It was found to be an excellent example mainly for the amount of levels it works on; from the individual working alone in his domicile, all the way up to a global network of countries struggling with energy use, politics and climate change. It is only in engaging with this design that the user can begin to perceive the systems that we rely on to provide us with energy for our domestic appliances.

**Drone Aviary – Superflux**

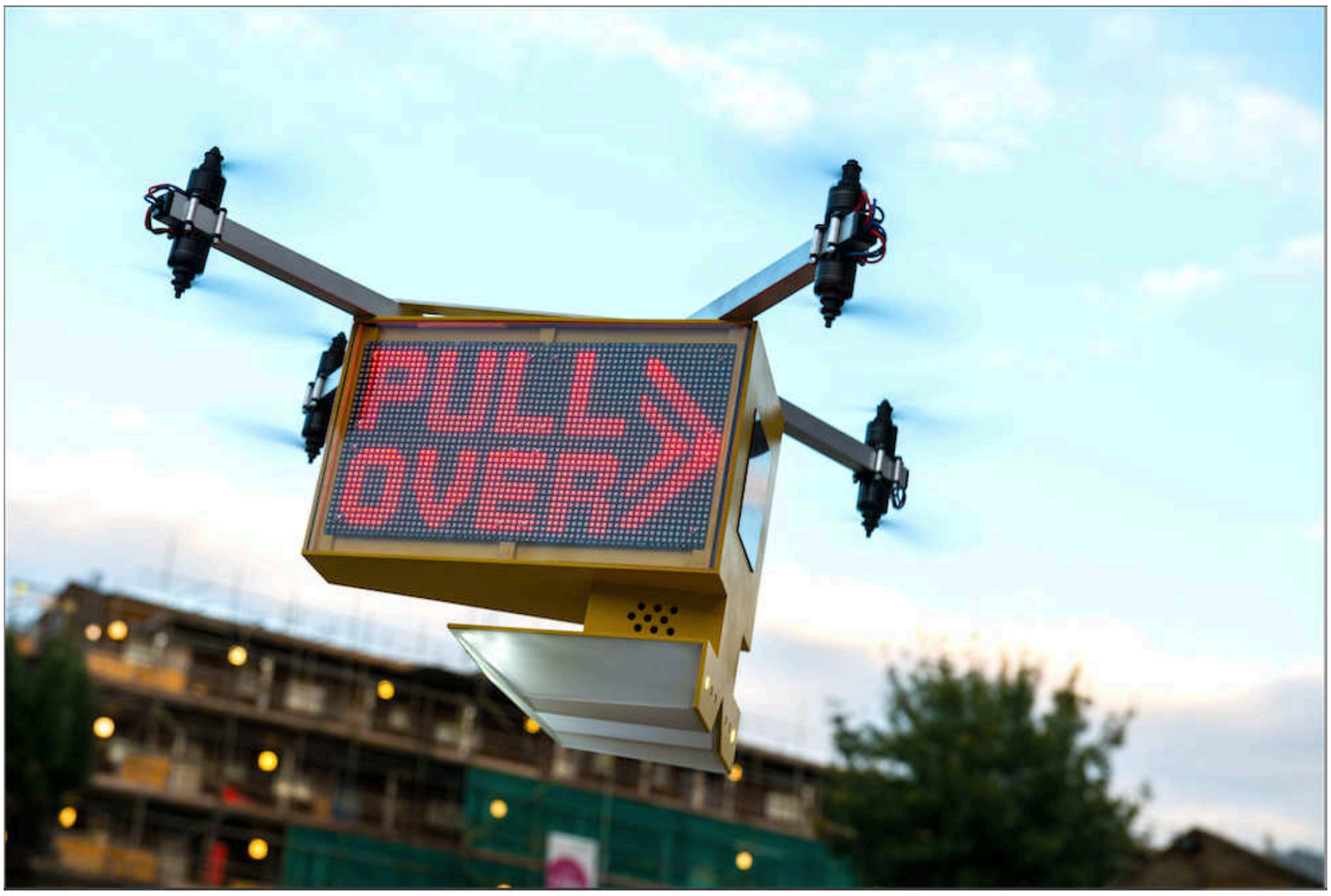

Figure 8: Police Drone (Source: Superflux)

www.superflux.in/blog/the-drone-aviary

The Drone Aviary is a good example of research by design. Not only were concepts explored, but actual high-fidelity prototypes were built and used to explore a variety of possibilities. The level to which Superflux went to with their designs creates an intriguing project which attracts and involves everyone in the discussion.

As a contemporary concern, drones are ranking high in the general public's minds. They have been the new face of warfare for more than ten years and are now, apparently, coming to our own backyards. Superflux tap into this by examining how drone and UAV technology might be exploited by commercial and government interests as part of the networked smart city. Although the different designs do not entirely live up to their descriptions, they have prototyped to a high enough level to make their existence palpable and, therefore, easier to debate and discuss.

At the advent of a new technology, it shows that we do not have to wait for its effect to become apparent before being able to understand and study it; rather we can use prototyping and speculation to look into the future.

**Sentient City Survival Kit – Mark Shepard**

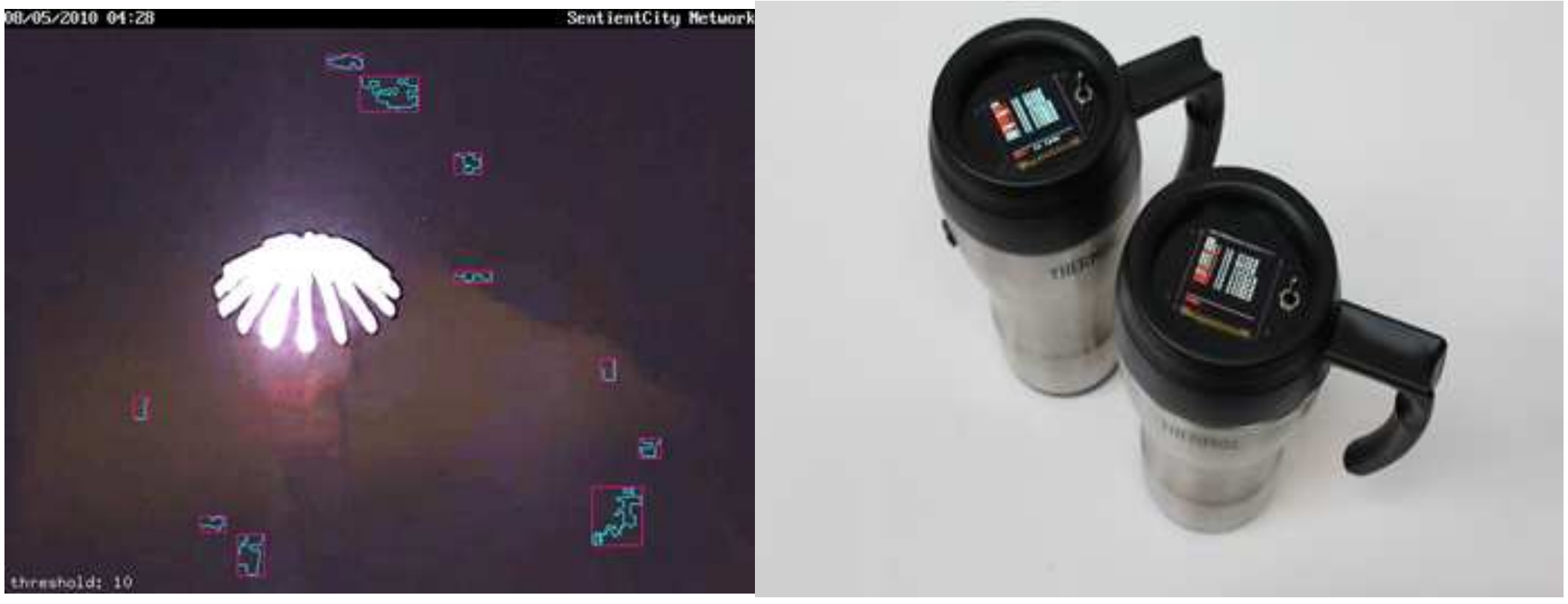

**Figure 9: CCD-me-not Umbrella & Ad-Hoc Network Travel Mugs (Source: sentientcity.net)**

*"The Sentient City Survival Kit probes the social, cultural and political implications of ubiquitous computing for urban environments. The project consists of a collection of artefacts for survival in the near-future sentient city."*
*- Mark Shepard (http://survival.sentientcity.net/)*

The Sentient City Survival Kit is a series of design interventions that highlight the surveillance aspects of the smart city. Shepard enters a speculative future where products need to be designed to allow people to take control of this environment and subvert it for their own privacy needs. These devices are designed to allow the user to escape from the feeling of being observed, while going about their daily lives. The project consists of four devices (so far); an umbrella which disrupts surveillance cameras with infra-red light, a set of wifi embedded travel mugs which have their own private communication networks, underwear which senses the presence of near-field communication sensors and a navigation app which takes the user on a more scenic route.

Explorations such as these provide a counterpoint to the optimism with which new technology is often presented. When raising concerns about ubiquitous computing, providing alternatives and ways for the public to regain control and challenge these initiatives are just as important as pointing them out in the first place.

**Blendie – Kelly Dobson**

Drawing on DiSalvo's own illustrations of Adversarial Design, consider *Blendie* (Kelly Dobson, 2007).

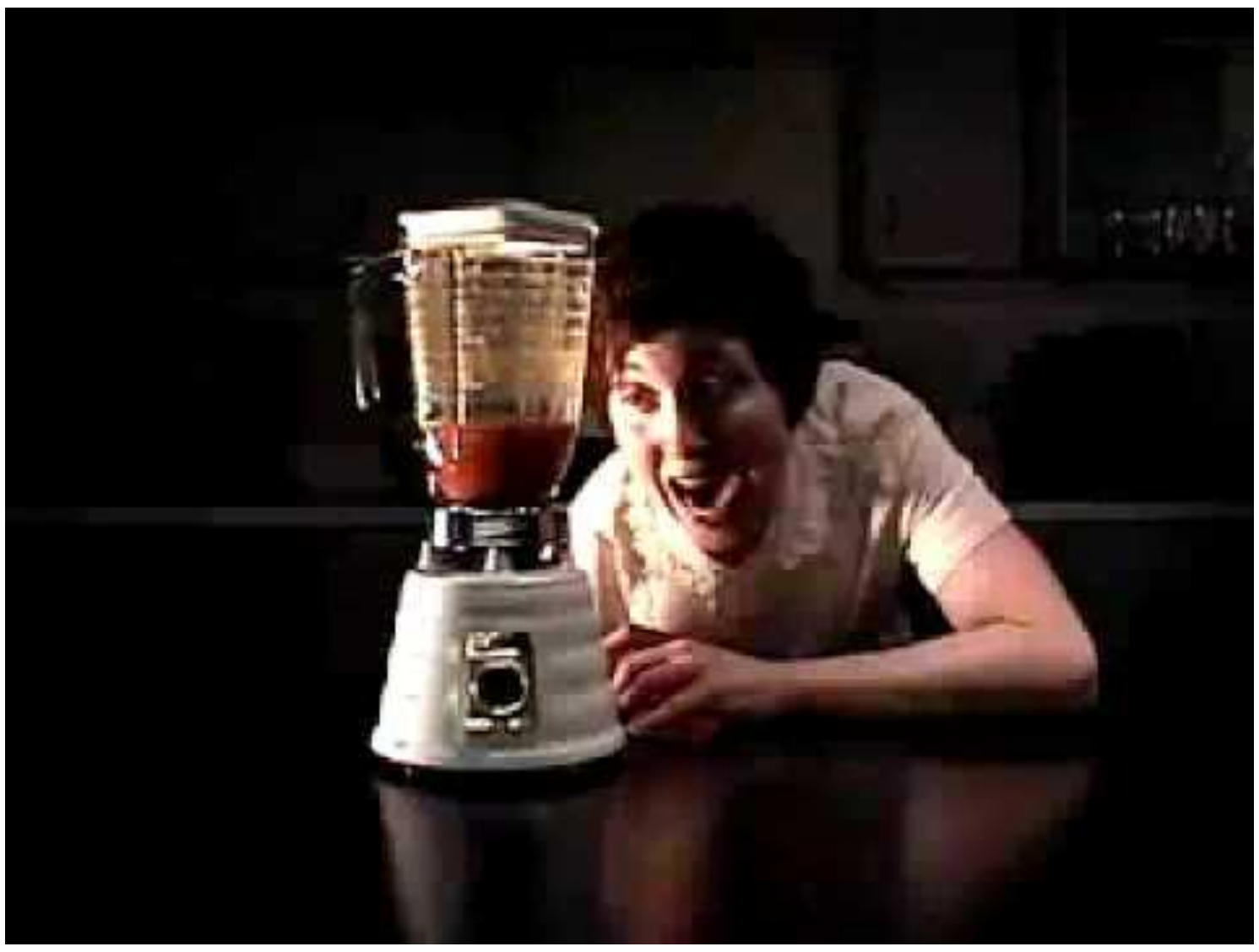

Figure 10: Blendie, by Kelly Dobson (2007) (Source: MIT)

*Blendie* is a domestic kitchen blender appliance which will only work if a user vocally mimics the sound of a blender. A higher pitch will make it blend faster and a lower pitch will run it slower. The effect is to bring the user into a sonic unison with the machine as it responds with its own motor noises. It is a dialogue between the user and machine where the user must embody machine qualities to make use of it, and the machine encourages this with its own 'voice'. A sort of feedback loop that challenges attitudes towards technology by direct participation.

This particular artefact showed how effective a strong interactive prototype helps to engage the public and involve them directly in the designer's intentions.

# 3. METHODOLOGY

## 3.1 Introduction

This chapter will provide a general summary of the expected process and the reasons behind these decisions. In the following chapters, the actual process, iterations and outcomes will be dealt with in greater detail and in chronological order.

## 3.2 Methodology Overview

As an initial task, the expectations and outcomes of the project needed to be reassessed, so that effective methodologies could be brought into use:

- An effective interactive prototype which would resonate with contemporary concerns surrounding ubiquitous computing .
- A means of engaging with the public that would provide a platform for discussing and exploring these issues.
- An exploration of contemporary smart technology and the generation of ideas for further prototypes that could examine these further.
- A way to visualise and map the outcomes of these engagements.

At this stage of the project, although there was an area that the project intended to focus on (ubicomp, smart/sentient cities), a means of public engagement was needed.

Based on experiences with User Centred Design and its participatory aspects, a good appreciation of the power of involving users and the general public in the design process had been gained. The use of cultural probes by William Gaver (Gaver, W., et al, 1999) was particularly interesting, but far too time-consuming within the scope of this project. As a more immediate alternative, it was decided to consider online surveys as a faster way to generate initial feedback for a subject matter to focus on.

Starting back in March, a basic survey was compiled and sent out on social media that sought to provide inspiration for an initial prototype of a design that addressed peoples' general concerns.

The survey consisted of three categories – interests, concerns and cares – and was designed to be quick to complete. A more detailed discussion about this survey can be found in Chapter 4.

After parsing the data, it was deemed relevant to focus on the more prominent financial and housing worries which participants indicated. This direction was also influenced by our interest in ubiquitous computing and how it will integrate itself within the home.

After some brainstorming and research of interesting projects and designs, the idea of a smart door lock was eventually teased out. A mid-fidelity prototype, a presentation and a video prototype were then produced as a means to convey the basic concepts. The lock itself was particularly designed to raise a number of controversies around ownership of property and the individual's role in the sentient city. This process and the prototype will be discussed in greater detail in Chapter 5.

Using these initial prototyped materials as a discussion topic in one-to-one sessions, it was determined that a larger group-based workshop would be the best way to explore this prototype further. This would help to get away from the critical design model of a single designer or design house producing such artefacts and simply presenting them in isolation for assessment. From experience of User Centred Design focus groups, and how effective participation can be in the design process, ways to help people invest in and explore this methodology were investigated.

At this point, with the first prototype ready, initial workshop design was assessed to see what activities, exercises and discussions would work best with the participants in exploring controversy and generating new ideas and designs. Much use was made of the experience and knowledge of acquaintances and local academics that already had experience in designing workshops for a variety of purposes.

As a starting point, an initial workshop structure was designed as follows:

- Introduction: A short talk about my agenda and role in the workshop.
- Video Prototype: viewing of prototype(s).
- Reactions & Discussion: an open forum for initial impressions and discussion.
- Role Play & Scenarios: Participants identify the actors involved, their own positions within this network and try to role-play from different perspectives.
- Brainstorming: working with a variety of materials to visualise effects within the ‘adversarial house’ model. Generating further scenarios, ramifications and designs.
- Finally, a follow-up question and discussion session.

It was hoped to run at least three of these workshops at various sites over the summer, using each workshop to generate new designs and refinements.

The outcomes and ideas that may be generated will be used to explore and provide data for visualisations and mapping. This may be shared online as a link between workshops and participants.

# 4. INITIAL RESEARCH

In this chapter I outline the use of an online survey solution (SurveyMonkey) to elicit a broad response from the public regarding contemporary concerns, in an effort to generate raw material for my initial design ideas. I talk about experiences with the survey platform itself, best uses for framing questions and how the responses were treated and visualised.

## 4.1 Online Survey

As an experiment to try to understand the general concerns and controversies amongst a broad cross-section of people, a short survey[11] was devised to elicit quick and short responses to simple questions.

The survey was designed to be shareable on social media and to be easy and quick to complete. The main concern was to do with experiences with other online surveys that expect a little too much from the user and often remain incomplete as a result. SurveyMonkey was chosen as the platform for the survey, mainly due to having experienced it as a user on a regular basis and from of seeing it advertised before.

To prepare the participants, the first page consisted of a short outline of my agenda for asking people to complete the survey and an overview of the three questions that would be asked:

> In this survey, I hope to gather some relevant raw material to kick-start my responses and to design future surveys and cultural probes.
>
> Over the next three pages I will be asking you to think about what matters to you. I will be using three themes:
>
> - Interests
> - Concerns

[11] https://www.surveymonkey.com/r/BKCP2BZ

- Cares

I'm using these themes to reflect different levels of importance:

Interests - general.
Concerns - more specific and important.
Cares - what you really consider important in your own life and globally.

You don't need to fill out all of the boxes and don't spend too much time, but do try to contribute something.

Thank you...

Over the next three pages, participants were asked to write down their Interests, Concern and Cares (up to five) in text boxes. They were open to write keywords or short phrases in each text box. Having tested the survey out on classmates, it was decided to add an extra page with a larger text box for any extra thoughts, feedback or questions that the participants might have. This helped to give people a bit of closure on the survey and provided an outlet for any questions the may have had. As it was an anonymous survey, this was a one way communication, but it was felt to be an important addition to the process.

## Interests

On this page, just write down any of your interests. Anything from hobbies to politics...

**1. Interests:**

1

2

3

4

5

Prev Next

## Concerns

OK, next level. Have a quick think about what concerns you in life. Money, government, environme

**2. Concerns:**

1

2

3

5

5

Prev Next

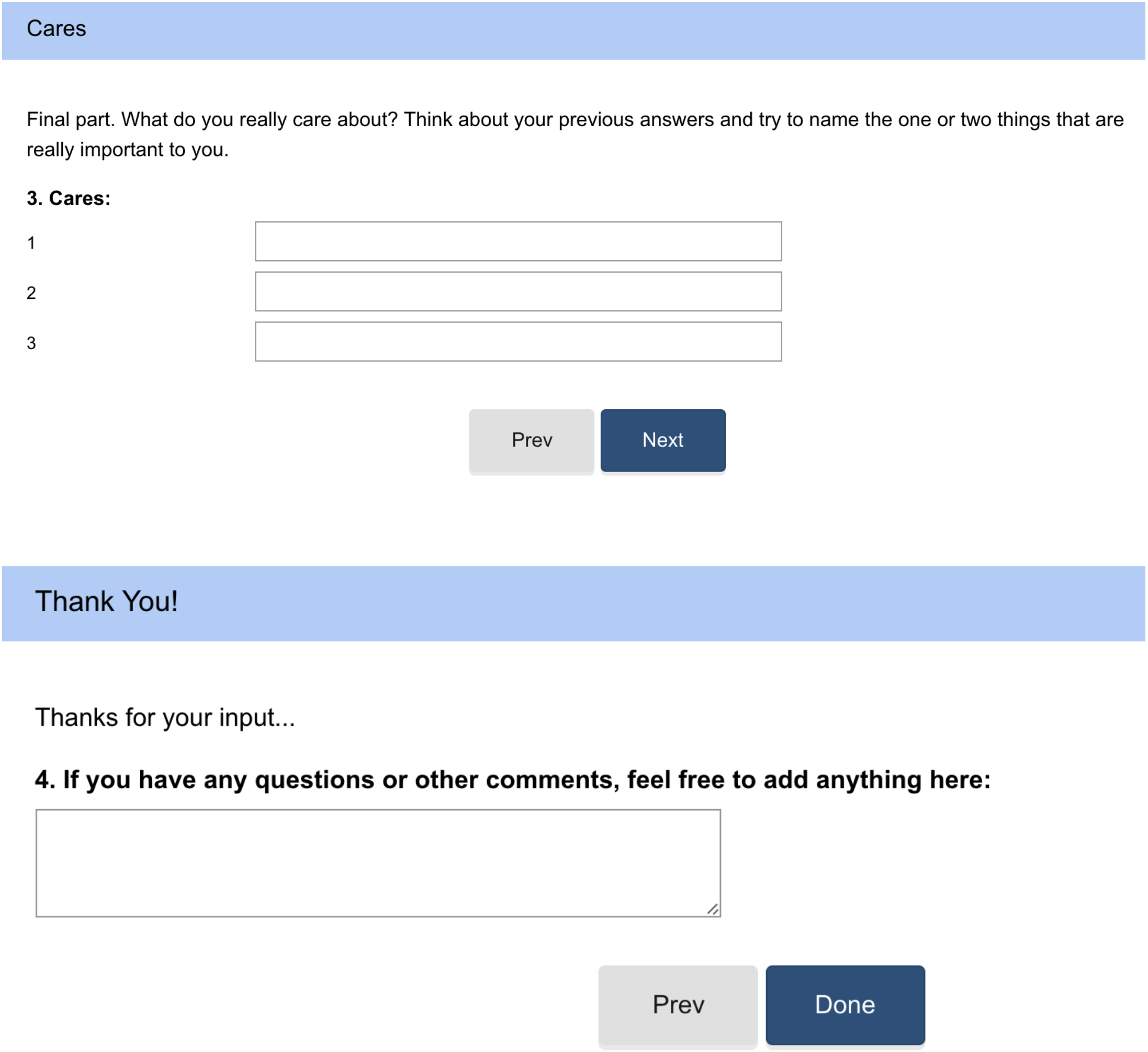

**Figure 10: Sequence of questions in the first survey**

Using an online survey helped to broaden the user group and break out of the academic environment. Throughout this project it is felt to be important to seek out a more representative sampling of society to allow for a wide spectrum of opinions and experiences. The ability to send such a survey out across social media and to encourage sharing increased the chances of hitting a broader user base.

## 4.2 Responses

The survey was launched on social media (Twitter and Facebook) and participants were encouraged to re-share the link. Some participants asked for extra information on the threads that allowed me to add some background and may have helped draw people in without overloading them with information.

Over two days, the survey generated over 100 replies and almost all of the questions were answered, with a considerable amount of extra input in the final comment page.

As a note for future online surveys, due to the 'freemium' nature of this platform, it was decided to investigate other platforms. This was mainly due to the extra cost involved to produce results in spread sheet format. Much time was wasted in transcribing the output into a usable state.

### 4.3 Usage

The output of the survey gave me a broad range of keywords and phrases reflecting the concerns of a good cross-section of Irish society, ranging from family, money, local politics and home to broader issues such as terrorism, the global economy and corruption. (See Fig. 11, for complete results).

The next task was to seek a means to visualise this in a way in which they could be used to generate ideas for an initial prototype to engage with and/or reflect some of these issues.

Having investigated the work of Venturini, et al., in mapping controversies with digital methods, the fractal-based visualisations of Jared Tarbell and the Raw visualisation applications at Density Design[12] , a few days were spent experimenting with attempts to represent the data in meaningful ways.

[12] http://app.raw.densitydesign.org/

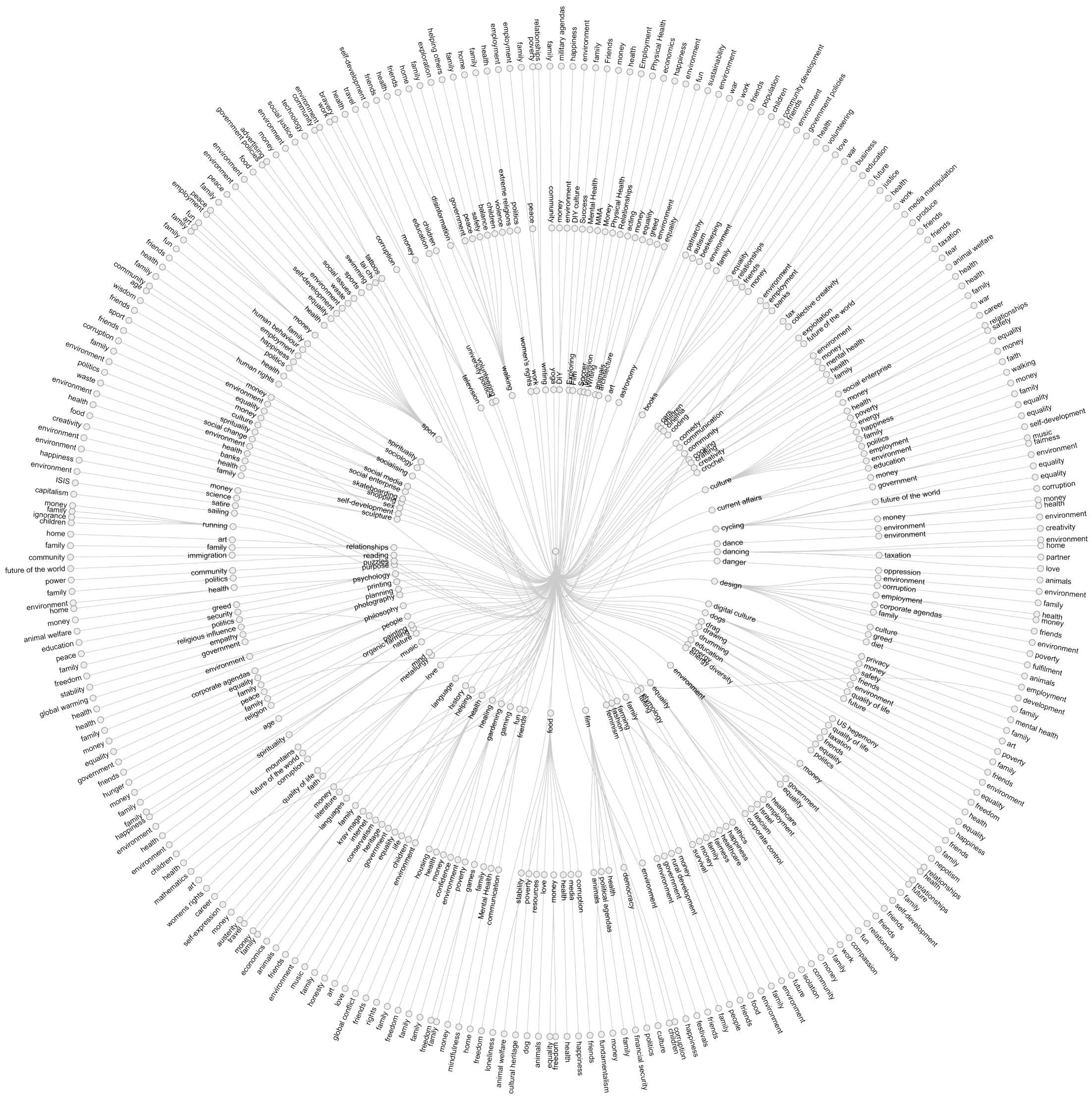

**Figure 11: Sample output from the Raw visualisation**

Although the relatively mature Raw visuals were impressive, it was felt that the survey itself would need to be redesigned to properly make use of these solutions. Notwithstanding, Raw will be investigated for future applications.

In the end, a solution of sorts was found using a Flash fractal library from Tarbell which produced  a series of unique arrangements of keywords. Although a little chaotic, the visualisations gave priority to the more popular concerns, but also allowed the rarer words to make appearances. As each visualisation was generated, the stronger responses tended to appear larger and more often, but less popular responses could also make an impression on the observer over time.

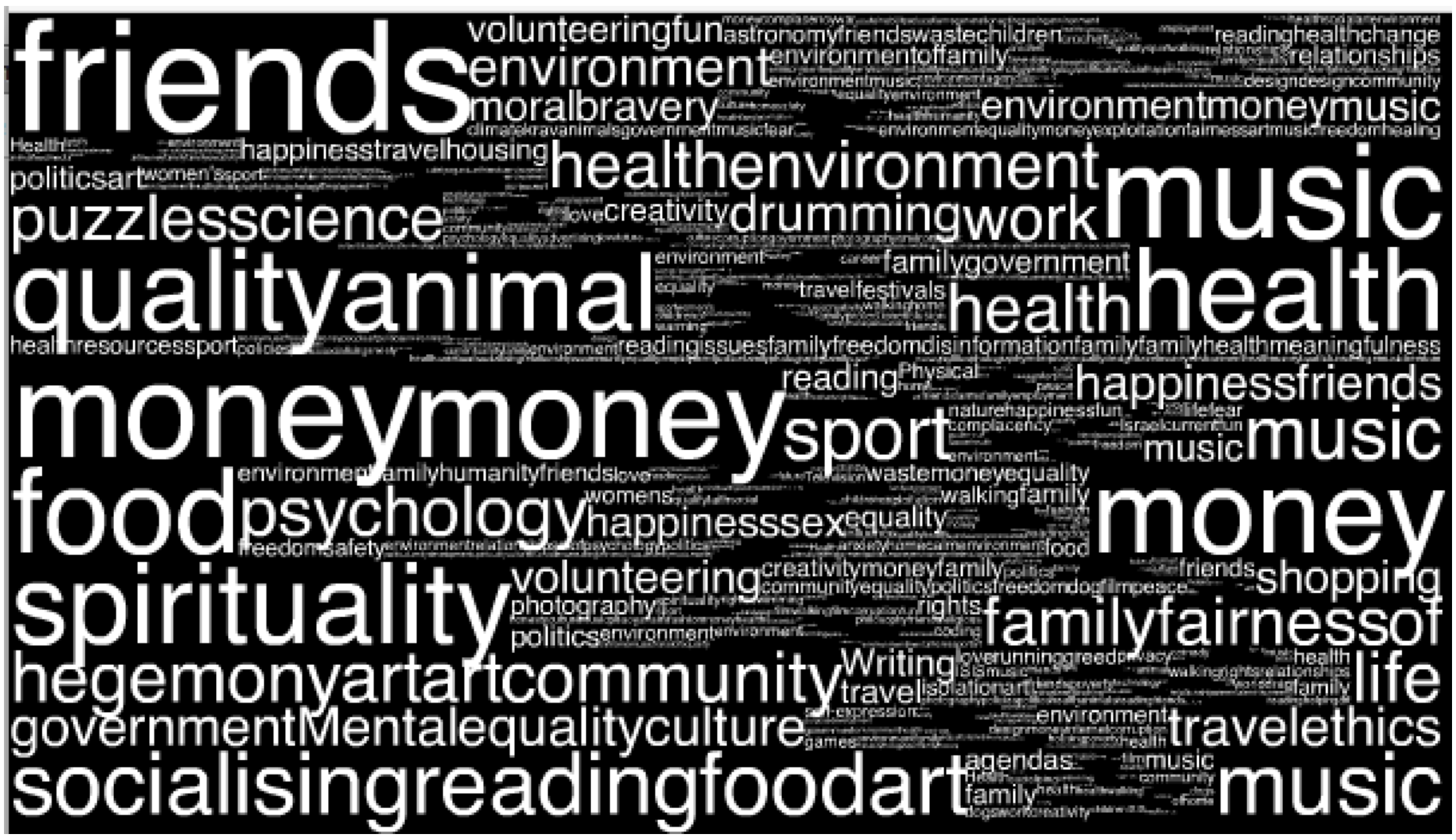

Figure 12: Screengrab of sample visualisation

This Flash app was uploaded to an online server both to allow myself to meditate on the outputs, and gain insights into the results, but also as a kind of response and closure for the survey participants.

It can be argued that this visualisation method has little grounding in any sound methodology, but it was felt to be very helpful in generating ideas for the next stage of the project. The process of reflecting upon these arrangements of keywords helped me to familiarise myself with the results of the survey without feeling the need to try to establish hard facts and links. Combined with earlier research it served as a filter for narrowing down areas for a possible prototype to address these concerns.

# 5. THE PROTOTYPE

In this chapter, I talk about the process of how I took the impressions from the first public survey and combined these with my own research to help flesh out an initial prototype. The objective of this process was to design a strong artefact which could serve as a focus for a workshop and allow participants to appreciate and discussion positive and negative aspects of technology.

## 5.1 Introduction

Returning to the original inspirations for this project in the fields of Critical and Adversarial Design, an on-going programme of research was being conducted into the histories and crossovers of these fields.

Taking pointers from the writings of Carl DiSalvo, Anthony Dunne and Fiona Raby, and working outwards through academic papers and design examples on the internet, a process of design-spotting and classification was begun as a list of clippings using the Evernote[13] application. As a parallel to immersing oneself in the concerns and cares of the survey, this exercise in studying areas of design which engaged in controversy and critical thinking contributed to an atmosphere which helped to inspire a prototype which was felt to reflect contemporary issues in interactive technology.

One of the main concerns which were returned to was the fundamental and contemporary theme of the home.

This was an attractive area to develop for many reasons. The home represents a basic social foundation that holds strong feelings and emotions for many people. It is a private space that is being slowly colonised by media and corporate interests through both passive and interactive technology. Also, since the most recent economic recession many preconceptions about ownership and the stability of the home are being undermined and questioned.

---

[13] https://evernote.com/

The themes and technologies of the ‘Smart City’ and the Internet of Things’ provided much food for thought in this search for an effective design, but the scope of this project began to be quickly overwhelmed by the scale of the possibilities in these broad areas. As the area of research was narrowed, the home, or house, seemed to work well as a microcosm of these themes and afforded a tighter focus for a design. Thus the concept of the ‘Adversarial house’ was conceived.

Exploring the spaces and interfaces of an average home, it was decided to focus on the first point of interaction: the front door. Specifically: the lock.

## 5.2 LOQ

Taking strong cues from the Addicted Toasters project by Simone Rebaudengo and Usman Haque (www.addictedproducts.com) a concept for an interactive lock was conceived.

As an initial practical design, it was to have the following features:

- A simple, unobtrusive design that could replace existing locks.
- Wireless connectivity to allow operation with a smartphone.
- Smartphone app interface.
- Open API for future connectivity.

Building on this basic technology, and implementing critical and adversarial aspects, more controversial qualities were explored. The idea of house ownership and what it means was addressed by giving the LOQ a degree of authority over the decision to open the door, or not, based on contemporary values for citizenship and social participation.

Giving the LOQ access to a user's online activities (browsing habits, purchasing patterns, social engagement and financial situation) served two different aspects of the design. One aspect was practical, security conscious and attractive; and another was to be more controversial.

On the practical level, having the ability to generate a digital (virtual) fingerprint for the user based on their unique digital habits provided a secure access token for their front door.

On another level, this information could also provide the LOQ with the ability to decide whether the user/tenant was in a financial/social/employment situation to warrant further ownership and access to the house.

In short, it was hoped that the LOQ design would serve as a constant reminder of the precariousness of home ownership and the underlying financial, political and social implications involved for many people. As a piece of design which could be appreciated as a natural progression of helpful technology, this underlying aspect was designed to help provoke users into questioning the ramifications and to create polarised standpoints.

To illustrate to possible outcomes of interacting with the LOQ, an initial three smartphone screens were designed, as follows:

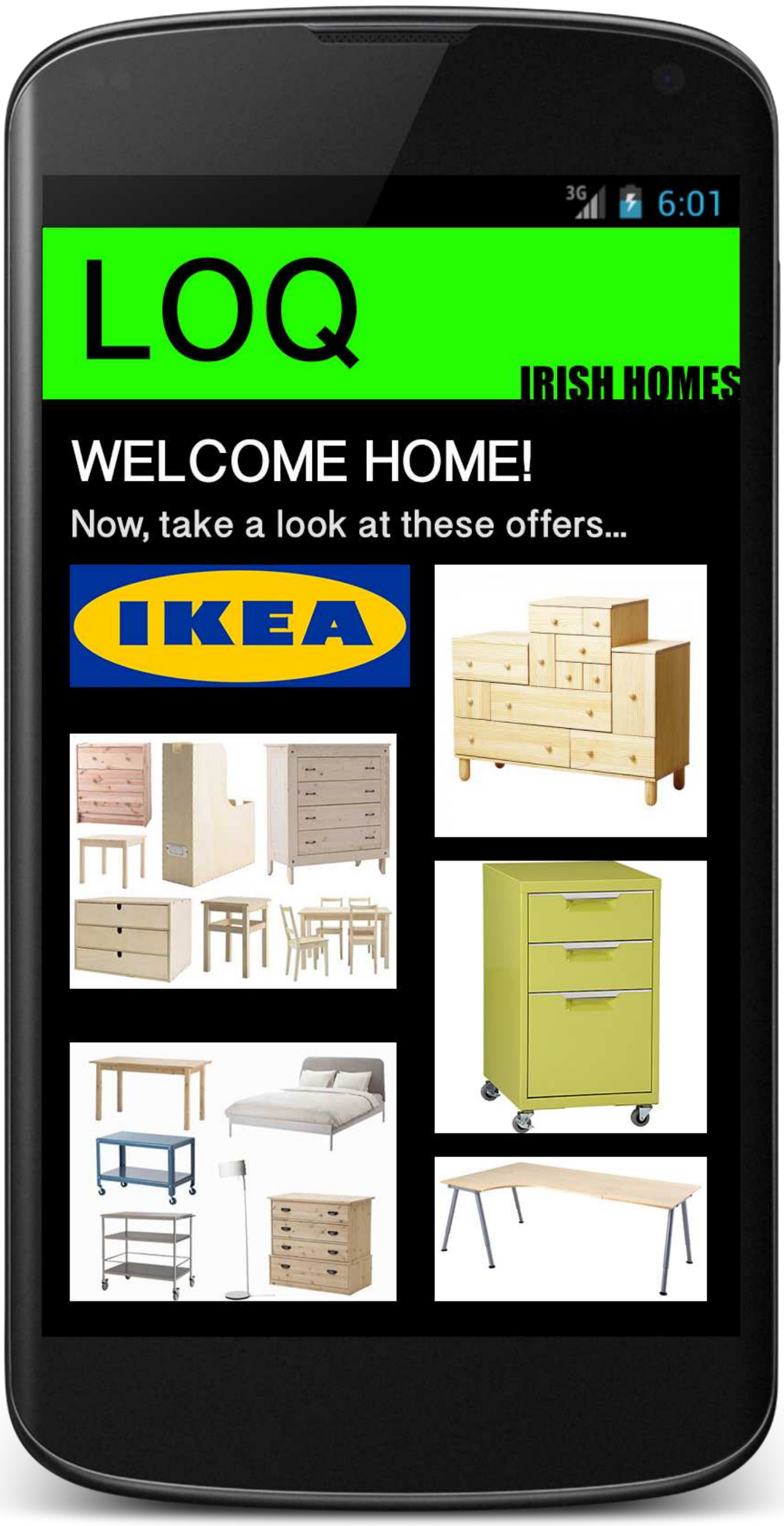

- Green: All is well with the users' finances and citizenship. The door opens and the app provides helpful advertisement for home products and consumables.

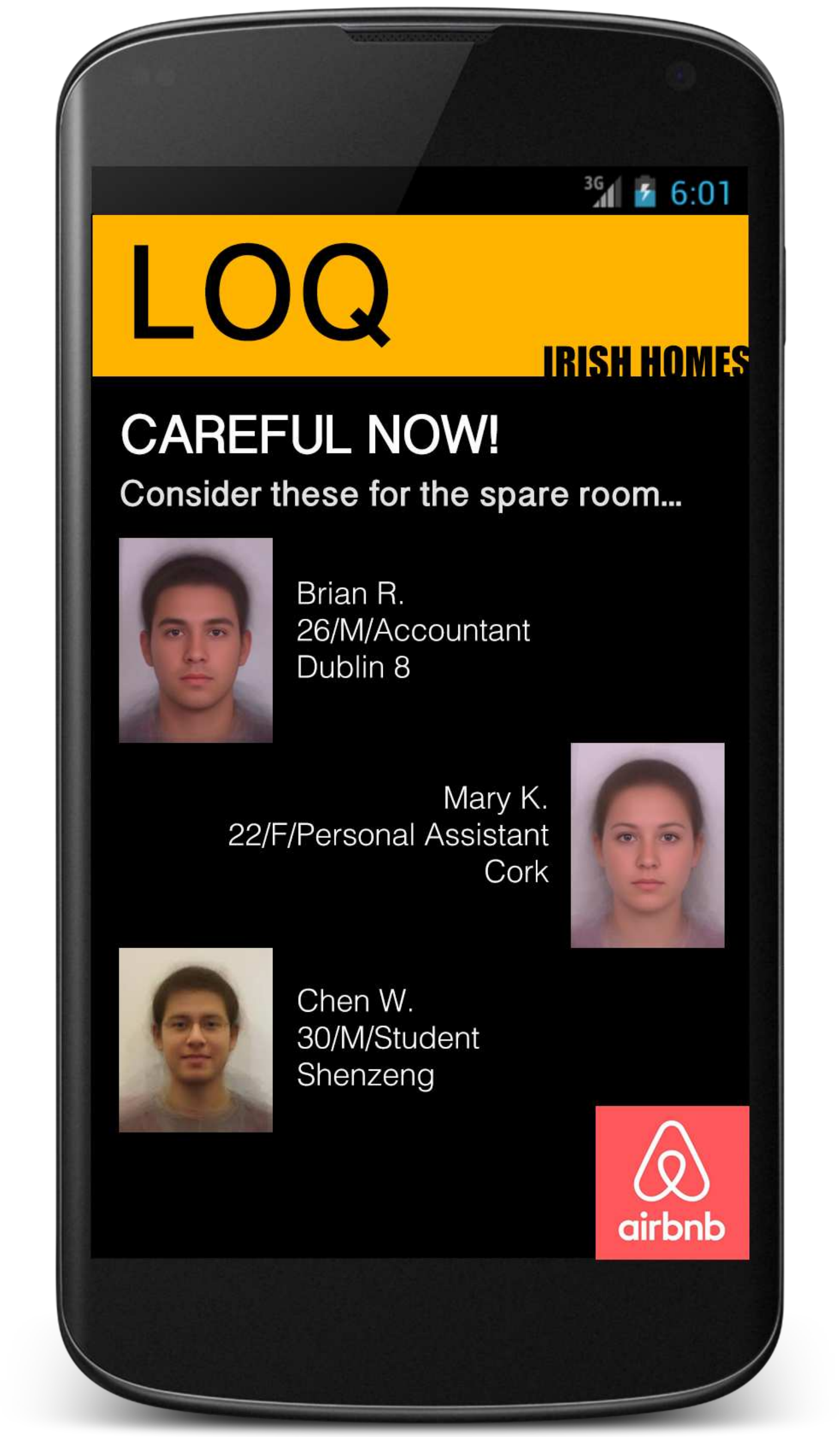

- Orange: The user may have some future financial issues. The app offers a choice of people who are looking for short-term shared accommodation in the area and arranges sublets based on the user's choice. The door opens, but certain parts of the house are now repurposed.

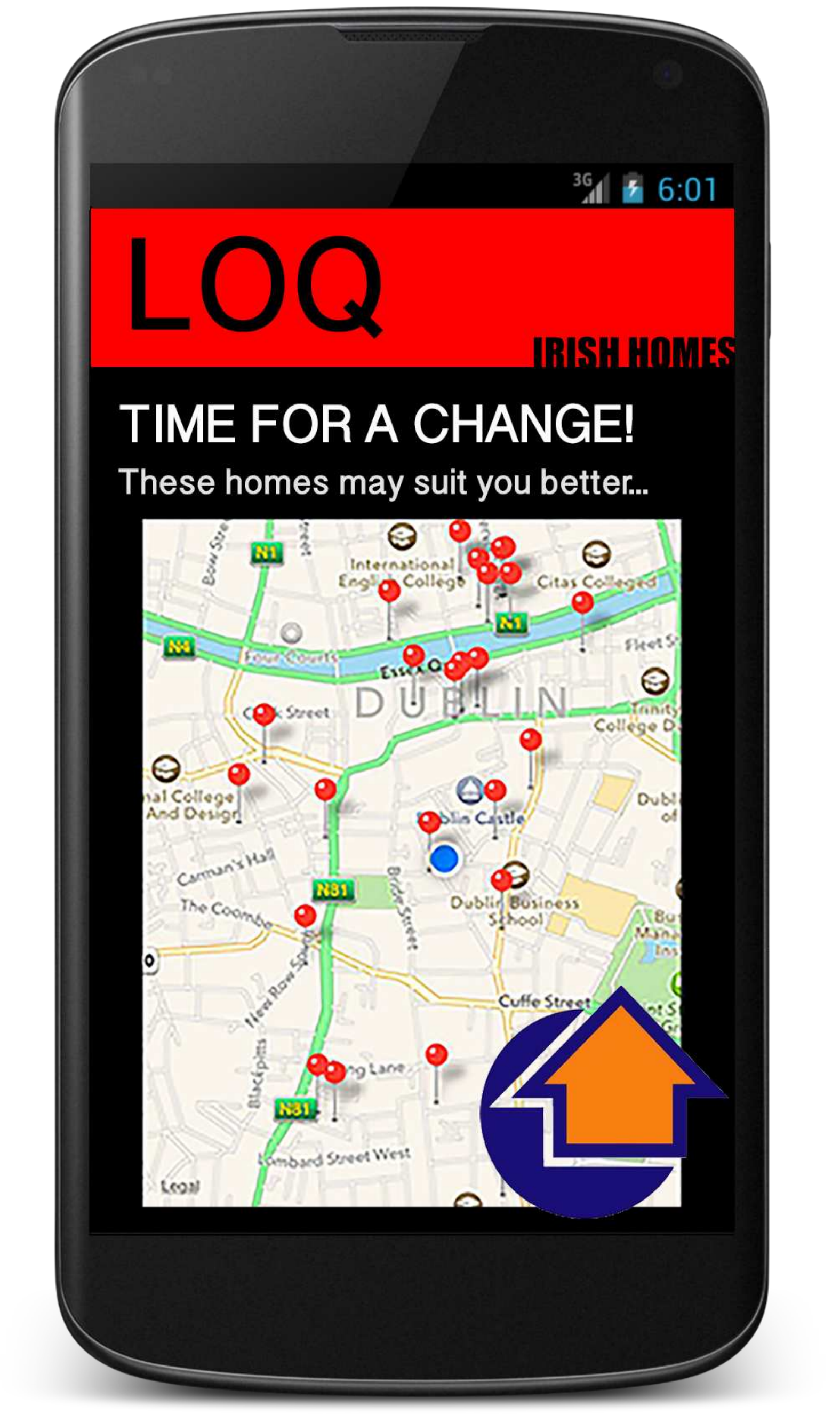

- Red: The user is in danger of falling behind on rent/mortgage payments. A map of smaller and/or more suitable accommodation in the area is provided. A moving company is notified and change-of-address procedures are begun. A 28-day move notice is given and the door opens.

For prototyping purposes, a simple RFID-based mock up was built using an Arduino and a RFID reader card. This served as a simple illustration of use with three LEDs as feedback (Red, Orange & Green) and as a proof of concept..

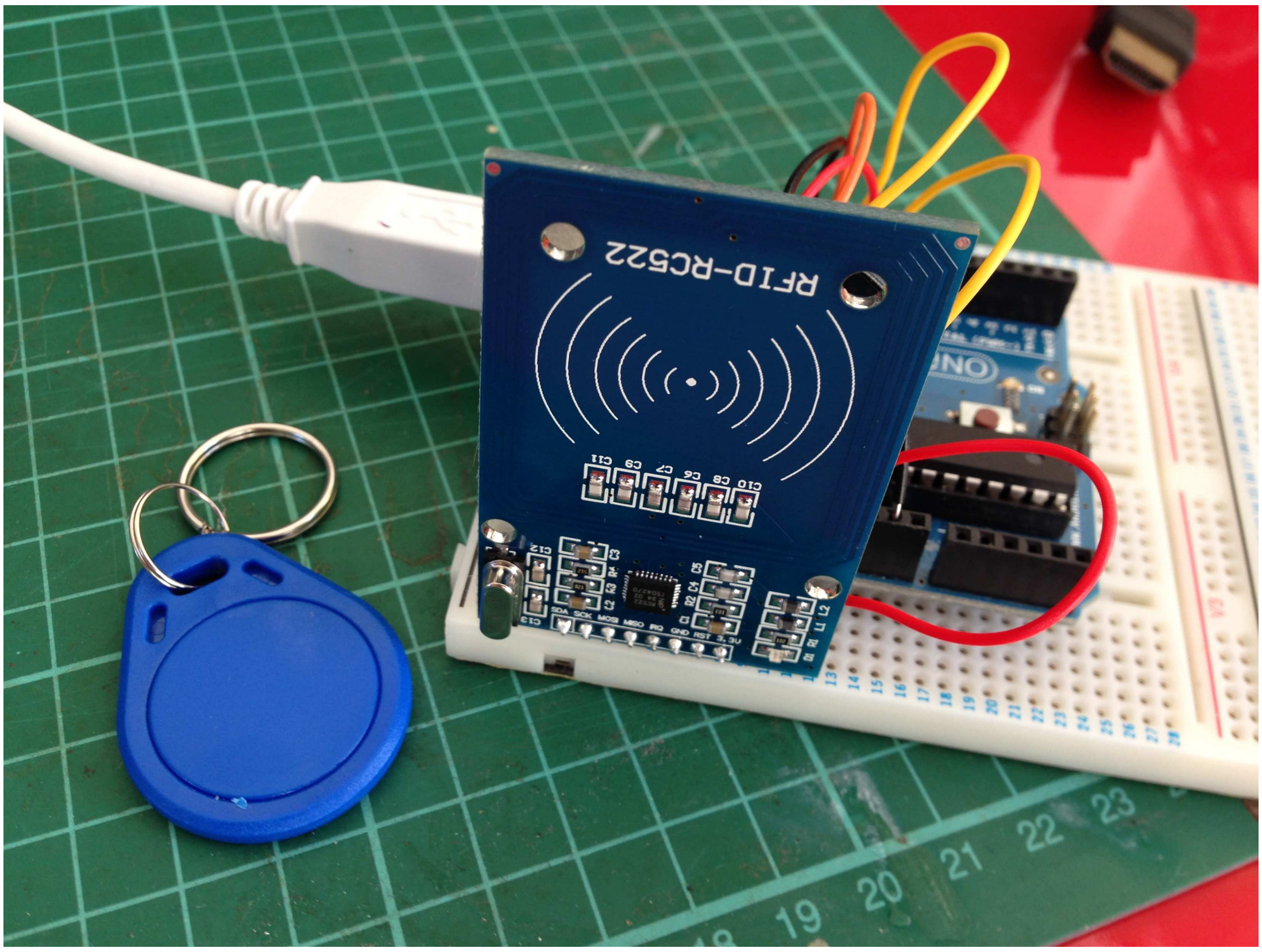

Figure 13: First prototype with RFID reader & tag

Though an interesting exercise for the designer, this prototype was felt to be of limited use in the next stage of workshop situations. On reflection, this artefact was not eliciting enough discussion without the designer having to spend significant time explaining its function. It was felt that a video prototype would convey the design intent much more succinctly and allow the hardware to be introduced as an interactive role-playing element at a later stage.

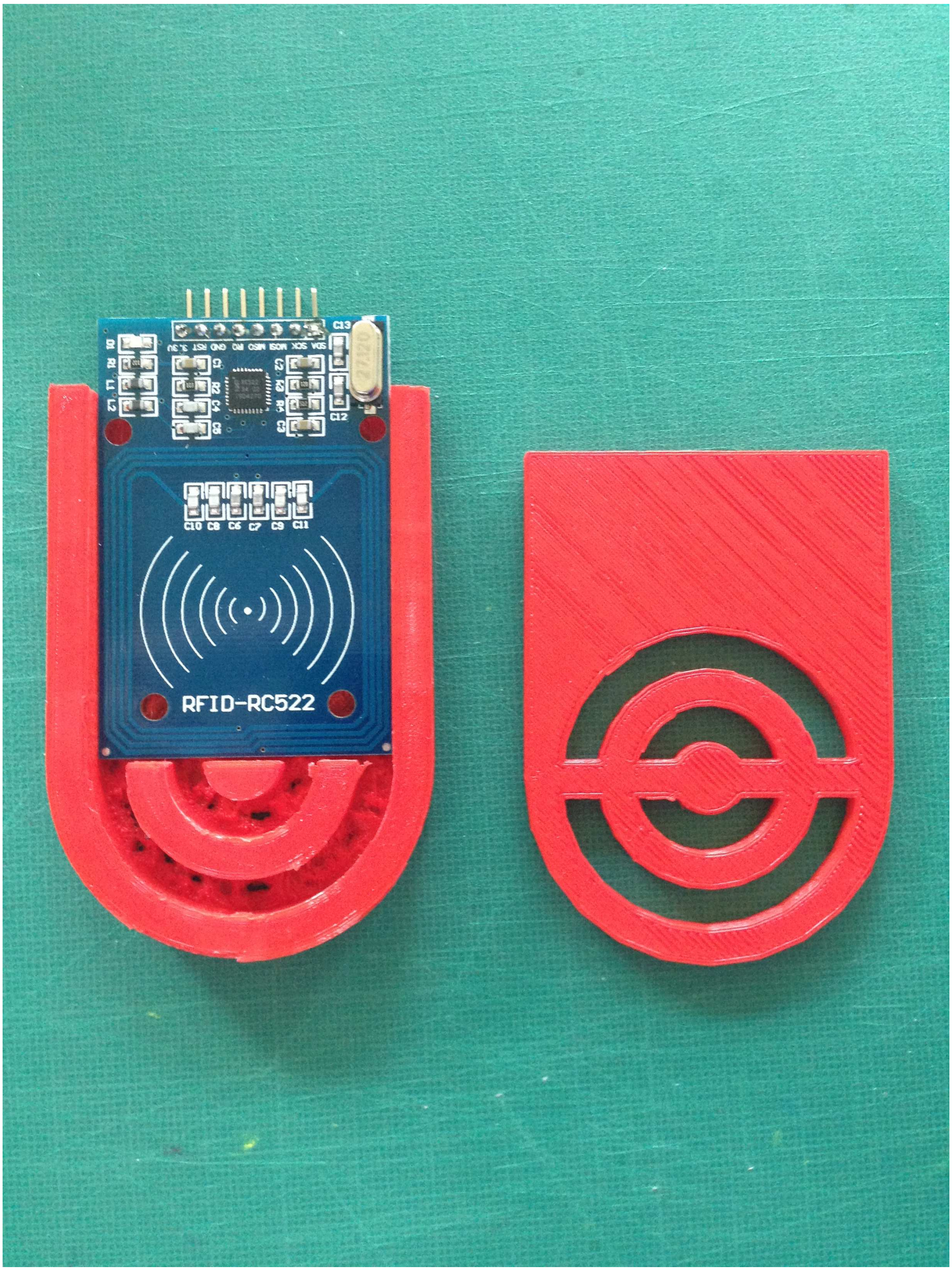

Figure 14: 3D printed housing for LOQ hardware

Taking inspiration from the plethora of Kickstarter product videos on the internet, a voice-over script was written and recorded. This was used to structure a storyboard for a one-minute advertisement as an introduction to the LOQ as real product, ready to be rolled out in the next year. A combination of live action footage illustrating a person using the LOQ, 3D modelling and presentation-style imagery was used to give a good overview of how the LOQ would be used in a familiar situation. (https://youtu.be/0km7o17uSRY).

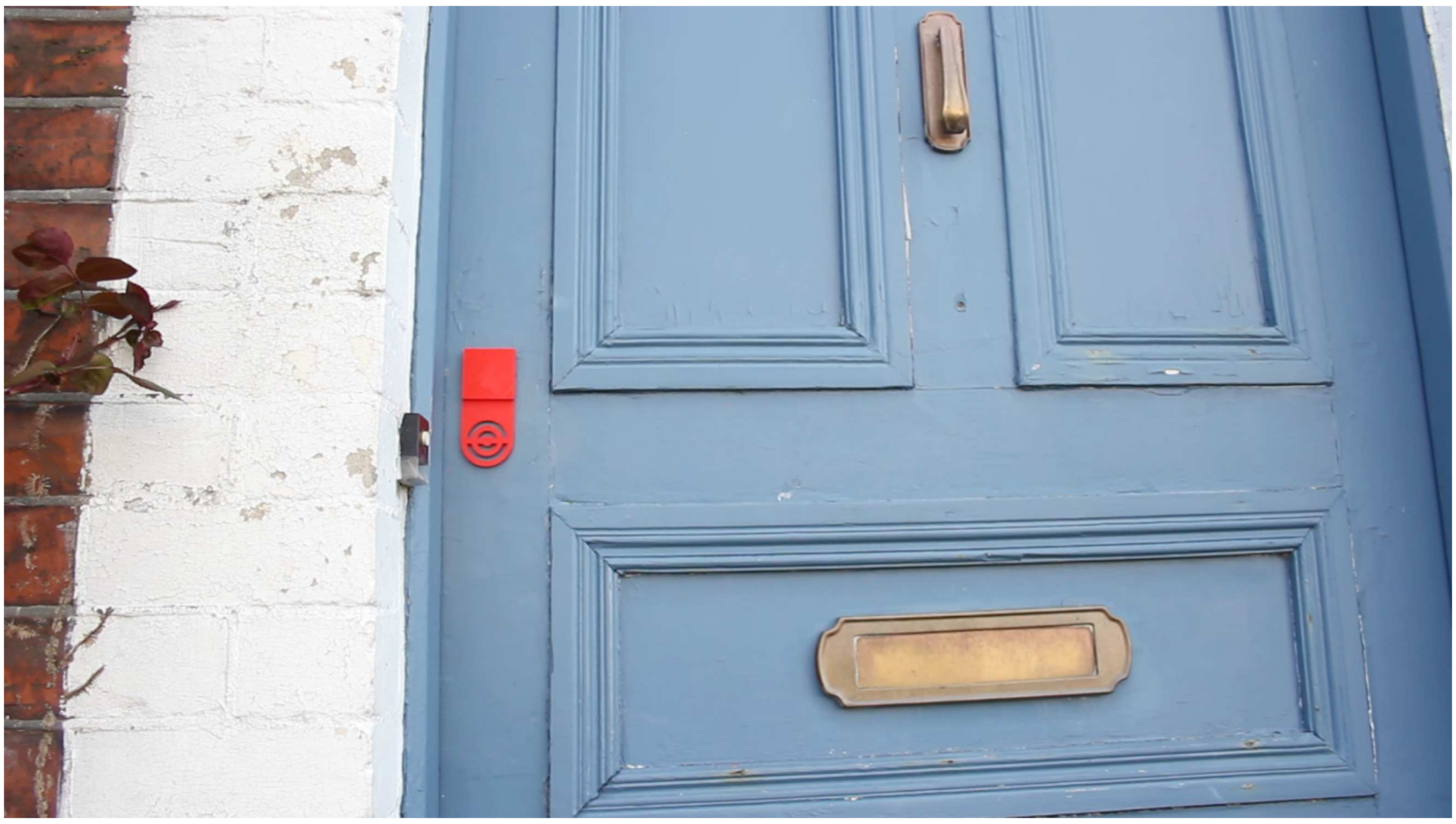

Figure 15: Still from Video Prototype, 3D-printed LOQ

As part of the video imagery, a 3D print of the LOQ concept was made and used show the product installed on a door. Photo-realistic renderings of the design were also created in SolidWorks to further enhance the believability of the LOQ.

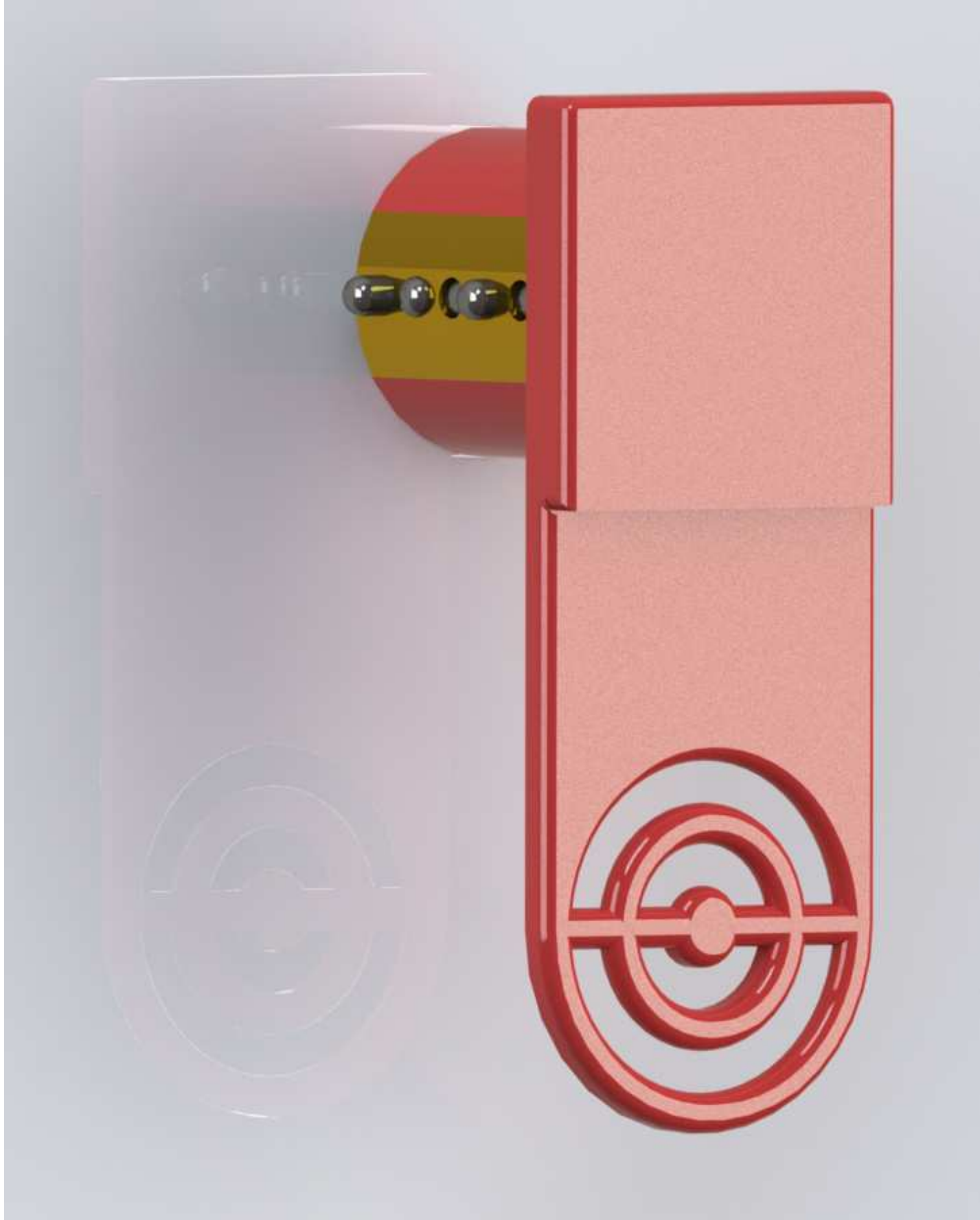

Figure 16: Photo rendering in SolidWorks

At this point, the prototype LOQ was ready to be incorporated into a workshop situation as a starting point for discussion and activities.

# 6. WORKSHOP 1

The following chapters describe the actual workshop processes and iterations. These workshops are experimental in nature and are facilitated and studied in an attempt to understand how participants would engage with the first prototype. Particular attention is paid as to how the structure and activities of the workshops may need to be augmented, or removed, to facilitate a more fluid experience for participants and to generate more constructive results for the designer.

Returning to the research question, the workshop model is essentially trying to achieve the following results:

- A good understanding of the prototype.
- Initial responses and reactions to the prototype.
- Questions about the design, followed by a participant-led discussion about the controversial aspects of the prototype.
- Further exploration of the prototype as it would affect the participants in real-world situations.
- An exploration of other smart technologies in the home and beyond.

## 6.1 Introduction

The first workshop was arranged for a midweek afternoon. As it was to be the first workshop, it was treated as a test run with a very open programme and an adaptable selection of activities. The most concern was with the reception of the video prototype and with feedback relating to how well the subject matter was conveyed and understood.

A small number of classmates and the supervisor were invited and the venue was the iMedia Design Lab at the University of Limerick. The lab was deemed most suitable as it provided good facilities with a whiteboard, projectors and a large table.

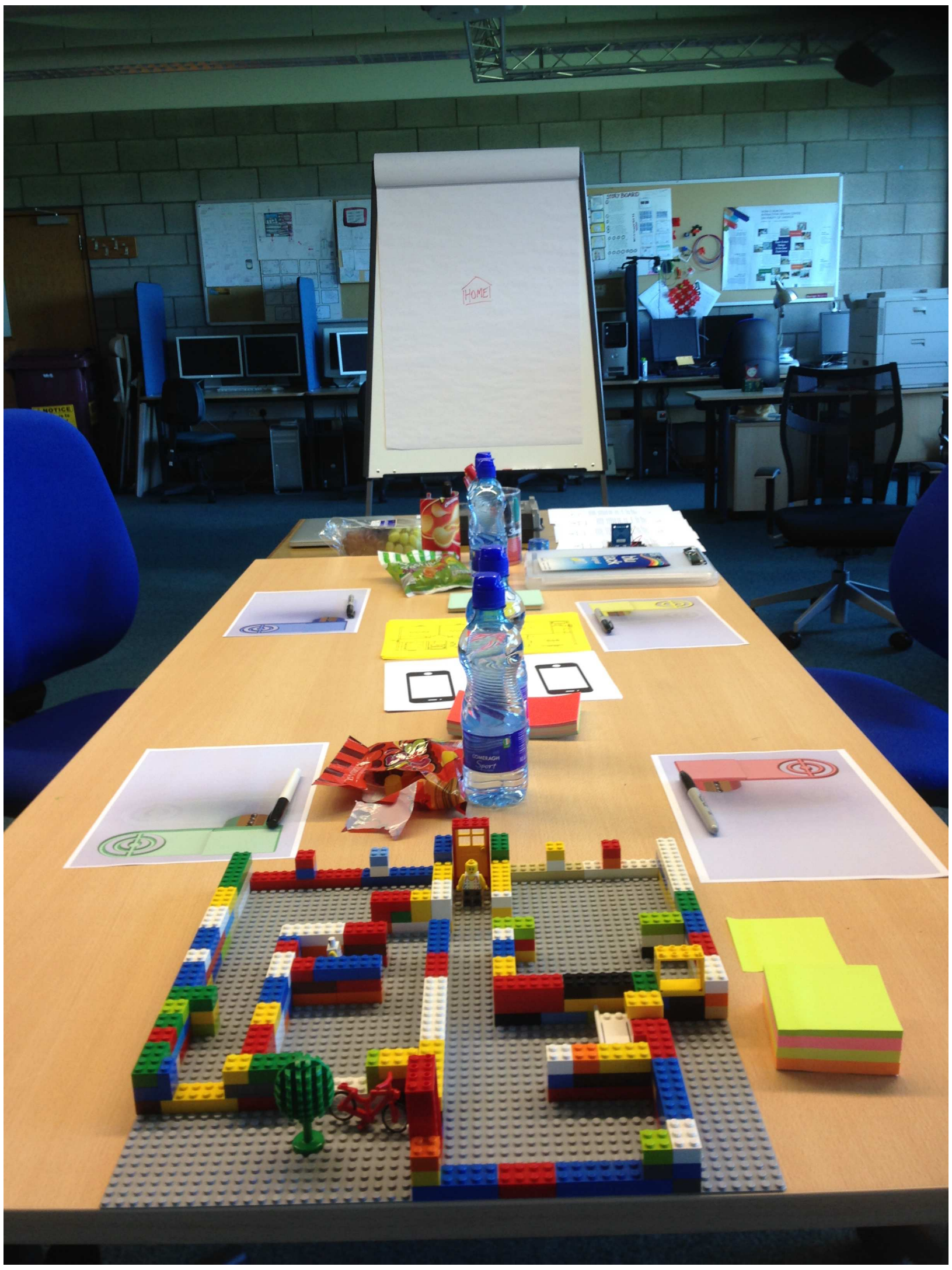

Figure 17: Workshop ready for participants

The workshop was programmed as follows:

- A short introduction by the facilitator, outlining the agenda and expectations.
- The video prototype. (https://youtu.be/qsvRjDLhgbg)
- A powerpoint presentation that would delve into the design in more detail.
- Question time relating to the design and any misconceptions about its functions.
- Activity: participants would generate keywords relating to the home.
- Open discussion relating to possible scenarios and ramifications of the LOQ.
- Activity: participants explore the actors and networks behind the LOQ.
- Activity: Participants would role-play random actors and explore their relationship with the LOQ. The facilitator could act as the LOQ itself.
- Final question & answer session.

This workshop was filmed for later analysis.

As it happened, there were a number of late cancellations by participants and the workshop comprised of just two members of the university faculty. It was quickly decided to go ahead with the video and presentation and then open it up to general discussion and feedback, without breaking it up with activities.

## 6.2 Evaluation

While the workshop felt like a failure at the start, the feedback on the prototype proved invaluable for redesigning and reassessing for the next iteration.

The small number of participants allowed for a continuous discussion beginning during the presentation. This helped to see what kind of questions were arising and where things could be clarified in the next version.

One of the first issues that arose was that the prototype revealed its negative qualities too early. A discussion followed that the LOQ should be introduced in a much more positive light initially, with the introduction of its more controversial facets as the workshop progressed. This would allow participants to invest in the design and discover their own issues as the workshop progressed. The general consensus was that the LOQ appeared to have little attraction to the participants and not much positivity. This feedback forced the facilitator to begin to explore justifications for the use of such a design and helped to flesh out the character and background of the LOQ.

The first half-hour of the workshop was spent fielding questions and building up a stronger profile of the design. Applications were discussed and scenarios involving government housing and the LOQ as a force to stabilise the housing market were considered. Issues of privacy and corporate-government partnership were raised and the discussion felt like it was returning to the grounds of the original research question.

Once the concept and ramifications of the LOQ were detailed to the participants satisfaction, the design began to be discussed as an accepted reality. At this stage the conversation broadened and swung away from the central premise of the workshop, but returned occasionally as it progressed.

Without detailing the range of topics discussed (from psychology to corruption and globalisation), here are the main areas that could be taken on to the next workshop iteration:

- The video prototype would need to be made more positive, initially.
- The initial questioning helped to create a stronger story for justifying the use of such a design as the LOQ.
- Focusing less on the negative and controversial, a number of benefits to using the LOQ were realised, which would help to 'sell' it to future participants. A more balanced presentation would help participants to make up their own minds.
- Plausible scenarios for the application of such technology were explored; such as for council housing and cooperative housing.
- Finally, as a vehicle to get people to question a solutionist design, it was felt to work well.

Overall, it was an important exercise that helped to create stronger material for the next workshop. The following time was used for a series of casual one-to-one discussions with various participants relating to their impressions of some of the finer details that had been raised in this workshop/discussion.

# 7. WORKSHOP 2

## 7.1 Introduction

Building on the experience of the first workshop at the University of Limerick, and the casual discussions since, it was decided to hold the next version in the city centre, where more participants could be accessed. Going with the ‘Adversarial House’ domestic theme, participants were invited to the facilitator’s home with the premise of a workshop followed by the ‘reward’ of a specially cooked dinner. This proved to be a much more successful model and variations of this will be used in future.

Five participants attended, representing a broader section of society with varying incomes and accommodation interests. Particular attention was given to selecting participants based on their relationships with their own accommodation situations. They varied from a renter, to a homeowner, a person engaging in renovating a house for let, and an artist who lived in government subsidised housing. These participants will be referred to as the Renter, the Homeowner, the Renovator and the Artist, respectively.

For this iteration, the workshop took place around a dining table with a large second screen to show video and presentations and a subtler recording set up. At this stage the elements of the workshop had been assembled into a portable kit with a small Flip video camera and a handheld audio recorder. All other materials could be transported in a bag. The table was prepped with an A1 sheet and a variety of writing materials.

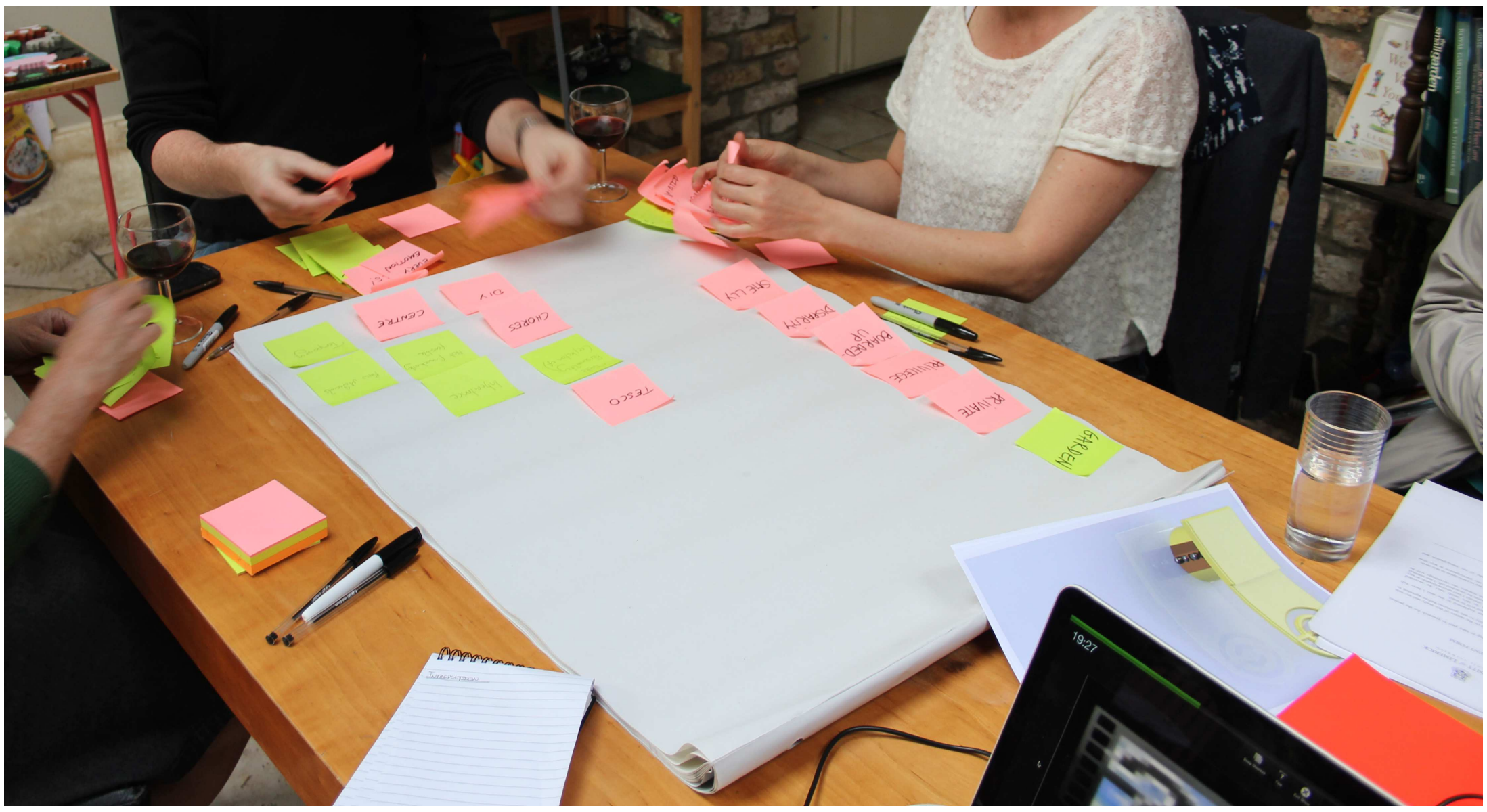

Figure 18: Generating keywords

This workshop was programmed as follows:

- Introduction: Participants had a short introduction to the agenda of the workshop and were informed that they would be doing much of the work, with the facilitator taking a more passive role. (See Appendix A: Workshop Introduction Transcript)
- Activity: Each participant was given a stack of small post-it pads and they were given 5 minutes to produce keywords relating to their impressions of what a home means to them (See Fig. 18). This served the purpose of initiating the participants into the subject area and also generating material for future iterations of the workshop. The participants then arranged their post-its on an A1 sheet that was hung on the wall in a visible position.
- Video Prototype: An edited version of the original video was shown. The edits were very subtle: the background music was changed to a less distracting version, and the reference to the 'Red' smartphone screen was removed. This contributed to presented the design in a more neutral light and allowed the participants to speculate by its omission. (https://youtu.be/1t54R9yFyAg)

- PowerPoint Presentation: The facilitator talked through the details of the LOQ, taking care to show its applications and to provide an unbiased information source. (See Appendix A: Presentation Screenshots)
- Short Q&A: This short session gave time for the participants to ask questions and clarify their understanding of the LOQ.
- Activity: Participants write short phrases on cards about their impressions and feelings around the LOQ.
- Short discussion, followed by the facilitator looking through the cards and raising some of the responses for discussion by the group. This exercise allowed the facilitator to reintroduce some of the more interesting responses to the group and allowed them to explain their impressions to the rest of the group.
- General discussion: An open conversation amongst the participants, exploring their realisations and explaining these to each other. A good time for the facilitator to just listen.
- Activity: Participants are left to draw out a mind-map of what actors they can think of who would be involved and how they are related to each other. A very short introduction to the idea of systems and stakeholders precedes this.
- Short discussion about the mapping and reasons for inclusion of certain actors.
- Activity: Dice are thrown by each participant that gives them a certain response from the LOQ (on a printed card). Lower numbers produce a red card and higher numbers produce green. They are then paired with an actor from the previous activity and each takes turns role-playing how they would relate to the LOQ's response and/or how it would affect them.
- Dinner: a more casual discussion and debriefing.

## 7.2 Evaluation

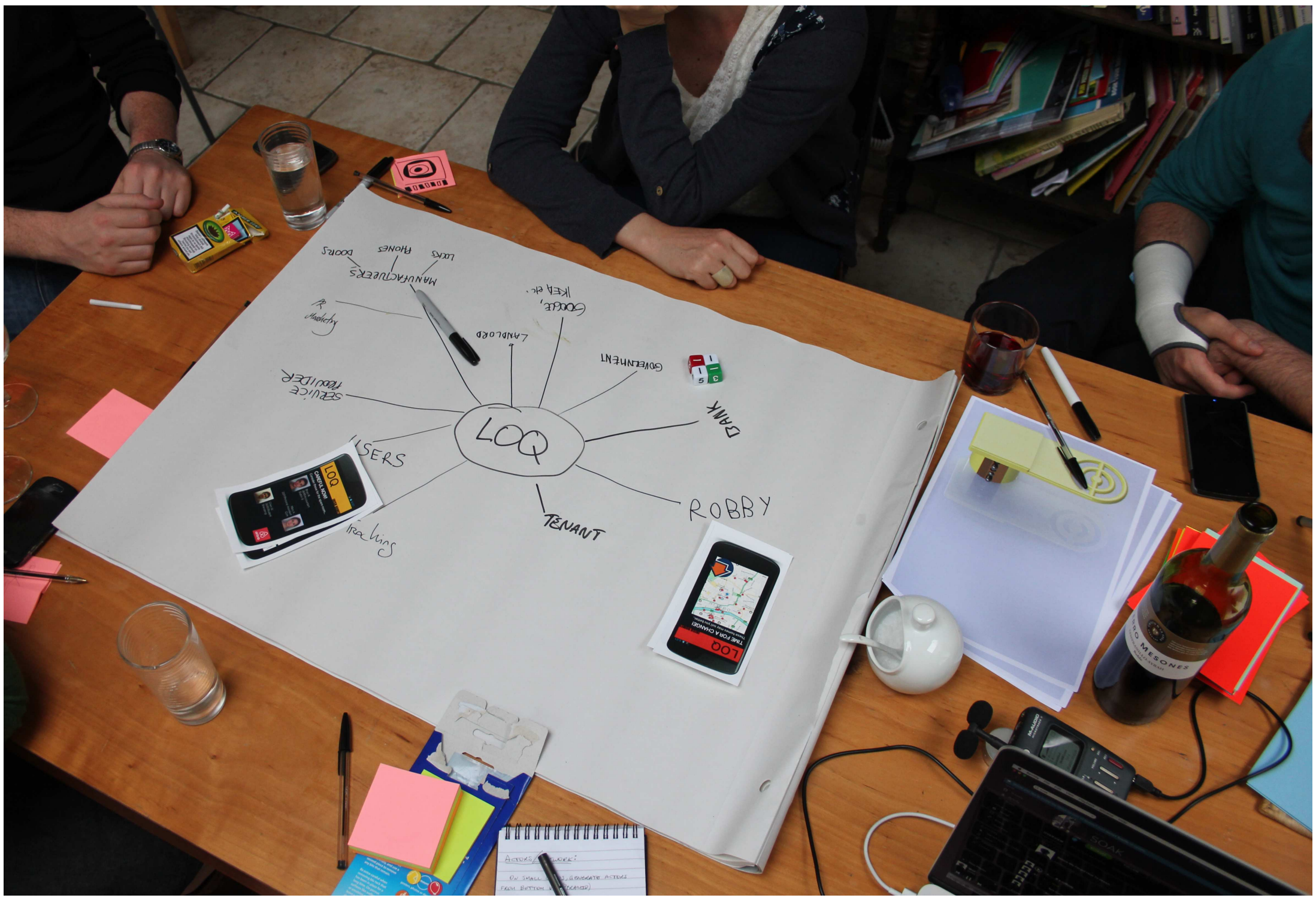

Figure 19: Mapping and roleplaying with the cards

This workshop reaped the benefits of its predecessor and proved to be much more structured and engaging for the participants. Holding it in a domestic environment with the element of cooking allowed the facilitator to disengage from the group during activities and also provided the participants with the space to engage more directly with one another.

The first activity served a certain purpose, but didn't really feed back into the workshop itself. It did serve to get the creative juices flowing and get participants minds focused on the tasks ahead, but the exercise itself was too broad and not directly relevant to the prototype. On reflection, though, these words may well provide a base level for mapping and contrasting with any data generated after the participants are exposed to the LOQ; hopefully highlighting the adversarial properties of this artefact.

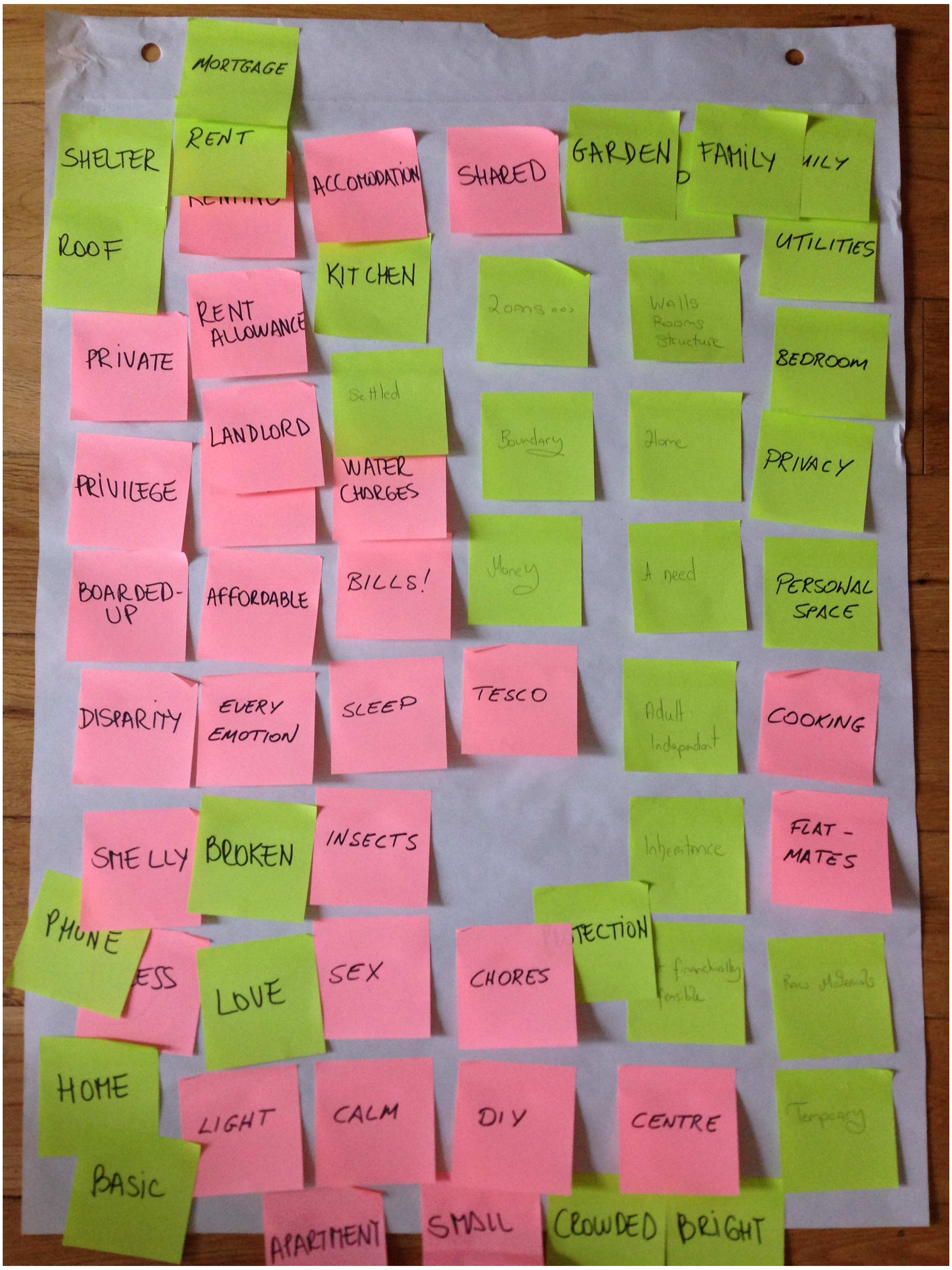

Figure 21: Initial Keywords

The initial introduction was much more concise and the participants seemed to accept the concept of the design much quicker, and with less questions, from the re-edited video and presentation. Presenting the LOQ in a neutral but useful light, and taking on a role of a representative for the design, engaged the participants and allowed them to take up their own positions with respect to its benefits or otherwise.

The LOQ was explained as a benign device that was created to make sure that everyone had a home to live in. It would be funded by in-app advertisers and property businesses, which would ensure that the home dwellers would incur no cost. Even if the LOQ decided that you had to sublet, this was presented as a form of 'car-pooling' and would only be a temporary measure to allow the tenant to get back on their feet again. Many of these details came from questions asked during the first workshop and helped to answer many of the participants' questions before they arose.

The main question that did arise was to ask about what else the LOQ did beyond just being a door lock and what its association with the internet might be. This quickly turned into a discussion about how it could know your habits by tracking entries and exits from the house, which led on to associating this fact with potential advertisers and how important this kind of tracking is to targeted advertising. Without prompting, the participants had already begun to understand and raise some of the hidden factors of the LOQ and also compared it to contemporary 'Internet of Things' products such as the Google Nest[14] smart thermostat. This was an important point where the participants started to create their own picture of a network behind the device and began to perceive the actors involved. This also highlighted the benefit of striking a balance between pre-empting some questions and also allowing enough room for the participants to engage and come up with their own.

The next activity was much more successful. By getting participants to commit their impressions in writing straight after the video and presentation, and then revisiting them, they then began to take stances on their issues that engendered a good adversarial discussion.

---

14 https://nest.com/

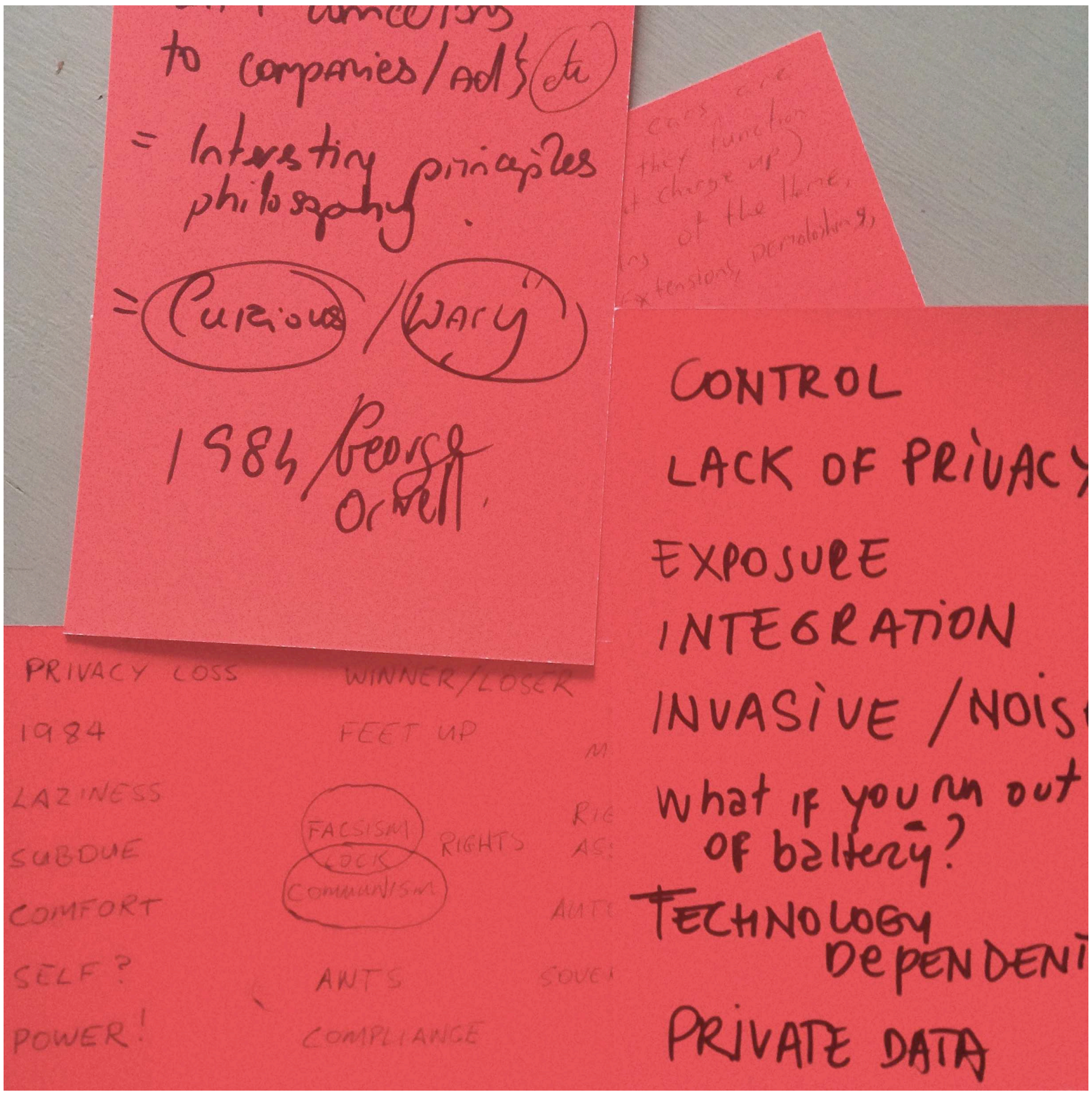

Figure 20: Reactions to the LOQ prototype

During this phase, the group were generally split between its benefits and its controversial aspects. Some made suggestions about further benefits that the LOQ could provide, while others maintained a negative opinion.

A satisfying result was that, even though opinions were polarised, both sides appreciated the others' points and began take these into account. This represented an important point; rather than having participants taking a polarised antagonistic standpoint in defending their opinions, they engaged in agonism and sought to understand others' concerns whilst maintaining their positions. In most cases the middle ground between contrasting opinions was explored as a possible amelioration of controversies. By the end of the workshop the participants were all exploring the respectful middle ground while maintaining their contrasting opinions.

For example, the two participants (Artist and Renter), who did not own property, both brought up issues of privacy, invasiveness, disenfranchisement and dystopia. The Homeowner saw some potential in the device for its "*…interesting principles and philosophy…*" – combined with a mix of curiosity and wariness – but was concerned about the connections with commercial interests and perceived Orwellian connotations. Finally, the Renovator/sub-letter took a much more receptive view and saw the LOQ connecting with many facets of the house and providing the user with a *"…consultation service…"* or personal assistant which would help them save money, extend the building and reduce their mortgage over time by advising a savings regime: *"it might say, 'get a tenant into your house and it will save you two years on your mortgage, etc.'"*

The Artist, who had raised the most issues with the LOQ, conceded that this was a very *"…rational…and logical…"* outlook with appreciable benefits, but that there needed to be more consideration for the user. Both the Artist and the Renter raised the issue of how the LOQ, and the advice it might provide, would affect people from different income backgrounds. The Renter said that, *"if I had a huge house, it would be great"* and even "*fun",* but coming from a less stable financial and renting situation, the LOQ represented a more frightening and controlling prospect where the advice would have much more serious connotations for the future. This also raised the question of whether the LOQ, and the house itself, had the users '*best interests*' in mind or would have more loyalty towards commercial interests and advertisers. Taking these points on board, the Renovator and the Homeowner both conceded that, although they saw potential in the design, that there would need to be safeguards to prevent the LOQ from enforcing economic classism and to protect the financially vulnerable.

Another common area which was explored was the concern with surrendering control and free-will to technology in exchange for convenience. This led the participants to question the concept of technology supporting laws and the removal of a citizens ability to break or challenge these laws. The house with a LOQ installed represented an impersonal solution to a very personal situation.

The mapping activity was a little forced, but served to generate some exploration of stakeholders. The facilitator was getting the sense that the participants had reached their energy limits and it was decided that this would be the final activity. The role-playing activity was started, but postponed until the next workshop.

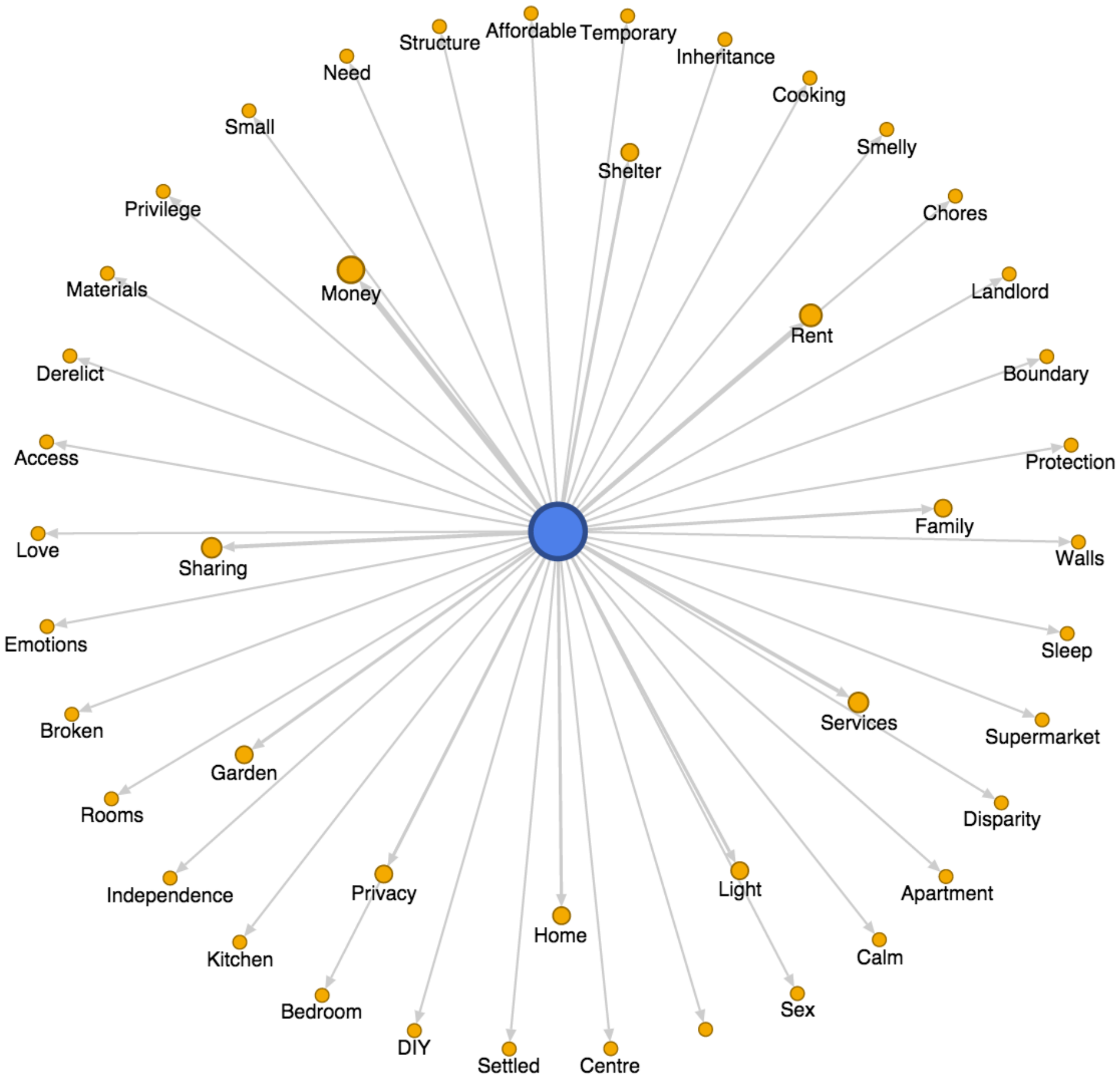

**Figure 21: Google Fusion Tables mapping of 'Home' keywords**

Overall, this workshop led the participants into many of the areas in which this project is concerned and produced lines of questioning that the participants had not previously considered. The structure allowed the facilitator to step back considerably and let the participants run the discussion for the most part. There are still refinements to be made, but this would be the essential model to move forward with.

## 7.3 Issues Raised

| **Pros** | **Details** | **Participant(s)** |
|---|---|---|
| **LOQ does its best to keep user in accommodation system** | There are no 'homeless' outcomes in the LOQ. Rather it will keep rehousing the user until a suitable accommodation situation is achieved. It will naturally move you to whatever home you can afford. | Renovator |
| **It encourages competition between service providers** | If the LOQ can be connected to the home's heating, internet and electricity services; it can also advise on the best service providers in its efforts to save the user money and maintain their housing situation. This would encourage providers to introduce more competitive packages. | Renovator<br>Renter |
| **LOQ can be a financial assistant** | As above, as the LOQ connects to other parts of the house and builds up data on the tenants, it can seek out more efficient ways to pay for services such as: heating, internet, television channels, home upgrades, appliances, etc. | Renovator |
| **Encourages saving money** | The constant reminder that LOQ provides the user that their relationship with their current home is based on their own financial security, would encourage people to think more deeply about how they treated their finances. | Renovator<br>Homeowner |
| **Users could trade tracking (data) for house payments** | In the event of a tenant's financial difficulty, the LOQ could offer them the opportunity to offset payments by allowing their online activities and habits to be tracked more aggressively. They could also be obliged to be exposed to a higher level of advertising and to | Renter |
| | provide feedback on its effectiveness. | |
| **LOQ appeals to the organised** | The participants who reacted more positively to the LOQ tended to be the ones with a more rational and organised lifestyle. The constraints and 'guidance' that the LOQ provided tended to support their outlook. | Renovator<br>Artist |
| **It could support and enforce rent control** | In a housing initiative that had standardised accommodation for different living situations (single, sharing, family, retired, etc.), the LOQ would encourage standardisation of rents/mortgages and automatically assign people to housing which suited them best. | Artist<br>Renter |
| **Useful applications in government and co-operative housing projects** | As above, a LOQ-enforced housing project could ensure that people stayed within the system as their needs and situations changed, and also ensure that space would not be wasted. | Artist<br>Renter |

| **Cons** | **Details** | **Participant(s)** |
|---|---|---|
| **Citizens treated as data** | As the funding for the LOQ is expected to come from targeted in-app advertising, the house in which it is installed will treat its residents as monetisable data sources. | Homeowner<br>Renter |
| **Smartphone would become a 'universal identification' card** | If the LOQ is to be 'rolled out' nationally, then every citizen would also need to be provided with a smartphone to be able to use it. Essentially, these phones would become their trackable ID and a central point for data gathering. | Renter |
| **Loss of privacy and autonomy** | The influence that LOQ would have on the user's financial decisions would pervade much of their lives, from spending money to socialising. The knowledge that the LOQ was monitoring them would constrain their real-world decisions and movements. | Artist<br>Renter |
| **Government and commercial invasiveness** | The fact that the LOQ might have to be a government/commercial partnership suggests that there would have to be a much higher trust in these entities than there is now, for people to feel comfortable about sharing information with them. | Renter<br>Artist<br>Homeowner |
| **More digital 'noise' in the personal space** | Adding this extra layer of technology to everyone's life introduces more distractions and another level of decision-making. | Renter |
| **No accounting for alternative lifestyles** | The sentience provided to the house by the LOQ would not necessarily take people at the fringes of society into account. If all citizens are considered the same, or within statistical norms, the LOQ may adversely affect those who choose to live their lives differently, or even those who suffer from mental illness. The rationality of the LOQ does not make room for irrationality. | Renter |
| **Possible impact on personal credit rating** | Considering LOQ's partnership with the banking system, falling foul of the LOQ could easily be used to lower a person's credit rating. Already, Facebook friend data is being considered as a means to judge a users' credit rating.[15] | Homeowner<br>Renovator |

15 How Facebook could affect your chances of getting a loan (Toronto Star) - http://on.thestar.com/1UzmDGL

| | | |
|---|---|---|
| **Loss of integrity** | As an artist, who strives to remain independent from commercial influence, this participant suggested that they would feel that his integrity would be compromised by having to fit in with how the LOQ would impose advertising and accommodation on their life. This could be considered relevant to many creative lifestyles. | Artist |
| **Affects the poor more than the rich** | This was one realisation that was arrived at by all participants. The LOQ would appear and act benevolently to those who could afford their lifestyles; but, as the participants placed themselves in the position of people on a lower financial level, they saw that the LOQ would have a much more serious and imposing relationship with them. | Renter<br>Renovator<br>Artist<br>Homeowner |
| **Ethical issues concerning advertisers** | Considering that the monetisation of citizens would be a fundamental aspect of the LOQ, participants were concerned that the advertisers would not have the best interests of the people in mind. They could use their position and influence to act unethically in trying to extract money or force financial decisions on people. | Renter<br>Artist |
| **Government surveillance** | Concerns about government access to such levels of data were seen to be beneficial to more authoritarian types of government. | Renter<br>Homeowner |
| **Compromises the right to assemble** | As above, being able to track a fundamental thing like whether a person is at home, or not, could help a government to enforce laws and dissuade citizens from assembling to challenge these laws. | Artist |
| **Favours certain commercial interests** | Whatever commercial interests become involved in a partnership with LOQ would maintain a very strong position of influence which would become difficult to challenge by other companies. Effectively a monopoly. | |

### 7.4 Outcomes

The only aspect that still remains missing from the workshop model is how to generate further prototypes from the discussions. Although other interactive technology within the home has been brought up, this area needs to be addressed and developed for the next workshop.

Along with the finalisation of the video and hardware prototypes, designing this into the next workshop will bring the project onto the next stage and further.

# 8 WORKSHOP 3

## 8.1 Introduction

This workshop was structured to be a means to address some of the areas of the previous workshops that it was felt hadn't been explored properly. Specific focus was placed in the areas of mapping, role-play, exploring other areas of ubicomp in the home and further prototype directions. This was also a chance to revisit a richer prototype, and for more individual interviews. Continuing with the domestic theme, this workshop was also facilitated in a home environment and ended with dinner for the participants and facilitator. To clarify the dialogue and aid with following the conversation more clearly, some of the participant's attendance times were staggered.

As the workshop structure was still not felt to have been fully explored or finalised, the same participants were re-invited and the focus was initially maintained on the LOQ prototype. The first part of the workshop was based around the two participants who had a more polarised outlook on the LOQ (the Renovator and the Renter), and the second part involved the Homeowner and the Artist.

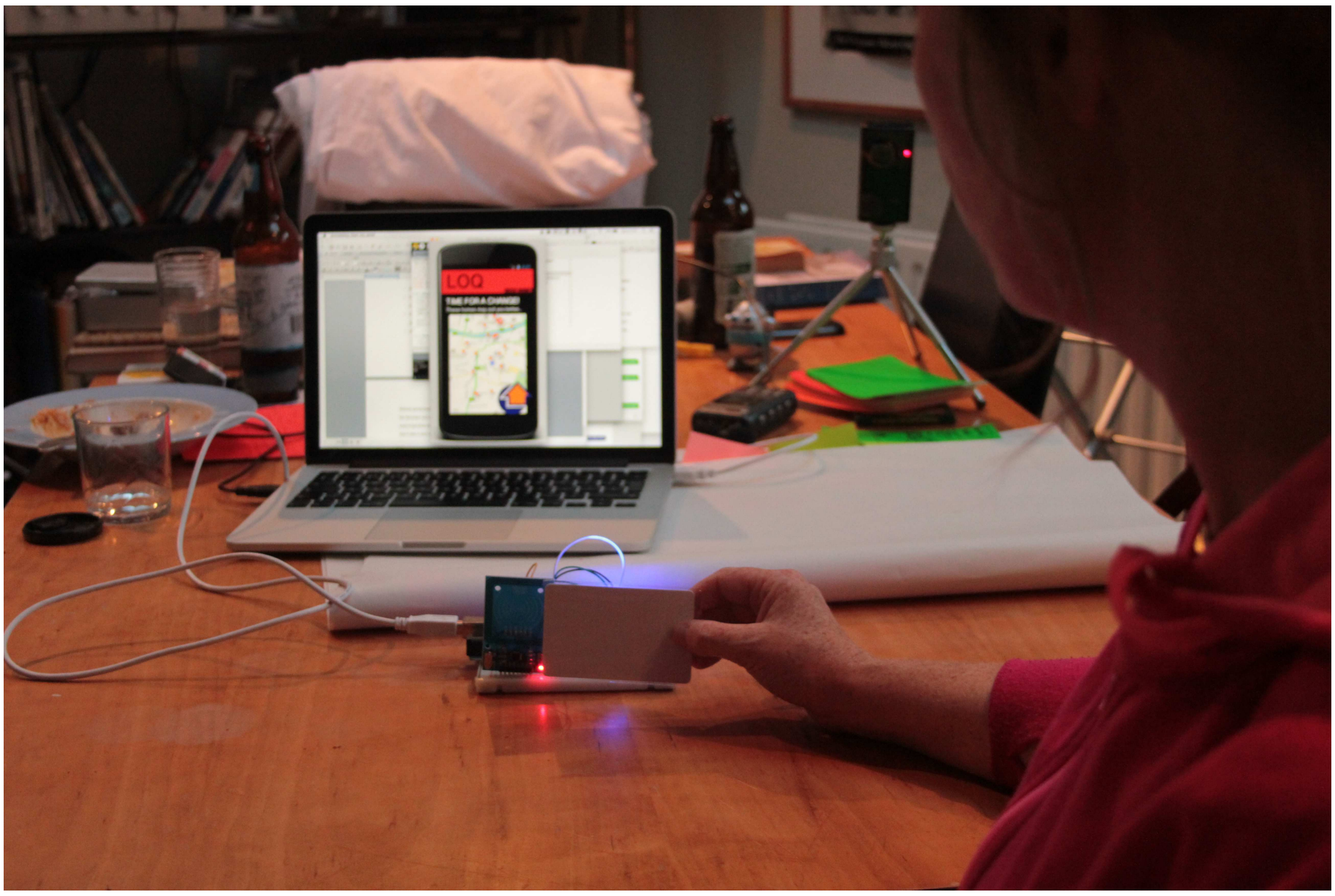

**Figure 22: Participant using the hardware prototype**

As a means to involve the participants more directly with the experience of using the LOQ, and to help with a role-playing aspect, the hardware prototype was introduced. This comprised of an RFID reader that was connected through an Arduino to a laptop. Participants could use RFID-embedded cards (representing smartphones) to engage with the LOQ. Responses from the LOQ were then displayed on an adjacent screen (See Fig. 22).

The first section (Renovator & Renter) of this workshop was structured as follows:

- A short recap of the LOQ, using the hardware prototype to remind the participants of how the LOQ would function and what kind of responses it might give.
- Engaging the participants in a short discussion about possible stakeholders involved in the LOQ. The example of an iPhone was used to get the conversation started; where stakeholders from Apple, phone resellers, service providers and repair services led all the way to mining companies who extracted minerals from the earth as raw materials for the iPhone's electronics.

- When it was felt that the participants had grasped the idea of stakeholders, they were given small post-it pads on which to write out their ideas of who or what might have a stake in the LOQ (See Fig. 23).
- After ten minutes of this activity, the participants were asked to place their stakeholders on a large sheet and to introduce them.
- As a break from the structuredness of the workshop up to this point, the participants were left to look over the keyword and mapping exercises from the previous workshop and to have a general discussion about the LOQ. The hardware prototype was also available to help trigger talking points.
- Roleplaying exercise: Once again the dice were introduced. Participants rolled dice on top of the sheet of stakeholders. Where the dice landed would suggest a stakeholder, and what number the dice rolled suggested a type of response from the LOQ (Red – 1 or 2, Orange – 3 or 4, Green - 5 or 6). This exercise was intended to make the participants try to create links between disparate stakeholders and the realities of the LOQ.

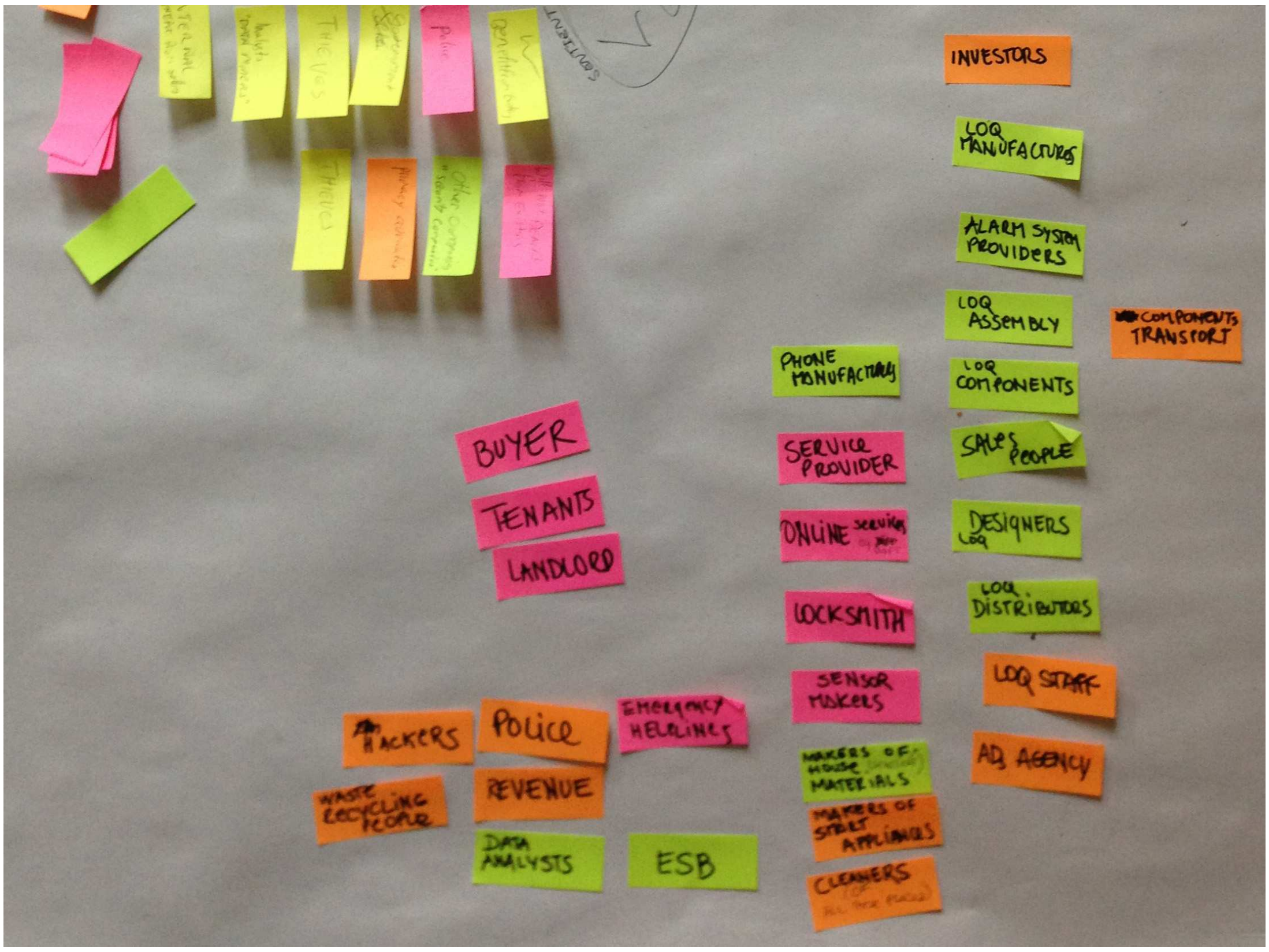

**Figure 23: Participants exploring stakeholders**

The second section (with the Homeowner & Artist) of this workshop was structured as follows:

- The participants were asked to explain to the facilitator, and each other, what they understood of the LOQ, and how it might affect them. This served to reassure the facilitator that the controversies and wider ramifications that had been raised in the previous workshop had been appreciated.
- The hardware prototype was reintroduced and the three LOQ responses were discussed. The participants were each asked to think of the kind of person and life situation that might produce these responses.
- A sheet with three blank smartphone screens printed on it was given to each participant. One for each LOQ response – Red, Orange and Green. They were then asked to use the blank screens to write or draw alternative responses from the LOQ. Any suggestions from serious, to humorous, to harsh or kind would be valid (See Fig. 24).

- Finally, a Lego model of an apartment/house was used to ‘walk’ through the various rooms and to prompt suggestions for other technologies (existing or speculative) that might be found in the home (See Fig. 25).

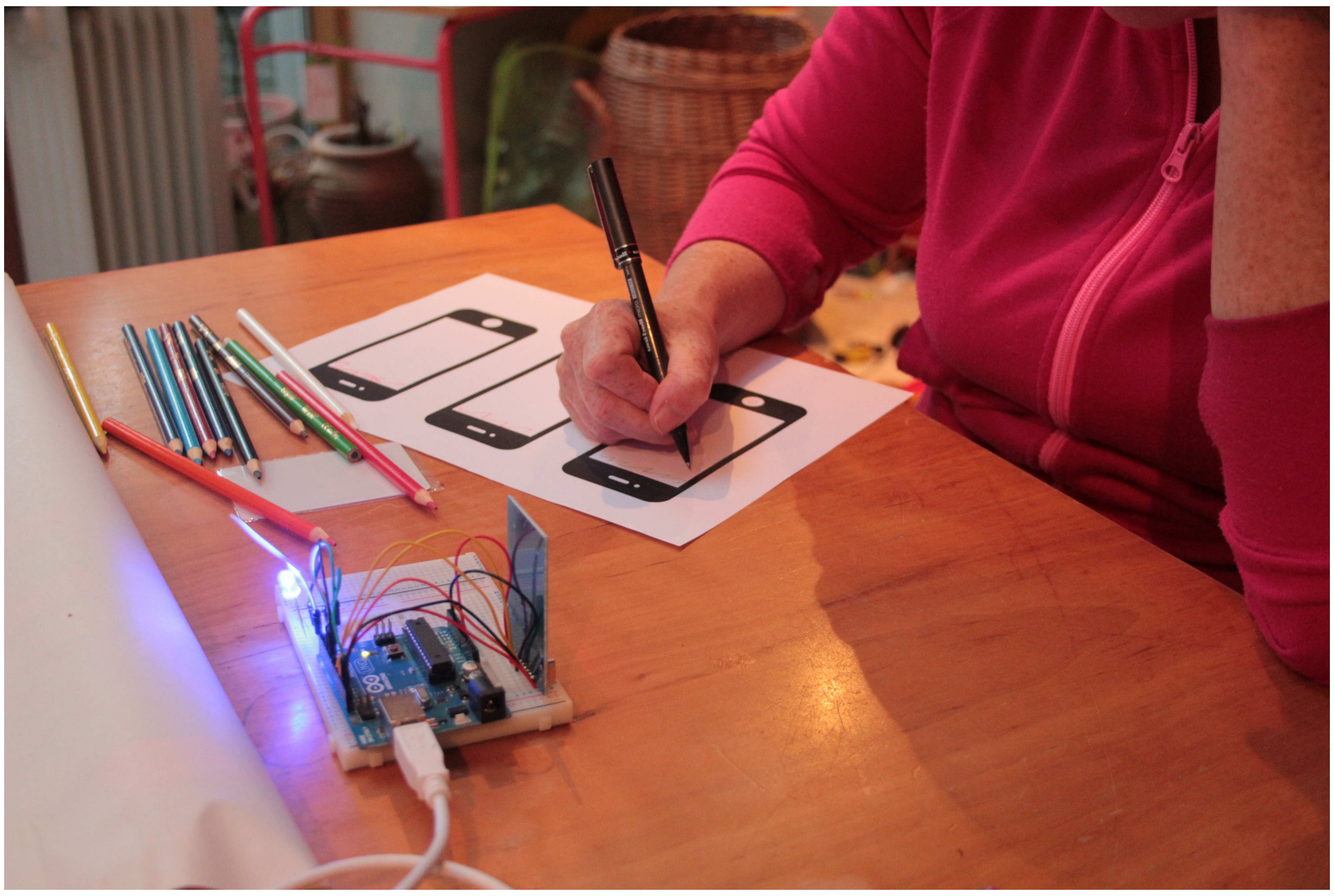

Figure 24: Participant suggesting LOQ responses on blank smartphones

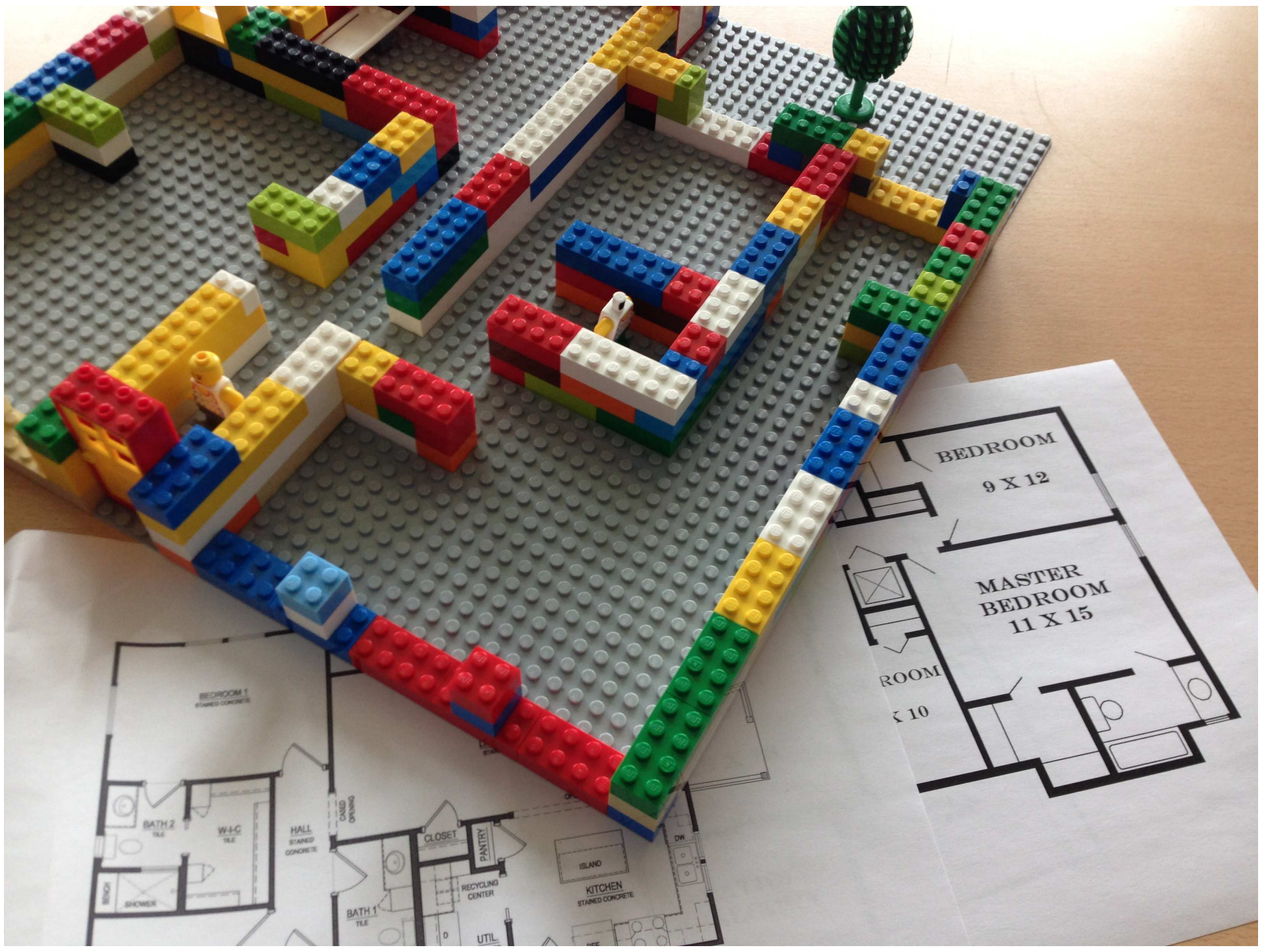

Figure 25: Lego house for exploring other smart technologies

## 8.2 Evaluation & Outcomes

Although the previous workshop provided ample material for this project, this workshop was a good opportunity to try out some unresolved activities and to gain a sense of how effective the presentation of the LOQ concept had been for the participants. Also, having a general time constraint due to having the second pair of participants arriving at a certain time, helped to move things along and ensure that any discussions could be finished before they became too broad.

Having the hardware prototype that could be interacted with was especially helpful for quickly engaging the participants and in helping them recap what they understood of the LOQ. Although the video prototype provided plenty of information to start a discussion, the interactivity of the hardware was felt to provide more investment of the participants in the design. Rather than the process being an exercise in imagining the realities of the LOQ, participants were able to appreciate how the interaction would work and how they would feel about the responses on the screen.

The introduction of the idea of exploring stakeholders was an attempt to get the participants to explore wider effects of the LOQ, beyond themselves and beyond the more immediate stakeholders that had been mentioned in Workshop 2. This was also a chance to generate further data points and nodes for possible mapping exercises in the future. This exercise was deliberately kept short and light hearted as the participants found it a little broad and the facilitator did not want to make them feel imposed upon. In future workshops, it was felt that the facilitator could draw their own actors from participant discussions and draw attention to them in a way that might encourage participants to begin pointing out these actors themselves as the workshop progressed. However, even though this activity was a struggle and provided only a small amount of data, it was still seen to be a good starting point for using a simple mapping that could be produced in further workshops to help initialise more discussion in this area.

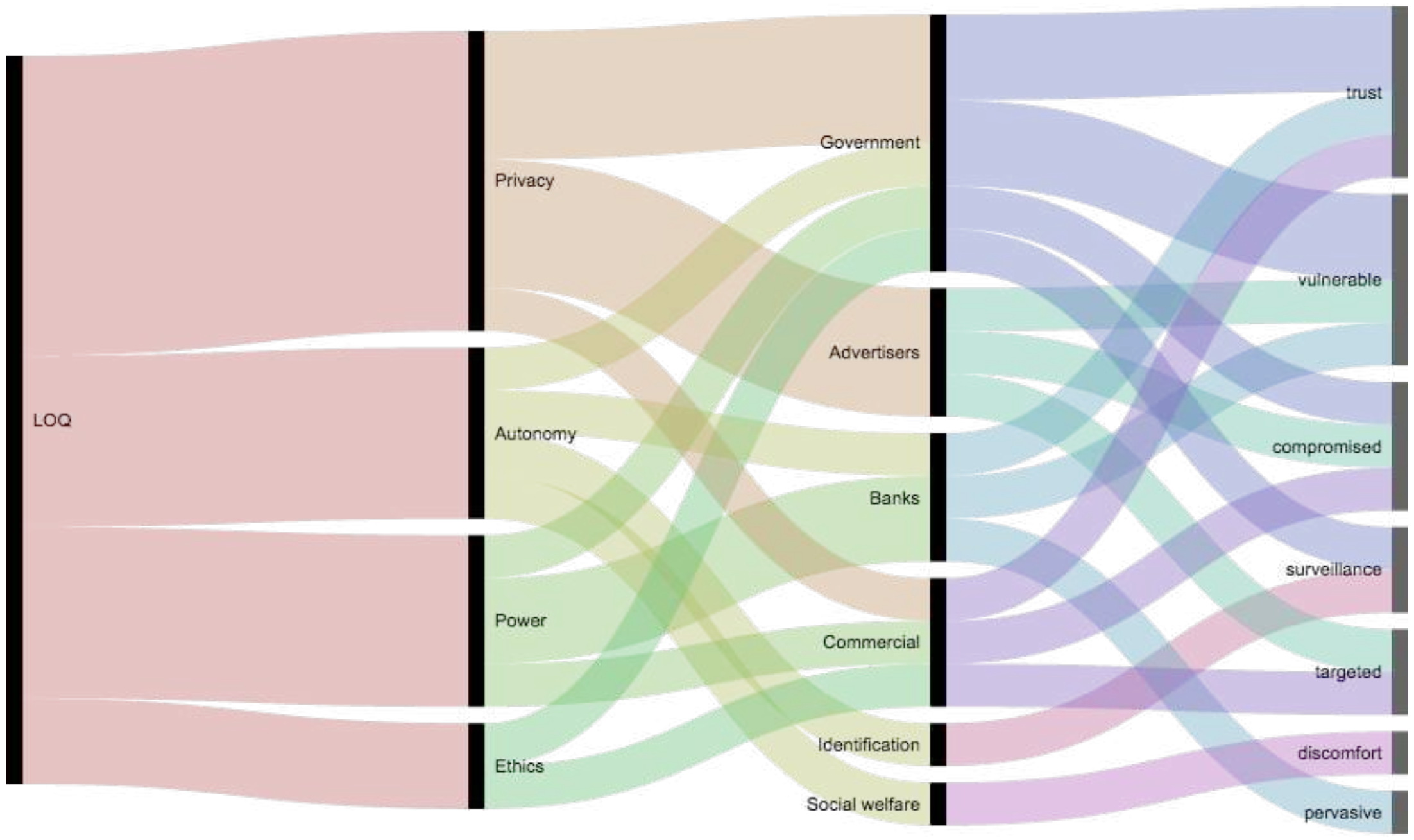

**Figure 26: Alluvial Mapping Diagram of perceived controversies, actors and emotions**

Allowing room for discussion, at this point, helped to relax the participants and bring them back into more familiar areas of discussion. Although the conversation generally recalled the pros and cons of the LOQ that had already been explored, some interesting ideas and questions were raised:

- How could the LOQ be avoided? By leaving the city?
- Can it be disrupted? How secure would it be from hackers, etc.?
- A ‘Freemium’ model for the poorer which would trade rent and privacy for targeted advertisements.
- Medical sensors around the house to monitor citizens’ health. Saving on healthcare by catching illnesses earlier.
- Shouldn’t everyone be planning at least a month ahead? The LOQ would enforce this. (Renovator)

An interesting comment from the Renter was that, “*before I had a mobile [smart] phone, I didn’t need one. When you incorporate these things into your life, you do not have the same amount of control*”. This seemed to resonate with Morozov’s (2013) critical view of technological convenience, and also with the spirit of the research question, of how it is becoming important for users to question their assimilation of new technology and of how it may affect their quality of life.

The final activity of the first section was an assessment of roleplaying as a possible tool for the workshop. This activity started well, but came to an impasse very quickly. The impression that the facilitator had was that such an activity appealed to some participants more than others. The Renter, who came from a creative and artistic background, was more receptive, but the Renovator, who had a more technical and practical background, did not feel comfortable. Overall, both participants did not feel that it was relevant to their understanding of the LOQ and that having made use of the hardware prototype helped them to appreciate others' perspectives more effectively.

The next section began with the arrival of the Homeowner and the Artist. This provided an opportunity for the group to talk amongst themselves about their own understanding of - and reactions to – the LOQ in practice. The facilitator gave their full attention to the participants' conversation until it was felt that the group had a good appreciation of each other's perspectives and of the intent of the LOQ.

At this point, each participant was given a random RFID card which would generate a different response from the LOQ hardware prototype and asked to describe what kind of personal situation might engender such a response from the device. Although the facilitator was actually looking to the participants to suggest effective personalities to be associated with the RFID cards (for public exhibition), this turned out to be an effective way to bring roleplaying into the workshop process and also flowed into the next activity.

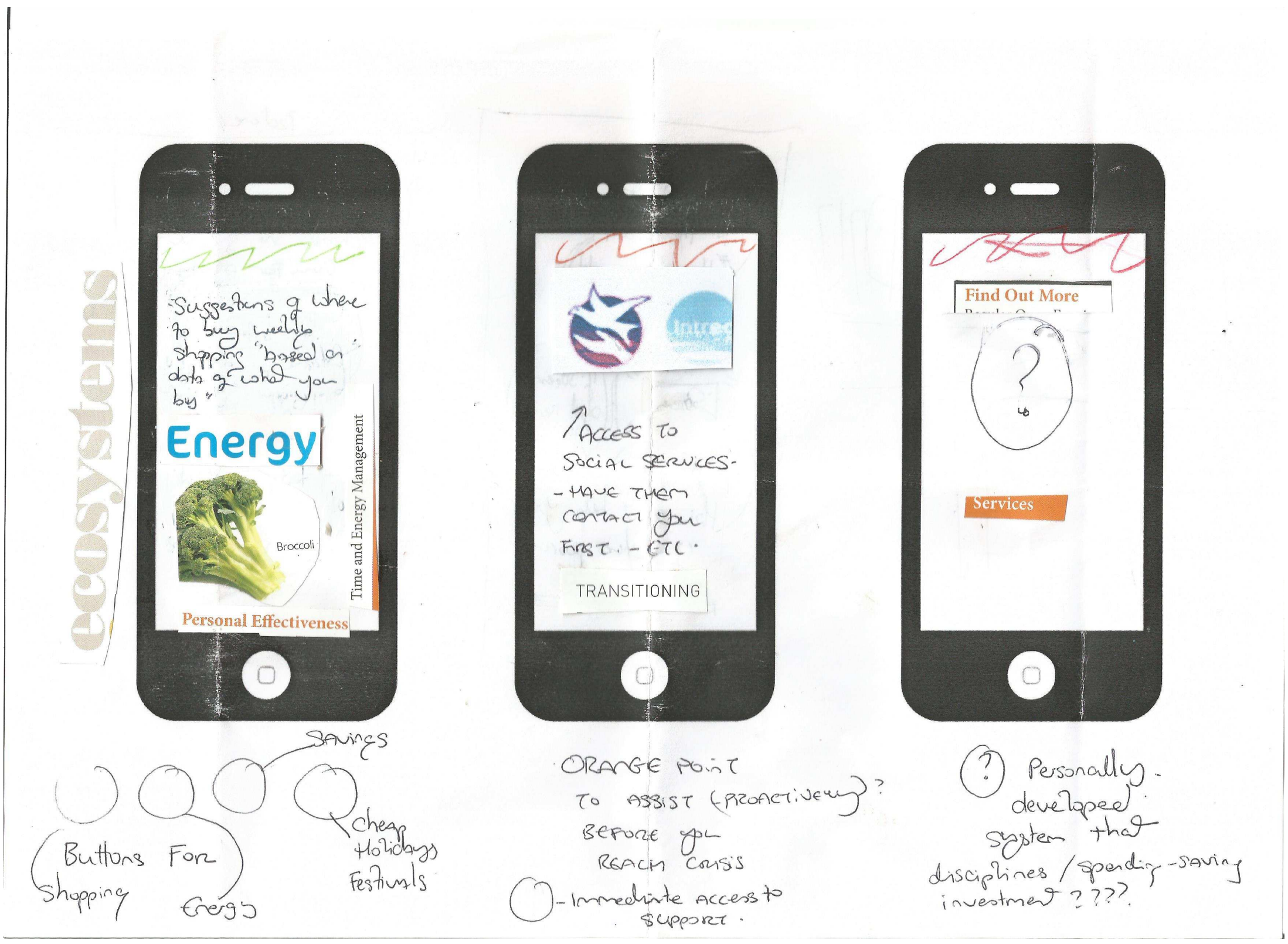

**Figure 27: Participant's response to alternative LOQ interface**

Having placed themselves in the position of affected users of the LOQ, each participant was given a sheet of paper with three blank smartphones on it and asked to spend a little time to come up with alternate responses from the LOQ interface. Some produced immediate responses, but others asked to take them home with them and think about it. As well as being good feedback for how people felt emotionally about the LOQ's 'personality', these responses also served to suggest a richer variety of possible outcomes from using it (See Fig. 27).

In an effort to keep the group focused, they were informed that there would be one last short activity. The facilitator introduced the Lego model of a single storey apartment, which was based on a generic floor plan. The facilitator moved a toy figure through the house and asked the participants to speculate on any smart or networked technology that they would hope or expect to be found in each area. Within this activity, the facilitator was looking for novel areas of the sentient house to explore in future iterations.

**Examples of responses:**

Living Room:

- Automatic temperature controller: Using the LOQ to assess how many people were in the house, the heating could be adjusted for occupancy of different rooms.
- Voice activated appliances: Lighting and volume controlled by stress levels in occupants voices.
- The Google Nest thermostat was also mentioned as a contemporary example of networked, smart technology.

Bedroom:

- A clothes advisor: a device that tracks all of the users clothing and can provide recommendations for what to wear and how to match various pieces. Also a humidity sensor to keep clothes in good condition.
- Wireless phone charger embedded into various pieces of furniture. Allowing the user to place his phone near or further away from the sleeping area.
- Automatic lighting and dimmable windows that would delay or advance dawn and dusk, depending on the season. This would help maintain sleep patterns.

Bathroom:

- An automatic tracking monitor for supplies: Supplies, products and medicine.
- A toothbrush that would notify the user when it needs to be changed.
- Hot water controller that would adapt to the users' habits.

Outdoors/Garden:

- Automatic insect repellent.
- Smart windows which would open and close to adjust the inside atmosphere.
- Automatic lawn mower.

## 9. CONCLUSIONS

In conclusion, this thesis hopes to contribute some new knowledge to the generally subjective area of designing for discussion and to engaging the public in the approaches of Critical and Adversarial Design. It has helped to form a framework for the artist-designer to check in with the very public that may be influenced by their designs; to ensure that they are both understood and conducive to debate, a sharing of perspectives and to a questioning mind-set. On a personal level, it has moved this designer from a position of isolated production into more participatorial and socially engaged directions, which can only be beneficial for such public-facing design approaches.

As technology pervades our everyday lives and infrastructure, it is becoming more important for some design approaches to set themselves in a position to reveal and question the wider ramifications of this, and of how an informed public can make their own decisions on how to receive or amend these advances. It is hoped that this thesis has, at least, provided some inspiration and direction towards this approach.

## Design

Although Participatory Design methods do seem to promise a possibility of user designed artefacts, this project showed that the responsibility for designing these initial critical artefacts is on the designer's own shoulders. What is meant by this is that, although the designer may use the public to generate areas of controversy in which to focus their design interventions, it is felt that the they should then parse and filter this data through their own knowledge and experience to produce the initial critical artefact.

At this point, the focus-group approach of the first workshop can then be leveraged to gain impressions from an impartial public and to see if the designer's message is being relayed and appreciated effectively. This approach can be used iteratively to sharpen and enrich the design to more effectively raise issues about its implementation in the real world.

## Workshop Structure

As a newcomer to the workshop approach, this designer used the thesis project as an opportunity to try out a wider variety of approaches and activities than would usually be attempted. Working with willing and supportive participants, this gave the designer some leeway in being both experimental and productive. As a result, activities that were more effective in creating a reflective and questioning atmosphere amongst the participants could be kept and improved; and less effective approaches could be discarded or reassessed for more appropriate applications. In the example of the role-playing activity, it was seen that a combination of unreceptive participants and awkwardness in the facilitator served to undermine the activity, but the use of the hardware prototype produced similar results with better engagement from all participants.

Overall, the workshop and focus-group approaches provided a revelatory and thought-provoking amount of feedback; both on the design itself, the controversies surrounding pervasive and authoritative technologies, and on the participants ability to learn to question and explore new concepts and realisations in unfamiliar territory.

## Mapping

The mapping exercises during this project have shown themselves to be, potentially, an excellent method of tracking and displaying the outcomes of such subjective discussions and controversies. This has been an experimental approach that has allowed the designer to experiment with a wide variety of visualisation methods. In retrospect, with such a narrow pool of participants during the workshops, the scale and detail of this mapping could never be representative of the public at large.

On a more optimistic note, the act of keeping the intent of mapping in mind, helped to focus the facilitator on establishing and maintaining a – subjective – record of controversies, actors and emotions throughout the workshops. Notwithstanding, the importance of mapping throughout any similar project has been appreciated and will be implemented in more effective ways in future. It can be seen as an activity that serves as a resource for the designer and a visual record of progress for participants and other interested parties.

## Further Prototypes

Using experience of rapid-prototyping, a number of prototypes will be explored as a result of participant feedback from the workshops:

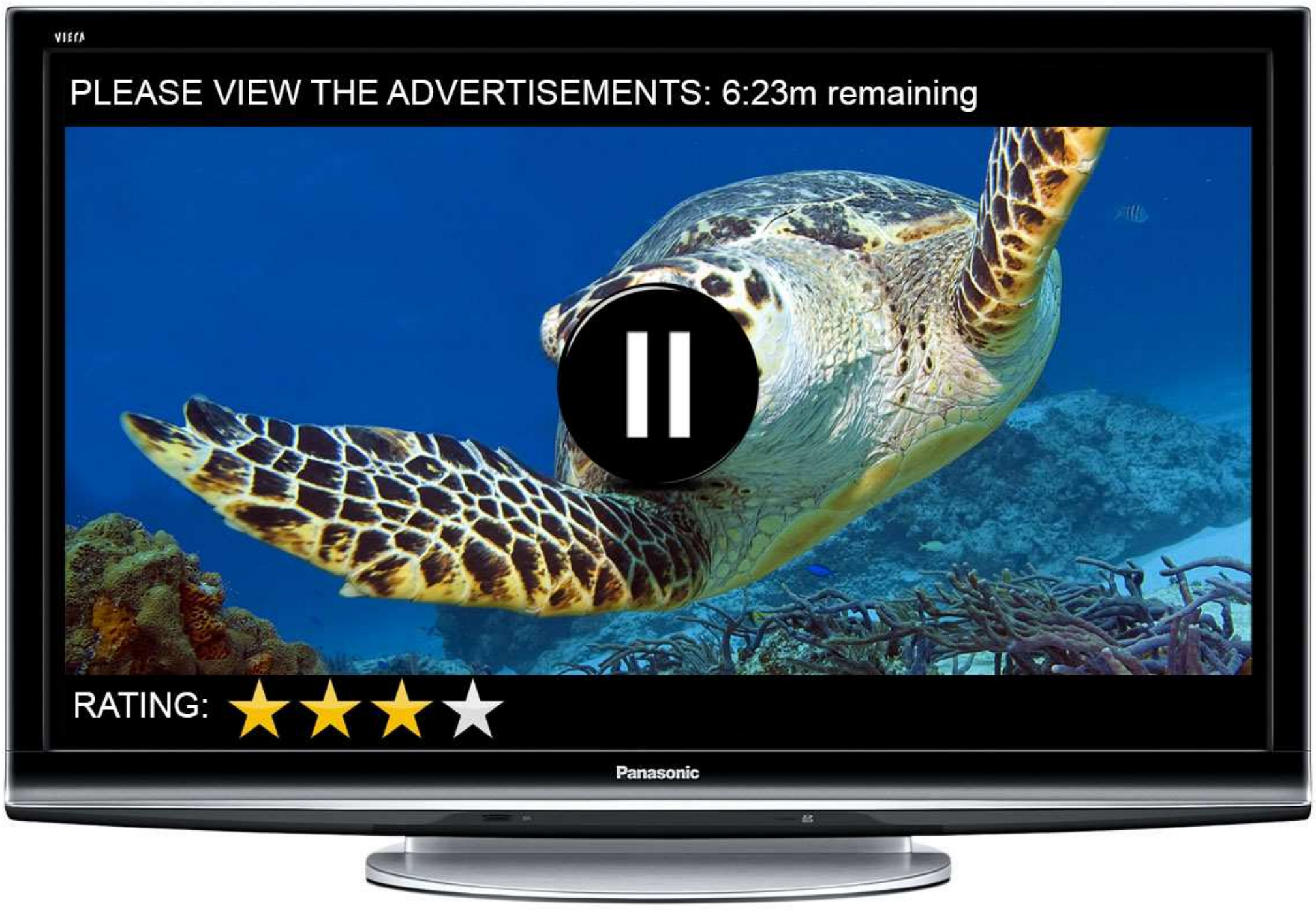

**Figure 28: Advertising TV mock up**

Enforced Advertising TV: A television that works in conjunction with the LOQ ensuring that advertisements take priority over content. This will be a TV with face-detection technology that obliges the user to watch all advertisements by pausing when the user looks away. The user's relationship with the LOQ (ability to pay rent/mortgage) will decide how much advertising must be watched before normal content is restored. This idea was inspired by the participants' suggestion of ways to offset payments by opening themselves up to more invasive tracking and advertising.

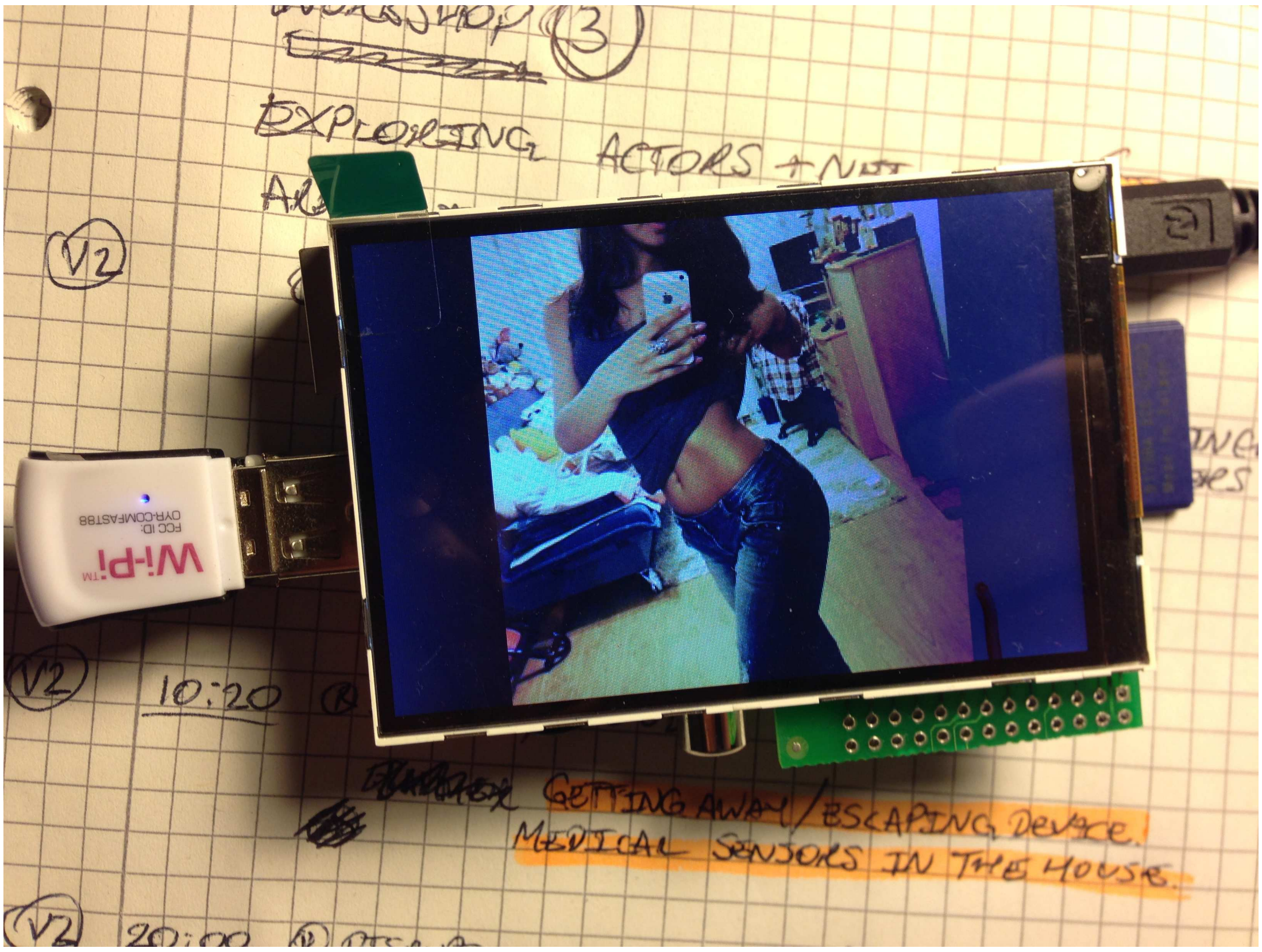

Figure 29: Picture Frame prototype crawling social media

Picture Frame: A series of picture frames, scattered about the house, which would display images from the occupants' social media feeds. This would serve as a constant reminder that all integration of networked technology often results in compromises in the user's privacy and trackability. This artefact was inspired by the participants' concerns about privacy and how intangible it is.

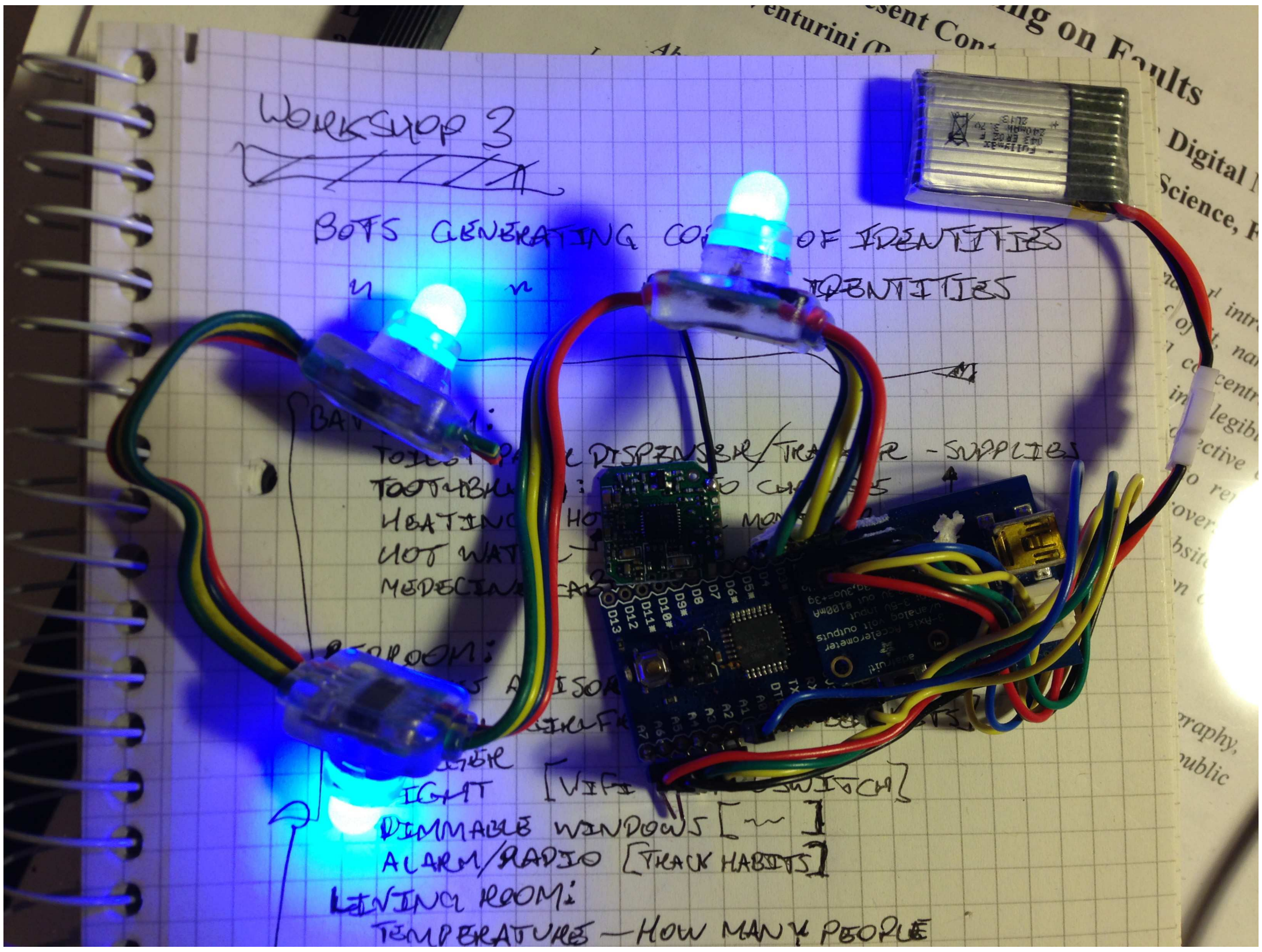

Figure 30: Proof of concept 'Switch Off' prototype

Switch Off: A portable device that allows the user to disable all connected devices in a 10-metre radius. As part of a reward system based on how the LOQ responds to the user, they will be allowed to actively disconnect from the internet and smart city for periods of up to 30 minutes. This is an empathetic response to the participants' weariness of 'digital noise' in their environment, and a way to raise questions about self-discipline in the networked society.

## Reflection & Future Directions

After the dust settled from this project, a clearer perspective of the process became discernable. In its relatively short period of time, much experimentation was had, and many things were learned about design, the public and the designer themselves. In the following paragraphs we will take a look at, and reflect upon, the lessons learned and where this approach can be directed and improved upon.

At the start of this project, it was felt that the designer needed to produce the first prototype themselves. This was seen as a means to begin what was to be a series of workshops which attempted to bring a sense of participation into the field of Critical Design. In retrospect, this approach fundamentally kept the project at a distance from Participatory Design and constrained the process for both the designer and the participants. To honour the concept of bringing Critical and Participatory Design together, the next iteration must seek to enable and entrust the participants with the knowledge and skills to generate their own concepts.

Coming from a more solitary practice, this designer used the thesis project as a means to explore a more outgoing design approach and to move into areas which were previously unexplored; such as workshops, participation and public dialogue. In the case of this designer, it served to highlight their own introversion, which calls for them to open up the facilitation to a more outgoing person who can help maintain the energy within the workshop environment and allow the designer to observe and learn. This will also help the designer to put a little distance between themselves and the designs.

Moving forward, the next session of workshops will begin with a cleaner sheet. The first workshop will be an introduction to Critical Design, with an exploration of examples, an investigation of what questions these designs raise and a look at how effective they can be. Through this, it is intended to sow the seeds of critical design thinking within the participants and to then take them into an exploration of themes that can be focused on. At this point, we can then start to truly engage with Participatory Design to co-create Critical Designs which address the aforementioned themes.

Using the experiences, mistakes and successes of this project, a new series of workshop-based sessions are being prepared, with a stronger focus on participation, energy and the public.

# Appendices

## A: Workshops

### Workshop Introduction Transcript

*“My name is Rob Collins and I am currently studying for a Masters in Interctive Media at the University of Limerick.*
*Today, I will be introducing a design which we will be exploring and discussing through this workshop over the next hour.*

*I am here to facilitate this exploration and to clarify any questions or misunderstandings which may arise regarding the design. On the other hand, I am not here to influence you, but rather to ensure the smooth running of the workshop and to allow yourselves to lead the discussion.*

*I am going to ask you to watch a short video about this product, after which I will do a short presentation; during which you can ask questions to help clarify your impressions about the product’s functions.*

*After this, you will be asked to take part in a series of activities and explorations related to the themes around this product with minimal participation from myself.”*

**Presentation Screenshots**

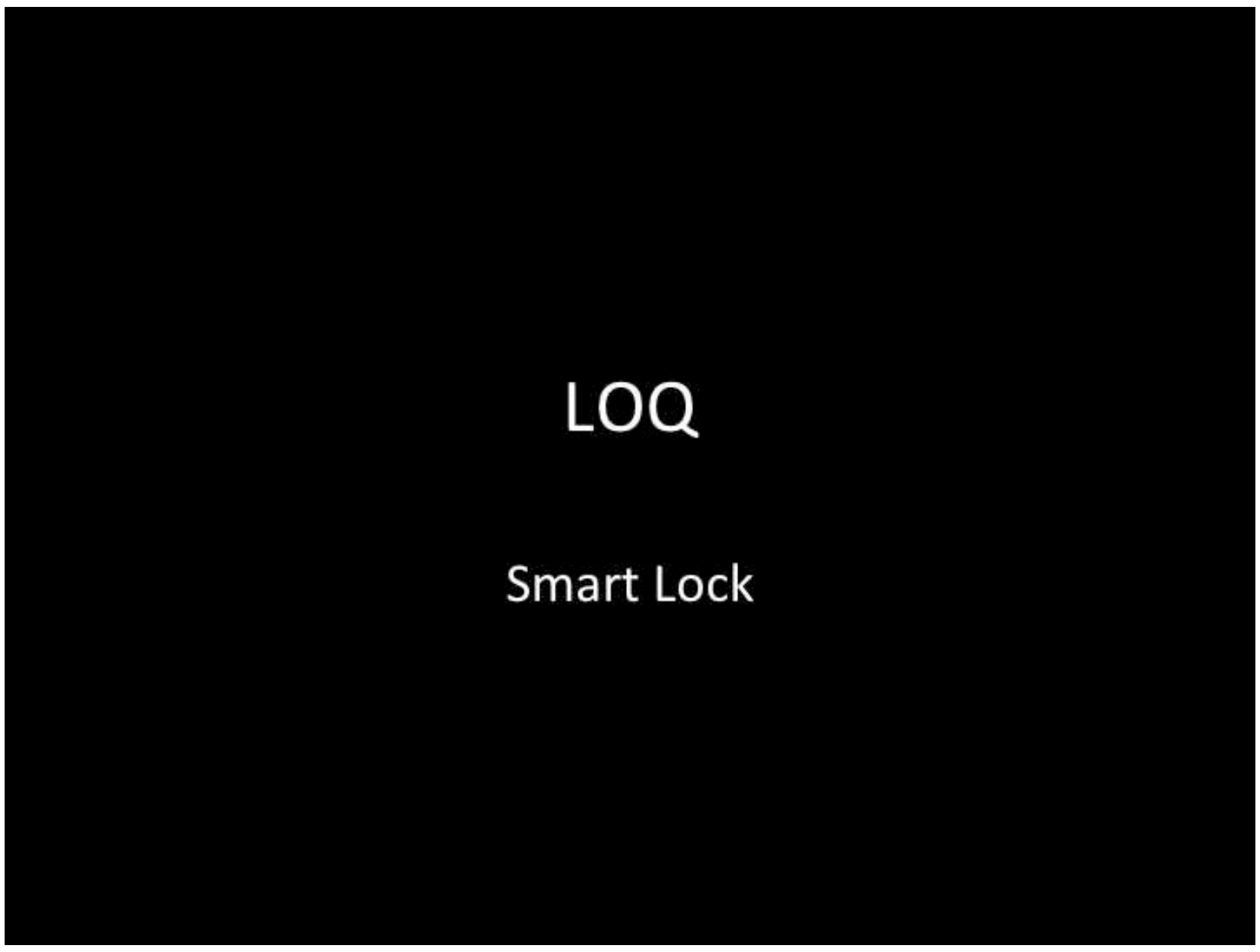

Figure 31: Intro Screen

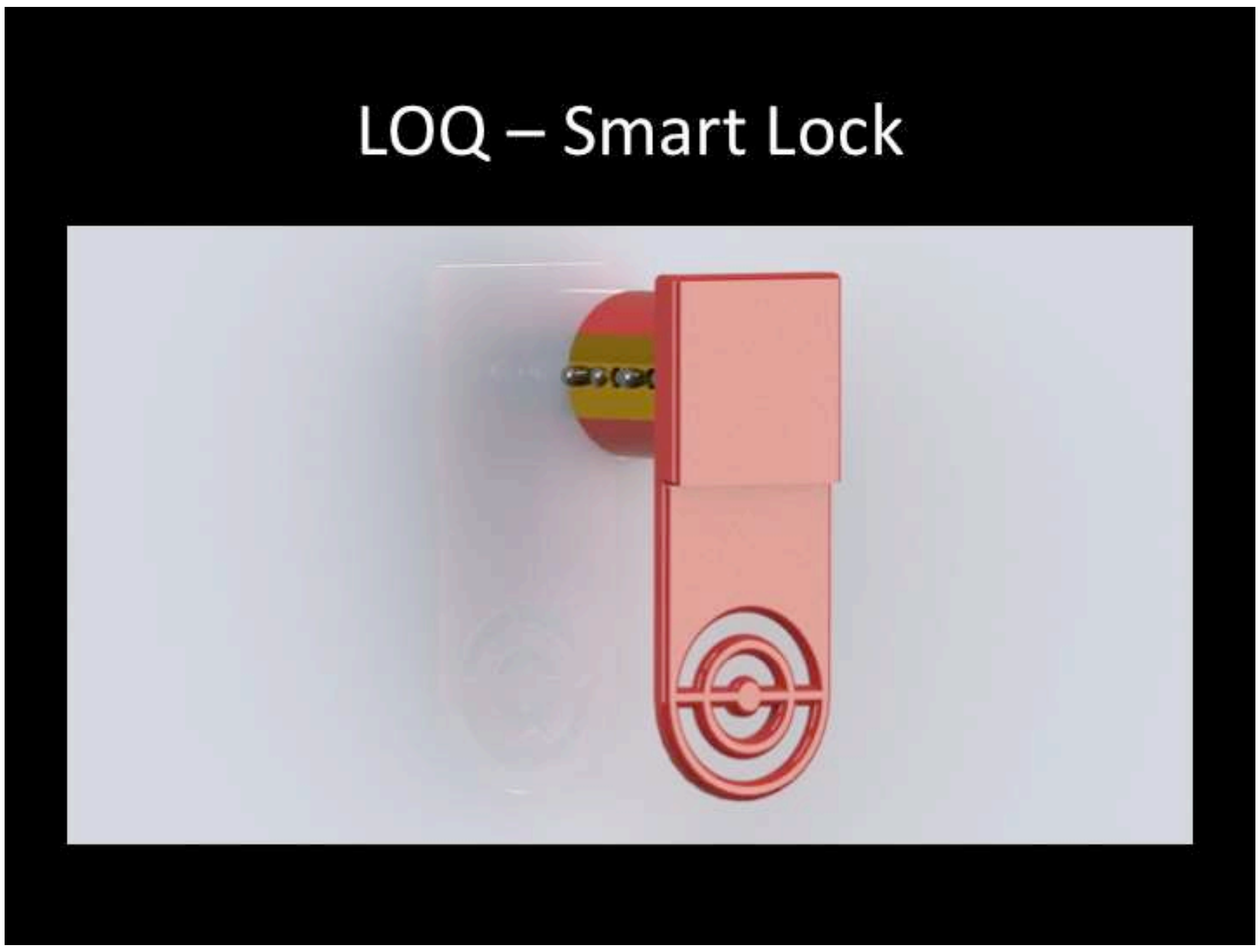

Figure 32: Introducing the design with 3D rendering

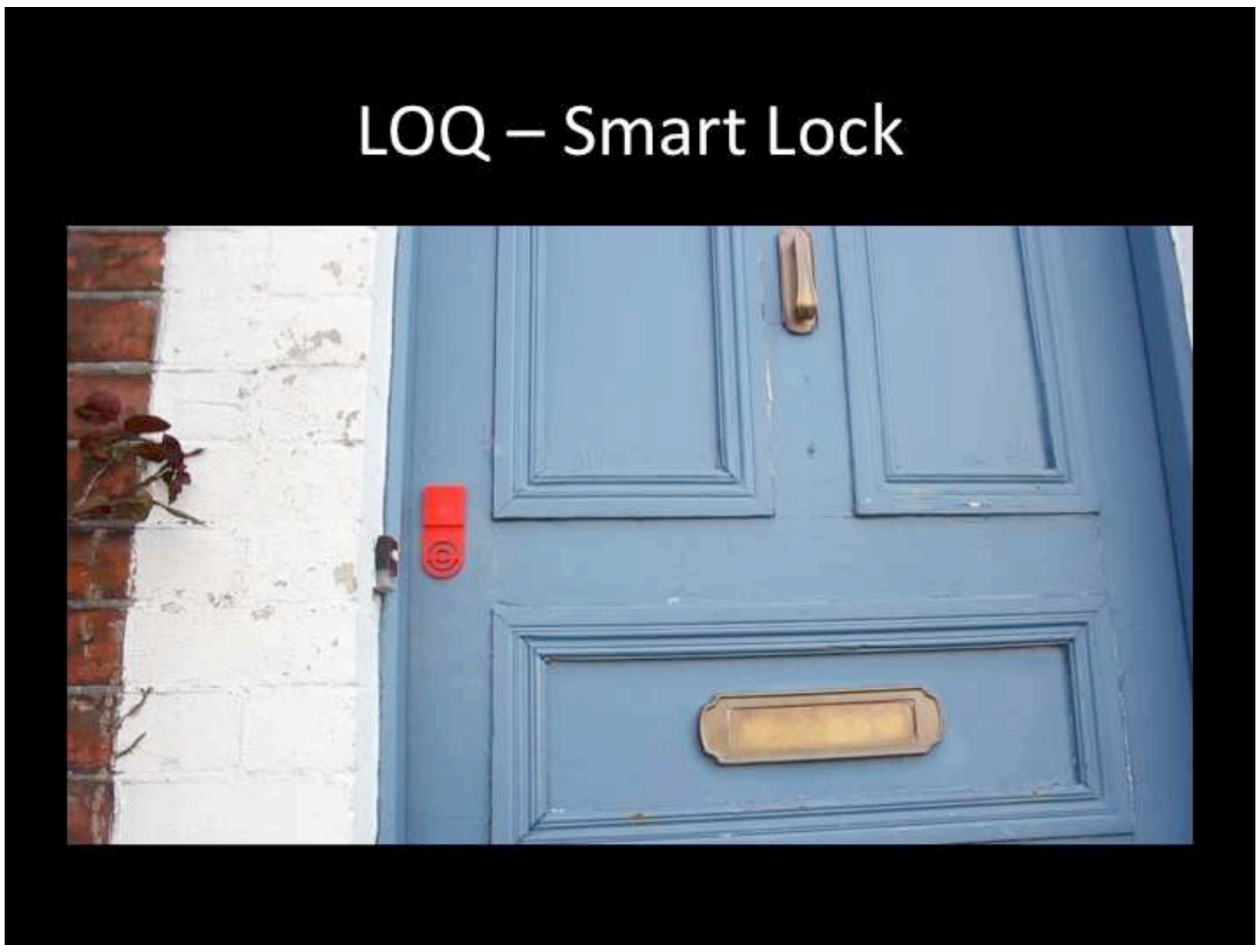

Figure 33: LOQ installed

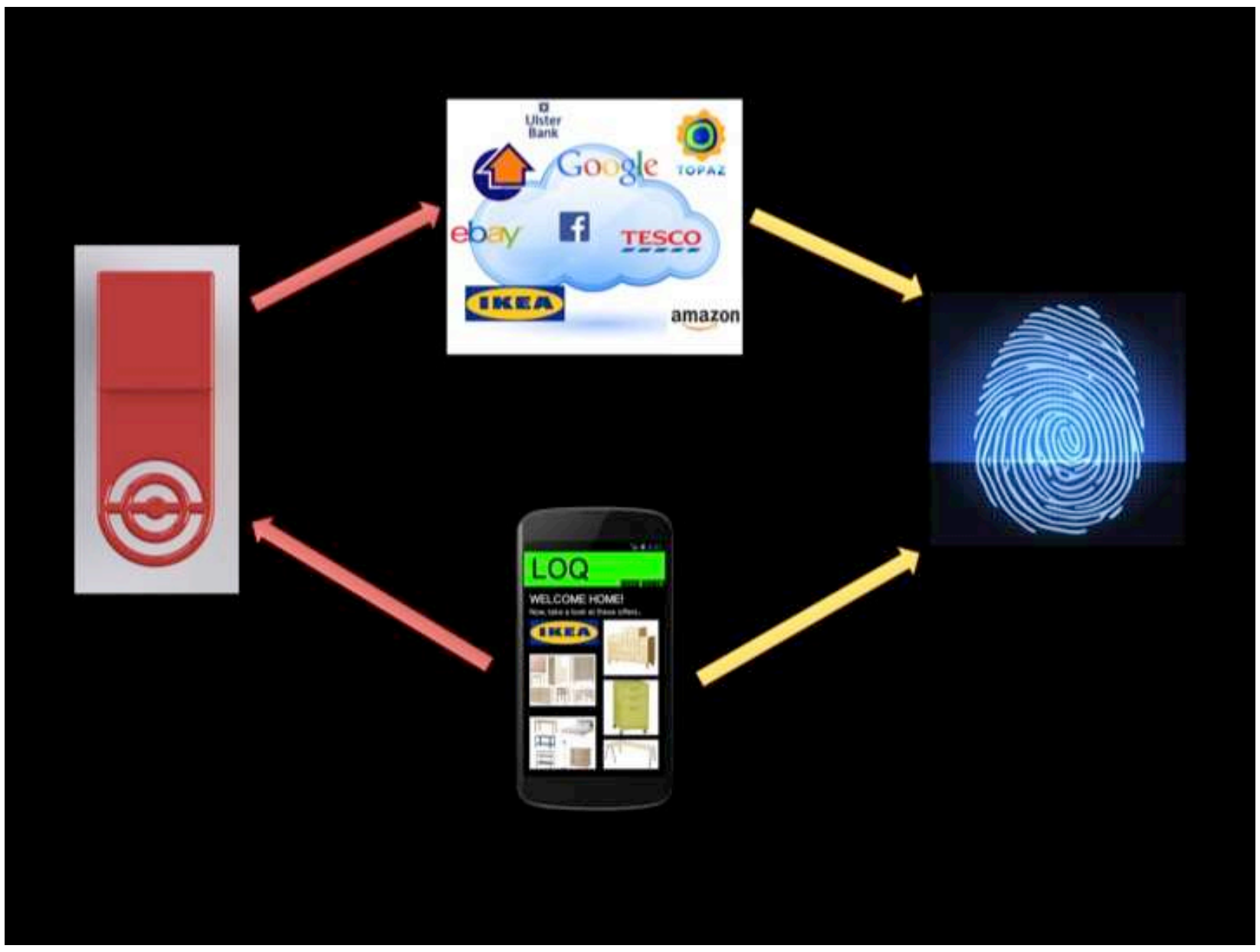

Figure 34: Illustration of how the LOQ relates to the phone and the cloud

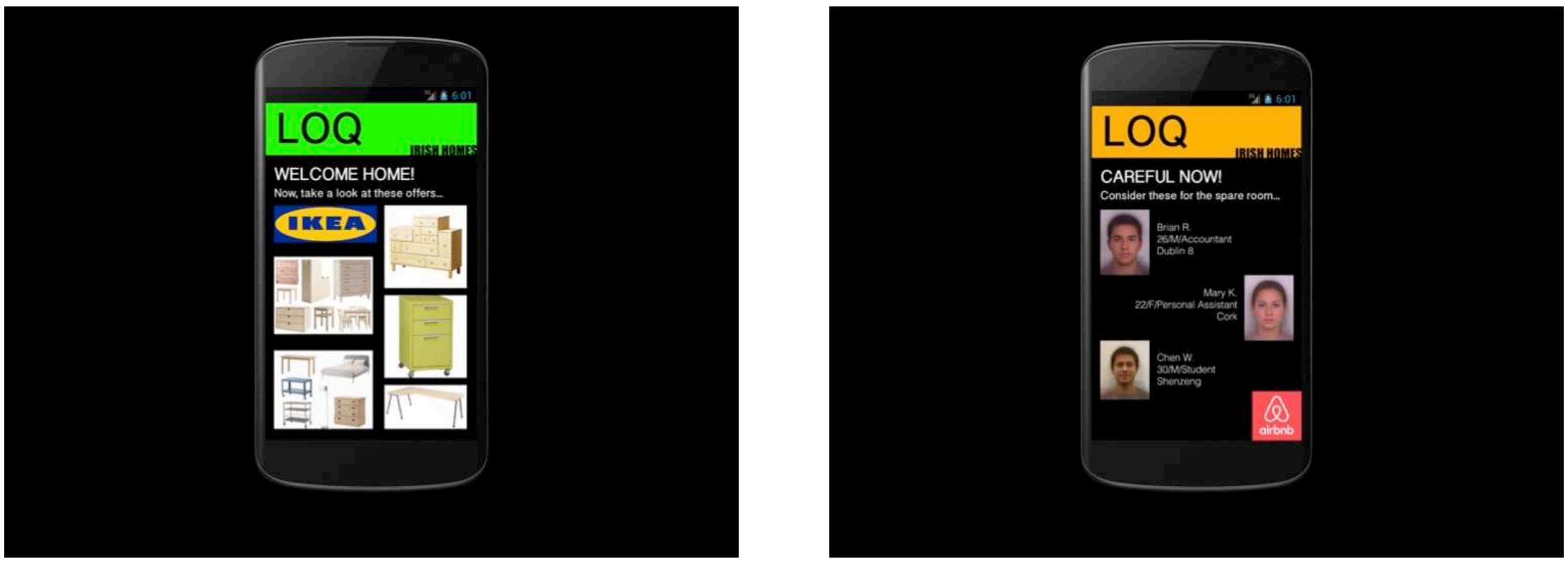

Figure 35: Possible smartphone outcomes

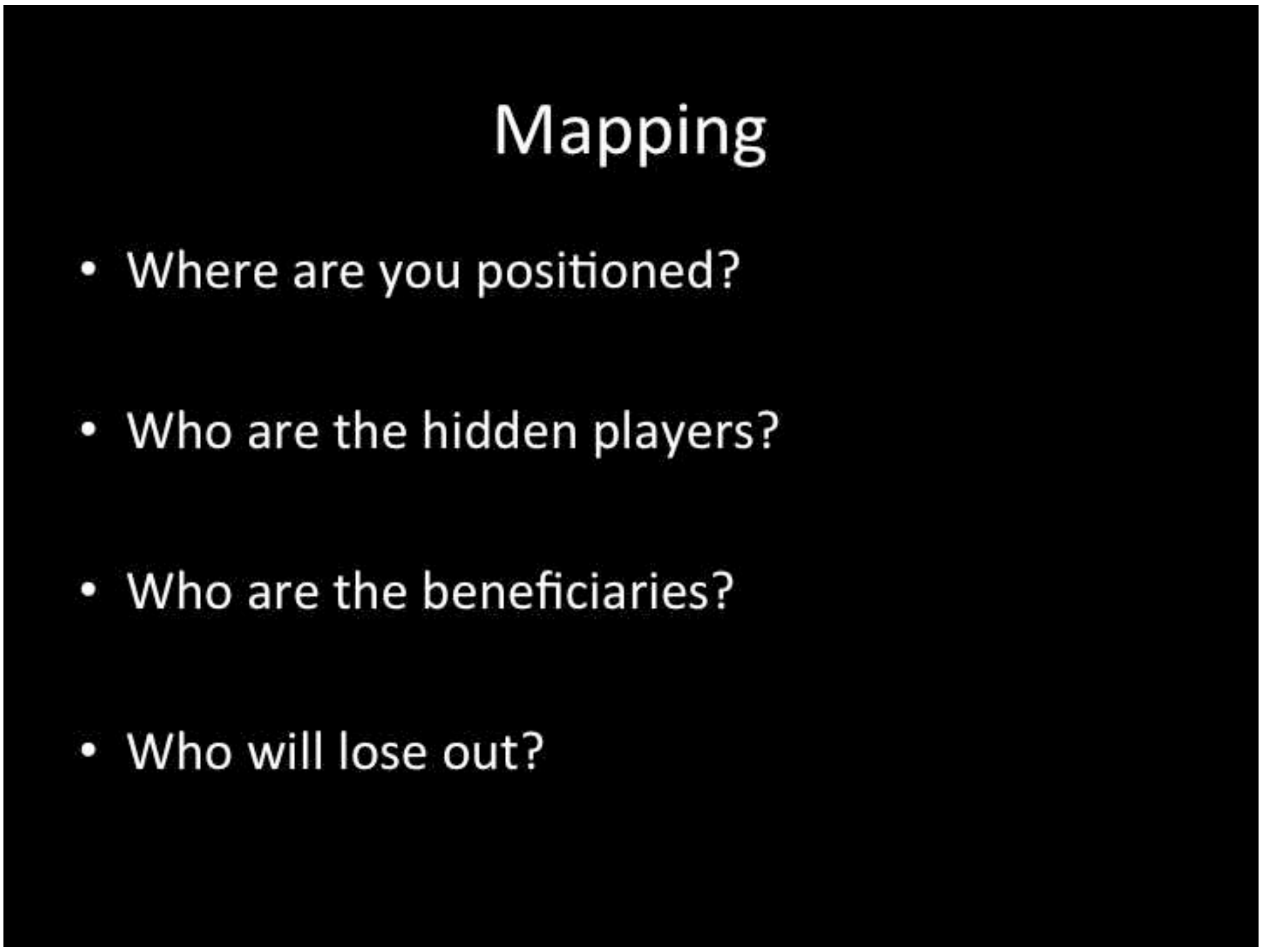

Figure 36: Thinking points for participants

## B: Code Example

**Fractal Visualiser Functions for first survey(Flash):**

Based on Emotion Fractal source by Jared Tarbell (levitated.net)

```
	// fractal function for keywords
	function fillFrame(x0, y0, x1, y1, d) {
		// pick one of the words and place
		n = random(wordArray.length);
		word = wordarray[n];
		// place the word object
		nombre = "word"+String(depth++);
		neo = this.attachMovie("mcWord", number, depth);
		// set word
		neo.setWord(word);
		neo.fitInto(x0, y0, x1, y1, d);
	}
	function fillFrameRequest(x0, y0, x1, y1, d) {
		// request has been made to recursively fill a region
		// only allow if reasonable
		rWidth = x1-x0;
		rHeight = y1-y0;
		if ((rWidth>2) && (rHeight>2)) {
			addFillRequest(x0, y0, x1, y1, d+1);
	//		fillFrame(x0, y0, x1, y1, d+1);
		}
	}
	function addFillRequest(x0, y0, x1, y1, d) {
		// queue up a request to fill a region
		var freq = {x0:x0, y0:y0, x1:x1, y1:y1, d:d};
		fillRequests.push(freq);
```

}

## C: Video Prototype Transcript

*"Your home is an extension of yourself.*
*It is the space that you inhabit.*
*It is the space in which you can be yourself.*
*It's the place where you feel secure.*
*In post-recession Ireland, your government wants you to be in the right place for your life and your dreams.*

*Introducing LOQ.*
*LOQ is the new interface for your home.*
*Integrating your home into the 'Internet of Things', LOQ provides you with an individual identity to access your home.*
*Based on your own habits within the 'smart city' and across the internet, LOQ integrates with your smartphone to give you the home that you deserve.*
*Basec on your lifestyle, your choices and your financial situation, LOQ will manage your ownership and accommodation needs.*

*As your fingerprint grows, LOQ will adjust your living arrangements to your lifestyle.*
*Expansion. Subletting. Downsizing.*
*LOQ will allow your home to develop and breath with your life situation.*

*LOQ will be rolling out to all registered houses across the country early next year.*
*Register now for smartphone grants and a choice of seven individual colours.*
*Irish Homes, a future for you."*